\documentclass[opre,nonblindrev]{informs3}

\DoubleSpacedXI % Made default 4/4/2014 at request
\OneAndAHalfSpacedXI % current default line spacing
\usepackage[flushleft]{threeparttable}

\usepackage{amsmath}
\usepackage{bbm}

\usepackage{setspace}
\usepackage{nomencl}

\usepackage[fleqn]{nccmath}
\usepackage{graphicx}
\usepackage{amsfonts}

\usepackage{apacite}
\usepackage{multirow}
\usepackage{tablefootnote}
\usepackage{placeins}

\usepackage{array}
\usepackage{booktabs}
\usepackage{makecell}
\usepackage[utf8]{inputenc}
\usepackage{blkarray}
\usepackage{hyperref}

\usepackage[ruled,vlined]{algorithm2e}
\usepackage[flushleft]{threeparttable}
\usepackage{algpseudocode}
\usepackage{endnotes}
\let\footnote=\endnote

\usepackage{natbib}
 \bibpunct[, ]{(}{)}{,}{a}{}{,}%
 \def\bibfont{\small}%
 \def\bibsep{\smallskipamount}%
 \def\bibhang{24pt}%
\TheoremsNumberedThrough     % Preferred (Theorem 1, Lemma 1, Theorem 2)
\ECRepeatTheorems

\EquationsNumberedThrough    % Default: (1), (2), ...

\begin{document}
%\doparttoc % Tell to minitoc to generate a toc for the parts
%\faketableofcontents % Run a fake tableofcontents command for the partocs
%\part{} % Start the document part
%\parttoc % Insert the document TOC
\bibliographystyle{apacite}

%\RUNAUTHOR{Ding, Ao, Duenas-Martinez, and Magnanti}

% Title or shortened title suitable for running heads. Sample:
% \RUNTITLE{Bundling Information Goods of Decreasing Value}
% Enter the (shortened) title:
\RUNTITLE{Decision-dependent Robust Charging Infrastructure Planning}

% Full title. Sample:
% \TITLE{Bundling Information Goods of Decreasing Value}
% Enter the full title:
\TITLE{\large Decision-Dependent Robust Charging Infrastructure Planning and Scheduling for Light-Duty Truck Electrification at Industrial Sites}

% Block of authors and their affiliations starts here:
% NOTE: Authors with same affiliation, if the order of authors allows,
%   should be entered in ONE field, separated by a comma.
%   \EMAIL field can be repeated if more than one author

\ARTICLEAUTHORS{Yifu Ding, Ruicheng Ao, Pablo Duenas-Martinez, Thomas Magnanti}
% \ARTICLEAUTHORS{%
% \AUTHOR{Yifu Ding}
% \AFF{MIT Energy Initiative, Massachusetts Institute of Technology, USA, \EMAIL{yifuding@mit.edu}} %, \URL{}}
% \AUTHOR{Ruicheng Ao}
% \AFF{Institute for Data, Systems, and Society, Massachusetts Institute of Technology, USA, 
% \EMAIL{aorc@mit.edu}} %, \URL{}}
% \AUTHOR{Thomas Magnanti}
% \AFF{Sloan School of Management, Massachusetts Institute of Technology, USA, \EMAIL{magnanti@mit.edu}}
% % Enter all authors
% } % end of the block

\ABSTRACT{Many industrial sites and digital logistics platforms rely on diesel-powered light-duty trucks to transport workers and small-scale facilities, which results in a significant amount of greenhouse gas emissions (GHGs). To address this, we develop a robust model for planning charging infrastructure to electrify light-duty trucks at industrial sites. The model is formulated as a mixed-integer linear program (MILP) that optimizes the charging infrastructure selection (across multiple charger types and locations) and determines charging schedules for each truck based on the selected infrastructure. Given the strict stop times and schedules at industrial sites, we introduce a scheduling-with-abandonment problem in which trucks forgo charging if their waiting time exceeds a maximum threshold. We further incorporate the impacts of overnight charging and range anxiety on drivers' waiting and abandonment behaviors. To model stochastic, heterogeneous parking durations, we classified trucks using machine learning (ML) methods based on contextual and time-location features. We then constructed decision-dependent, feature-driven robust uncertainty sets in which parking-time variability varies flexibly with drivers' charging choices. These feature-driven sets are applied to two robust optimization formulations with decision-dependent uncertainty (RO-DDU), resulting in distinct outcomes and managerial implications. We conduct a case study at an open-pit mining site to plan charger installations across eight charging zones, serving approximately 200 trucks. By decomposing the problem into a short rolling horizon or using a heuristic approach for the full-year or representative-day dataset, the model achieves an optimality gap of less than 0.1\% under diverse uncertainty scenarios. \\

}%


\maketitle

\maketitle

\section{Introduction}
\subsection{Background}

Many industrial sites, such as the mining and metals sectors, and digital logistics platforms, rely on diesel-powered light-duty trucks to transport and dispatch goods. Their use has produced significant GHG emissions. In 2022, light-duty trucks were the largest GHG emission source in the U.S. transportation sector, contributing 58\% of total emissions \cite{us_epa_facts_2024}. Battery-electric vehicles (BEVs) are a key solution for decarbonization. Currently, the BEV charging infrastructure can be categorized into two types. The first type is Level 1 and Level 2 AC charging (hereafter referred to as ``slow charging''), which generally requires between four hours and a full day to charge a vehicle battery from empty to 80\% state of charge (SoC). The second type is the direct current fast charging (hereafter referred to as ``fast charging''), which has been deployed at certain public stations. Fast chargers can recharge an empty vehicle battery to 80\% SoC in about an hour \cite{us_department_of_transportation_charger_2025}. However, fast-charging infrastructure entails considerable capital investment and installation expenditures, primarily due to the required power-electronics installations. Fast charging beyond 80\% SoC is strongly discouraged for safety reasons \cite{mussa_fast-charging_2017}. Another fast recharging option is battery swapping. Battery swapping enables EVs to replace depleted batteries with fully charged ones swiftly \cite{mak_infrastructure_2013}. However, compared with adding more trucks to the fleet, deploying battery swapping at industrial sites may yield limited benefits due to operational risks, including potential battery damage during swapping.

In this paper, we develop an integrated planning and scheduling model for charging infrastructure with mixed-type BEV chargers, leveraging contextual stochastic optimization and ML techniques. We comprehensively incorporate heterogeneous BEV driving behaviors in the scheduling, including charging abandonment, overnight charging, and range anxiety. To capture variability in parking duration and its dependence on charging decisions, we construct a feature-based, decision-dependent uncertainty set using ML-based clusters and apply it to two RO-DDU formulations with different assumptions and computational costs. We conduct a case study at an open-pit mining site with 200 operational diesel-powered trucks. Without modifying the GPS data, we extract parking and driving patterns relative to potential charging locations, estimate energy consumption, and efficiently compute the charging infrastructure plan and schedules.

\subsection{Related Work}

\subsubsection{Mixed-type Charging Infrastructure Planning}

Existing studies on charging infrastructure planning primarily focus on determining locations, quantities, and types of BEV chargers, as well as the associated charging schedules given selected installations. For electric buses and trucks at industrial sites, their charging operations typically rely on the opportunity or overnight charging, meaning that vehicles are charged during layovers and overnight parking periods \cite{abdelwahed_evaluating_2020}. After determining a finite set of locations for EV charger deployment, the numbers and types of chargers can be optimized based on committed trips and available buses. In prior studies, including \cite{hu_joint_2022}, \cite{legault_integrated_2025}, and \cite{gkiotsalitis_electric_2025}, the charging planning objective includes charging infrastructure costs, BEV purchase costs, and operational costs such as scheduling penalties and charging prices.  \cite{hu_joint_2022} model a joint optimization problem for locating fast chargers at selected bus stops and optimizing schedules during bus layovers, and the objective function incorporates the time-varying electricity price and penalty for the additional waiting times. \cite{legault_integrated_2025} formulates a multi-period integrated charging infrastructure planning and scheduling problem for an electric bus fleet. The model accounts for BEV and charger acquisition costs over multiple years, as well as the phased retirement of diesel buses. \cite{gkiotsalitis_electric_2025} optimize the mixed-type BEV chargers' locations and numbers by allocating bus trips to their nearest chargers when they are available, assuming that buses are charged at the depots or the end of their trips. 

\subsubsection{BEV Charging Scheduling}

The classical multi-depot vehicle scheduling problem takes a set of timetabled trips as input and requires that they all be performed by a single vehicle. The BEV charging scheduling problem is considered as a variant of this problem in which battery SoC constraints are imposed. The problem can be modeled as a network with nodes corresponding to discretized battery SoC levels and solved as a shortest
path problem \cite{de_vos_electric_2024}. The charging procedure can be viewed as a service scheduling if a fleet of vehicles is involved. The charging problem thus requires consideration of charger availability and waiting times (or queues) for charging services. For instance, the waiting time can be modeled as a monetary penalty in the objective function \cite{hu_joint_2022, tao_coordinated_2024, gkiotsalitis_electric_2025}. When the waiting time exceeds a threshold, the vehicle that initiated a time-critical task or trip may abandon charging and seek another charging spot, which can be modeled as abandonment. A unique aspect of BEV charging is range anxiety, the concern among drivers that their battery will deplete before reaching a charging station or their destination. Empirical studies show that range anxiety may incentivize the installation of additional charging stations or the adoption of larger battery capacities \cite{lim_toward_2015}. \cite{lu_coordinated_2025} model the impact of range anxiety by setting a threshold SoC before reaching the destination. \cite{tao_coordinated_2024} formulate the accumulated range anxiety level as a function of SoC depletion along the route and integrates this term into the objective function.

\subsubsection{Charging Event Characteristics and Uncertainty} Charging events are usually characterized by charging duration and charging power (or energy), and are correlated with diverse features including charging type, hour of day, day of week, location, and vehicle type \cite{cui_stacking_2023, zhang_analysis_2025}. \cite{cui_stacking_2023} processed and analyzed more than 220{,}000 real-world charging records in Beijing. The authors clustered these charging events based on diverse features and structured them as a behavioral tree of fast charging. These clusters are labeled as low-, medium-, and high-priority based on dwell duration. In BEV charging scheduling problems, charging and dwell durations are typically unknown at the time of scheduling. These uncertainties can be modeled using stochastic optimization (SO), robust optimization (RO), and distributionally robust optimization (DRO). Among these formulations, RO typically provides the highest level of reliability, making it particularly suitable for critical infrastructure planning. Different shapes of uncertainty sets, such as boxes or ellipsoids, lead to different robust counterparts and affect tractability \cite{bertsimas_robust_2022}. \cite{hu_joint_2022} model uncertainty in passengers' boarding and alighting times during bus trips using box and budget uncertainty sets in an RO formulation. \cite{chen_exponential_2024} consider EV charging scheduling with stochastic arrival times of different EV user groups and introduce an exponential-cone programming formulation to approximate an upper bound. \cite{chen_integrated_2023} develop a two-stage stochastic model for planning and operating shared autonomous EV systems and create scenarios for uncertain user trip requests; an accelerated two-phase Benders decomposition-based algorithm is proposed to solve the resulting problem. \cite{wang_featuredriven_2023} develop an ML-cluster-based, feature-driven ambiguity set for surgery scheduling using a DRO formulation; the clustering is based on classification and regression trees.

\subsubsection{Decision-dependent robust optimization for charging decisions} With electrification, parking durations that previously involved no charging activities are likely to change after charger deployment. Large-scale charging-installation trials conducted in multiple countries have demonstrated that charging decisions significantly affect BEV parking and waiting durations \cite{cui_stacking_2023, zhang_analysis_2025}. Parking (or charging) duration can thus be modeled using a decision-dependent uncertainty set that depends on endogenous factors, such as charging decisions. \cite{nohadani_optimization_2018} demonstrate that, for simple robust linear problems, RO-DDU is NP-complete and requires either global optimization or reformulation techniques such as the big-M method. \cite{luo_distributionally_2020} study decision-dependent DRO models with five types of ambiguity sets, where an instantaneous decision directly impacts right-hand-side uncertainty, including moments and the radius of ambiguity sets. \cite{noyan_distributionally_2022} propose a decision-dependent ambiguity set for a machine scheduling problem and show that introducing affine dependence on decision variables can substantially reduce the number of constraints induced by McCormick envelopes and big-M reformulations. However, existing studies are largely limited to settings with instantaneous decisions, whereas in practice, uncertainty is influenced by aggregated and expected decision outcomes.

\begin{table}[htbp]
\centering
\begin{threeparttable}
\footnotesize
\label{tab:literature_review}
\begin{tabular}{|l|c|c|c|c|c|c|c|}
\hline
 & \multicolumn{4}{c|}{Decision Variables} & \multicolumn{2}{c|}{User Behaviors} & \multirow{2}{*}{\makecell{Uncertainty\\formulation}} \\
\cline{2-7}
 & Location & Number & Scheduling & Type & \makecell{Waiting \\ time} & \makecell{Range \\ anxiety} & \\
\hline
\cite{abdelwahed_evaluating_2020} &  &  & \checkmark &  &  &  & D \\
\cite{hu_joint_2022} & \checkmark  & \checkmark  & \checkmark  &  & \checkmark  &  & RO \\
\cite{chen_integrated_2023} & \checkmark & \checkmark & \checkmark & \checkmark &  &  & SO \\
\cite{chen_exponential_2024} & &  & \checkmark &  &  &  & SO \\
\cite{tao_coordinated_2024} & \checkmark &  &  &  & \checkmark & \checkmark & D \\
\cite{lu_coordinated_2025}  &  &  & \checkmark  &  &  & \checkmark  & SO \\
\cite{legault_integrated_2025}  & \checkmark & \checkmark & \checkmark & \checkmark &  &  & D \\
\cite{gkiotsalitis_electric_2025} & \checkmark & \checkmark & \checkmark & \checkmark & \checkmark  &  & D \\ \hline
Our paper & \checkmark & \checkmark & \checkmark & \checkmark & \checkmark  & \checkmark  & RO-DDU \\
\hline
\end{tabular}
\begin{tablenotes}
\item \textit{Notes:} D = Deterministic model; RO = Robust Optimization; SO = Stochastic Optimization; RO-DDU = Robust Optimization with Decision-Dependent Uncertainty.
\end{tablenotes}
\end{threeparttable}
\caption{Literature Review Summary}
\end{table}

All the aforementioned work on charging infrastructure planning and scheduling is summarized in Table~\ref{tab:literature_review}. Compared with these works, our paper makes the following contributions:
\begin{itemize}
    \item We develop an integrated charger planning and scheduling model that optimizes charger locations, quantities, and types, and jointly determines schedules given the selected infrastructure. The model introduces a scheduling-with-abandonment formulation that reflects the industrial site's strict operational rules and captures the impacts of overnight parking, range anxiety, and waiting times.
    \item We develop a feature-based, decision-dependent uncertainty set to capture the variability of charging durations and their dependence on charging decisions. Leveraging ML techniques and contextual information, we restructure time--zone features to preserve the calendar effects of truck trips and group parking durations to efficiently set the bounds of the uncertainty sets.
    
    \item Given the limited discussion in the literature on dependency assumptions in decision-dependent uncertainty problems, we distinguish between two cases in which charging duration depends on either \textit{instantaneous} charging decisions or \textit{cluster-level} expected charging decisions. We derive tractable reformulations for both cases and discuss the resulting outcomes and managerial implications.
    \item Considering the high computational cost of solving the scheduling problem over a long optimization horizon, we design a ``fix-and-optimize'' heuristic to forecast parking intervals with potentially high charging demand. Using representative days or a rolling horizon, the model achieves an optimality gap of less than 0.1\% within a reasonable computation time under different scenarios.
\end{itemize}

\section{Problem formulation}

\subsection{Problem description}

In the joint planning and scheduling problem, we consider a fleet of electric light-duty trucks $i \in \mathcal{I}$ operating around an industrial site based on the fixed stop points and schedules during the time period $\mathcal{T}$.

\begin{figure}[h]
    \centering
    \includegraphics[width=5.5in]{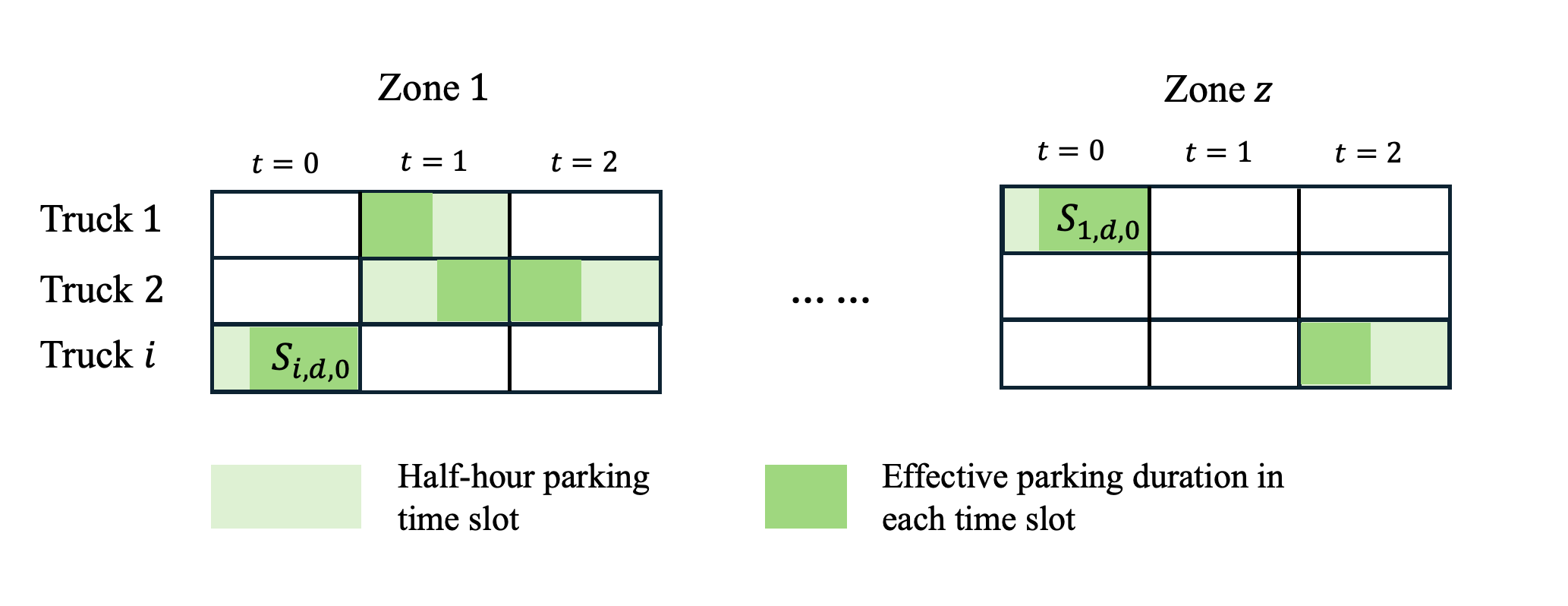}
    \caption{The studied multi-zone charging scheduling problem for the light-duty truck fleet. 
At any half-hour parking time slot \ensuremath{\mathcal{S}} shown in light green, 
A truck can choose to wait, abandon charging, or charge. 
The effective parking durations \ensuremath{pp} are shown in dark green.}
    \label{zone_wise_chart}
\end{figure}

As shown in Figure~\ref{zone_wise_chart}, a truck $i$ can either wait to be charged, abandon charging, or charge directly using one of the available charger types $j \in \mathcal{J}$ during its available half-hour parking time slots $\mathcal{S}_{i,d,t}$ at time $t$ on day $d$ in charging zone $z$. A truck cannot appear in more than one zone at a time. In each parking time slot, the effective parking duration is less than half an hour, which restricts the maximum amount of energy that can be charged during this window. Across the optimization horizon, the total energy scheduled for each truck must satisfy the truck's trip-related energy consumption, subject to the battery energy balance, battery SoC constraints, and charging power constraints. Traveling distance and energy consumption during parking slots $\mathcal{S}_{i,d,t}$ are not considered. Furthermore, at any time $t$ on day $d$, the number of chargers of type $j$ used simultaneously in zone $z$ cannot exceed the number of installed chargers of that type in the zone, $x_z^j$.

\subsection{Set, index and parameters}
\begin{itemize}
    \item \( \mathcal{J}, j\): Set, index of EV charger types.
    \item \( \mathcal{Z}, z \): Set, index of EV charging zones.
    \item \( \mathcal{I}, i \): Set, index of electric trucks.
    \item \( \mathcal{D}, d \): Set, index of days in the planning horizon.
    \item \( \mathcal{T}, t \): Set, index of hours in day \( d \).
    \item \( \mathcal{S}_{i,d,t} \): Set of discrete possible parking times for truck \( i \) on day \( d \) at the time $t$.
    \item \( E_i^{\text{bat}} \): Battery energy capacity (kWh) of truck \( i \).
    \item \( \text{SoC}^{\text{max}} \): Maximum SoC of EV batteries.
    \item \( \text{SoC}^{\text{min}} \): Minimum SoC of EV batteries.
    \item \( P^j \): Power rating (kW) of charger type \( j \).
    \item \( \eta^j \): Charging efficiency of charger type $j$.
    \item \( \rho_{i,d,t} \): Energy consumption usage (kWh) of truck \( i \) at time \( t \) on day $d$.
    \item \( C_j^{\text{install}} \): Installation cost for charger type \( j \).
    \item \( P^{\text{low}} \): Penalty for low SoC inducing the range anxiety
    \item \( P^{\text{charging}} \): Penalty for charging time of trucks.    
     \item \( \mu\): Mean of parking durations by truck $i$ at hour $t$ at zone $z$ across all the optimized days.
    \item \( \sigma\): Standard deviation of parking durations by truck $i$ at hour $t$ on day $d$ at zone $z$ across all the optimized days.
    \item $\Delta t$: Duration (mins) in each parking time interval. 
\end{itemize}

\subsection{Decision variables}
\begin{itemize}
    \item \( x_z^j \in \mathbb{Z}^+ \): The number of chargers of type \( j \) installed in zone \( z \).
    \item \( w_{i,d,t} \in \mathbb{R}^+ \): Accumulated waiting time for truck \( i \) at charger \( j \) on day \( d \) at time \( t \).
    \item \( b_{i,d,t} \in \mathbb{R}^+ \): Battery SoC level (\%) of truck \( i \) at time \( t \) on day \( d \).
    \item \( p_{i,d,t} \in \mathbb{R}^+ \): EV charging power (kWh) of truck \( i \) at time \( t \) on day \( d \).
    \item \( v_{i,d,t} \in \mathbb{R}^+ \): Battery SoC inadequacy below the low SoC threshold.
    \item \( T^\text{max}_{i,d,t} \in \mathbb{R}^+  \): Maximum waiting time before truck \( i \) abandons charging at time \( t \) on day \( d \).
    \item $\delta_{i,d,t}^{30} \in \{0, 1\}$: Indicator for the lower SoC bound of EV batteries inducing range anxiety.
    \item \( y_{i,d,t}^j \in \{0,1\} \): Binary variable indicating that truck \( i \) is charged with charger type \( j \) at time \( t \) on day \( d \) (1 for charging and 0 for not).
    \item \( a_{i,d,t} \in \{0,1\} \): Binary variable indicating that truck \( i \) abandons charging at time \( t \) on day \( d \) (1 for abandonment and 0 for not).
\end{itemize}

\subsection{Uncertainty variables}
\begin{itemize}

\item \( pp_{i,j,z,t} \in \mathbb{R}^+ \): Parking duration for truck $i$ at time $t$ in zone $z$ using charging type $j$. 

\item \( \bar{Y}_{c}^j \in \mathbb{R}^+ \): The expected cluster-wise charging decision with the  cluster $c$. 
\end{itemize}

\subsection{Objective function}

The objective function includes the charging planning and scheduling. The charger planning accounts for multiple charging infrastructure types, including slow and fast chargers. In addition to power capacity, we also model the SoC charging range for each type of chargers, which are detailed in the following section. The scheduling part includes penalties for low battery SoC, inducing the range anxiety and the total charging time.

\begin{align}
\min \; & 
\underbrace{\sum_{j \in \mathcal{J}} \sum_{z \in \mathcal{Z}} C_{j}^{\text{install}} \cdot x_{z,j}}_{\text{Installation cost of mixed-type EV chargers}} 
+ \underbrace{P^{\text{low}} \sum_{i \in \mathcal{I}} \sum_{d \in \mathcal{D}} \sum_{t \in \mathcal{T}_d} v_{i,d,t}}_{\text{Penalty for low SoC of EV batteries}} \nonumber  + 
\underbrace{P^{\text{charging}} \sum_{i \in \mathcal{I}} \sum_{d \in \mathcal{D}} \sum_{j \in \mathcal{J}} \sum_{t \in \mathcal{T}_d} y_{i,d,t}^j}_{\text{Penalty for EV charging time}} 
\end{align}

where the low SoC penalty is calculated as the sum of the soft violation variable $v_{i,d,t} := [0.3 - b_{i,d,t}]^+$. This variable is greater than zero (i.e., \(v_{i,d,t} > 0\)) if and only if the battery SoC level falls below 30 \%. The charging-time penalty is calculated as the sum of charging times across all zones and the time horizon to prevent low charger utilization (i.e., low average charging power).

\subsection{Constraints for charging planning and scheduling}

The following subsections detail constraints pertaining to charging planning and scheduling for the entire fleet of electric trucks. We also introduced three big-M constraints, and we discussed the range value of these $M$ in Appendix \ref{sec:bigM}.

\subsubsection{Truck charging constraints}
We impose the following constraints to govern the truck charging behavior:

\begin{align}
&\text{(Single charger usage)} 
&& \sum_{j \in \mathcal{J}} y_{i,d,t}^j \le 1 
&&  \forall i, d, t \label{const:single_charger} \\[0.5em]
&\text{(Charging only in parking windows)} 
&& y_{i,d,t}^j = 0 
&& \forall i, t \notin \mathcal{S} \quad  \forall j \label{const:parking_window} \\[0.5em]
&\text{(Low SoC indicator)} 
&& b_{i,d,t} - \frac{30}{100} \ge -M_1 \cdot \delta_{i,d,t}^{30} + \epsilon \nonumber \\
& 
&& b_{i,d,t} - \frac{30}{100} \le M_1 \cdot (1 - \delta_{i,d,t}^{30}) 
&& \forall i, d, t. \label{const:low_soc}
\end{align}

Constraint~\eqref{const:single_charger} ensures each truck uses at most one charger at any time.  
Constraint~\eqref{const:parking_window} enforces charging only during the scheduled parking window.  
Constraint~\eqref{const:low_soc} defines the binary indicator $\delta_{i,d,t}^{30}$ for the low SoC condition, where $\delta_{i,d,t}^{30} = 0$ if and only if the battery level exceeds 30\%. Here, $M_1$ is a large constant, and the effective range of $M_1$ is presented in Appendix \ref{sec:bigM}. The parameter $\epsilon$ is a small positive number that ensures the indicator equals 1 only when the battery SoC is strictly greater than 30\%, thereby guaranteeing a unique waiting time when the battery SoC is exactly 30\%.

\subsubsection{Abandonment and waiting time constraints}
We model two coupled phenomena during each half–hour parking slot $t$ on day $d$: (i) \emph{abandonment}—once a driver gives up waiting, no charging may occur in that slot; and (ii) \emph{waiting-time accumulation}—while a truck remains parked in the \emph{same zone} across consecutive admissible slots and is not charging, its waiting time increases by $\Delta t$; otherwise it resets to zero.

\paragraph{Abandonment}
Let $a_{i,d,t}\in\{0,1\}$ indicate that truck $i$ abandons in slot $(d,t)$ and let $y^j_{i,d,t}\in\{0,1\}$ be the charger-type assignment. We enforce
\begin{align}
&\text{(Once abandon, no charging)} & 
a_{i,d,t} \;\le\; 1 - \sum_{j\in\mathcal{J}} y^j_{i,d,t},
&& \forall i,d,t, 
\label{abandonment_no_charging}
\end{align}
so choosing abandonment excludes charging in that slot. During a continuous parked stretch (defined below), the abandonment variable is non-decreasing:
\begin{align}
a_{i,d',t'} \;\ge\; a_{i,d,t}, \qquad \forall (d',t') \text{ successor of }(d,t).
\label{abandonment_monotone}
\end{align}

\noindent\textit{(Persistence over a parked stretch: once abandoned, a truck remains abandoned in subsequent slots within the same zone.)}

\begin{figure}[t]
  \centering
\includegraphics[width=0.6\linewidth]{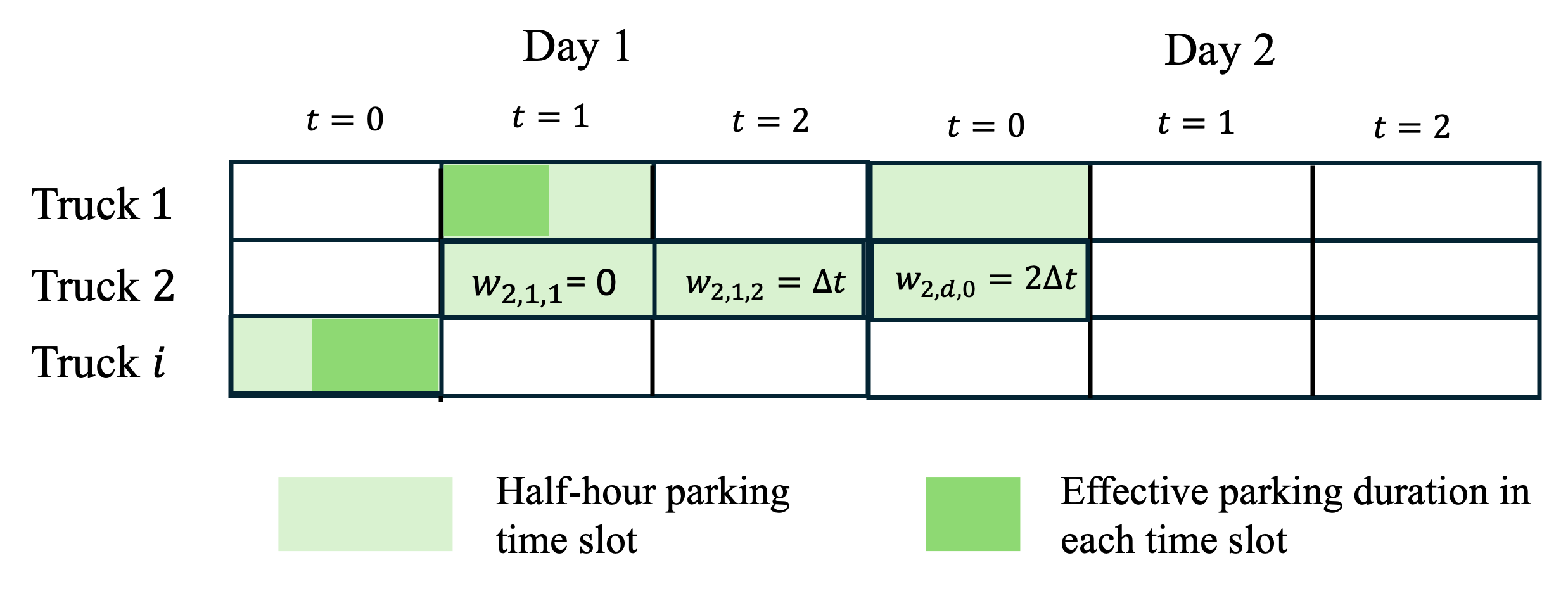}
  \caption{Day-to-day carryover of waiting time in the same zone. For Truck~2 on Day~1, the first admissible slot initializes $w=0$. The next slot in the same zone accumulates to $w=\Delta t$, and carryover to Day~2 in the same zone yields $w=2\Delta t$.}
  \label{fig:waiting-day}
\end{figure}

\paragraph{Waiting time}
Let $w_{i,d,t}\ge 0$ be the accumulated waiting time upon entering slot $(d,t)$. When two consecutive admissible slots for truck $i$ occur in the \emph{same zone}, waiting evolves as
\begin{align}
&\text{(Within-zone recursion)} &
w_{i,d,t+1} \;=\; w_{i,d,t} 
+ \Delta t \!\left(1 - \sum_{j\in\mathcal{J}} y^j_{i,d,t}\right),
&& \forall i,d,t \text{ with same-zone successor,}
\label{eq:waiting_time}
\end{align}
i.e., $w$ grows by $\Delta t$ only if the truck was \emph{not} charging in the preceding slot. Otherwise, $w$ resets:
\begin{align}
&\text{(Resets)} &
w_{i,d,t} \;=\; 0, 
&& \begin{array}{l}
\text{if $(i,d,t)$ is not an admissible parking slot;}\\
\text{or if the predecessor admissible slot is in a different zone.}
\end{array}
\label{waiting_reset}
\end{align}
Rule \eqref{waiting_reset} initializes $w$ at the start of any parked stretch and breaks carryover across zone changes; Equation \eqref{eq:waiting_time} propagates $w$ across day boundaries when the zone is unchanged (e.g., from $(d,T_d{-}1)$ to $(d{+}1,0)$). Two special cases for resetting the waiting time, the day-to-day carryover of waiting time in the same zone and zone-change reset of waiting time, are presented in Figs. \ref{fig:waiting-day} and \ref{fig:waiting-zone}, respectively. The full illustration of waiting time and abandonment is presented in Appendix \ref{full_illustrations}.

\begin{figure}[t]
  \centering
\includegraphics[width=0.6\linewidth]{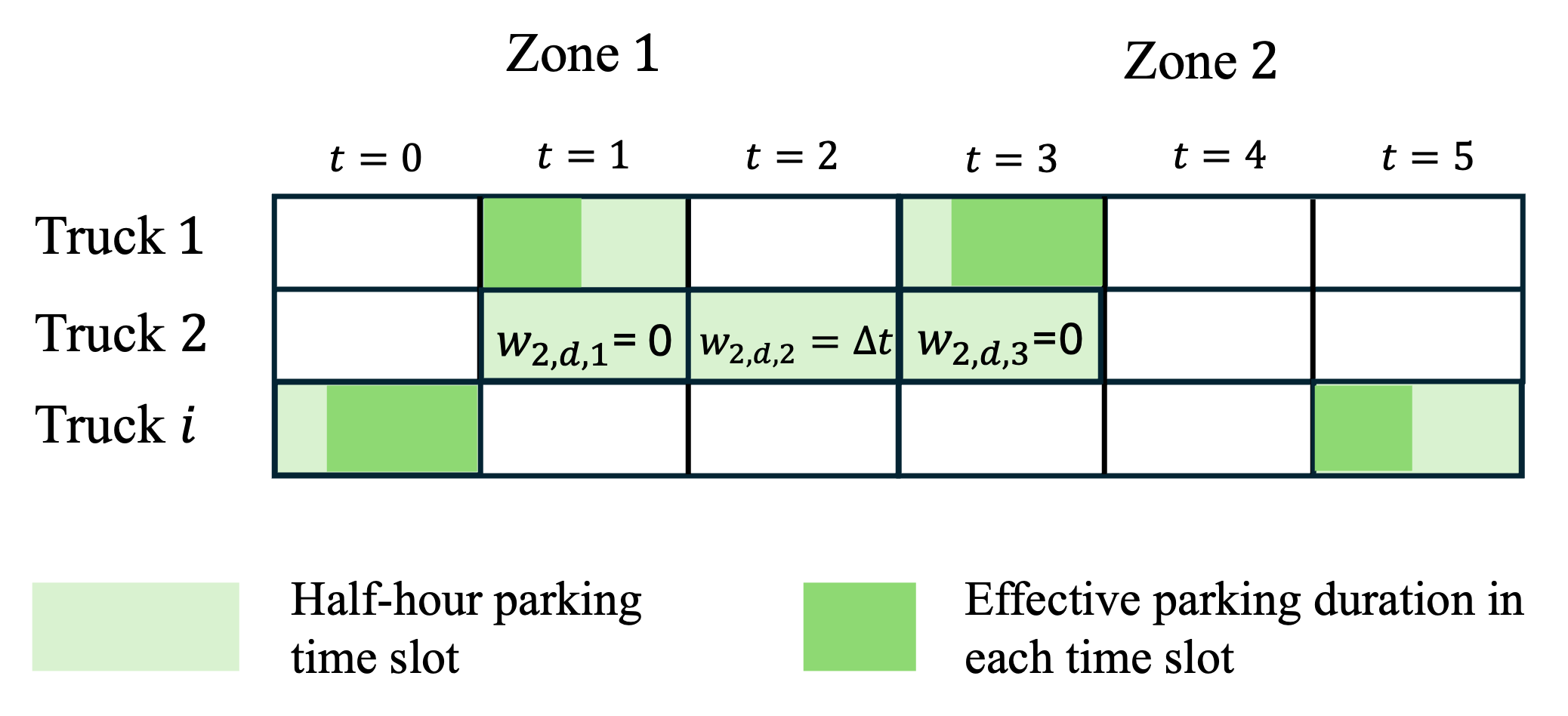}
  \caption{Zone-change resets of waiting time. For Truck~2, waiting grows from $w=0$ to $w=\Delta t$ across two slots in Zone~1. Transition to Zone~2 resets $w=0$, after which accumulation resumes in subsequent uncharged slots.}
  \label{fig:waiting-zone}
\end{figure}

\subsubsection{Charging abandonment rules}

At an industrial site, a truck abandons charging if its waiting time exceeds a maximum allowed waiting time \( T^{\text{max}}_{i,d,t} \), as constraint~\eqref{const:waiting_abandon}. \(M_2\) is a large constant to relax the waiting time constraint when \(a_{i,d,t}\) equals to 1, in other words, when the trucks abandon charging. Notably, the maximum waiting time varies with the truck's range anxiety and with whether it is parked in special zones that permit overnight charging. 

Based on the industrial partner's experience, we formulate the following waiting and abandonment rules, accounting for range anxiety and overnight charging in special zones as follows. As constraint \eqref{const_nl:tmax}, if the SoC is greater than or equal to 30\%, no waiting time is allowed as the driver has no range anxiety. If the SoC of a truck is below 30\%, this will result in the driver's range anxiety and would extend their waiting time to half an hour before abandoning charging. Let the set \(S^Z\) denote the parking time slots at any of the special zones for overnight charging. Constraint~\eqref{const:waiting_zero} sets the waiting time to be zero when the truck is outside the parking windows (i.e., no parking allowed) or in special zones (i.e., no waiting time limits). Constraint \eqref{const:nonnegative} regulates that the waiting time is always non-negative.

\begin{align}
& \text{(Abandon after the maximum waiting time)} 
& w_{i,d,t} \le T^{\text{max}}_{i,d,t} + M_2 \cdot a_{i,d,t}
&& \forall i, d, t \in \mathcal{S} \label{const:waiting_abandon} \\[0.3em]
& \text{(Maximum waiting time in non-special zones)}
& T^{\text{max}}_{i,d,t} = \Delta t \cdot \delta_{i,d,t}^{30}
&& \forall i, d, t \in \overline{\mathcal{S}^{Z}} \label{const_nl:tmax} \\[0.3em]
% & \text{(Maximum waiting time in special zones)}
% & T^{\text{max}}_{i,d,t} = L 
% && \forall i, d, t \in S^{Z}_{i,d} \label{const_l:tmax} \\[0.3em]
& \text{(Outside parking windows or in special zones)}
& w_{i,d,t} = 0
&&  \forall i,d,t \in \mathcal{S}^Z \cup \overline{\mathcal{S}} \label{const:waiting_zero} \\
& \text{(Non-negative constraint)} 
& w_{i,d,t} \geq 0
&& \forall i,d,t \in \mathcal{S} \label{const:nonnegative}
\end{align}

\subsubsection{Battery dynamics constraints}

We impose the following constraints to model battery SoC evolution and charging power limits:
\begin{align}
& \text{(Battery energy balance)}
& b_{i,d,t} E^{bat}_i  = b_{i,d,t-1} E^{bat}_i  + p^{ch}_{i,d,t} - \rho_{i,d,t}, 
&& \forall i, d, \forall t \geq t_0 \label{const:battery_soc} \\[0.3em]
& \text{(The initial and final SoC are equal)}
& b_{i,d_0,t_0}  = b_{i,d_e,t_e} 
&& \forall i, d, t \label{battery_soc_balance} \\[0.3em]
& \text{(Maximum charging power limit)}
& p^{ch}_{i,d,t} \le \sum_{j \in \mathcal{J}} \eta^j \cdot P^j \cdot \Delta t \cdot y_{i,d,t}^j \cdot pp_{i,d,t}, 
&& \forall i, d, t \in \mathcal{S}  \label{const:max_charging_limit}  \\
& \text{(Effective charging time limits)}
& \frac{5}{30} \leq pp_{i,d,t} \leq 1, 
&& \forall i, d, t \in \mathcal{S}  \label{const:charging_limit}  \\
& \text{(Battery SoC limits)}
& \text{SoC}^{\text{min}}\le b_{i,d,t} \le \text{SoC}^{\text{max}} 
&& \quad \forall i, d, t \label{const:soc_limit} 
\end{align}

Constraint \eqref{const:battery_soc} models EV battery dynamics considering charging power and energy consumption, and it should be noted that we do not use a binary variable to model battery discharging status, as it shall not happen during parking.  Constraint \eqref{battery_soc_balance} sets the initial SoC at $t_0$ and final SoC at time $t_e$ of the optimization horizon to be the same. Constraint \eqref{const:max_charging_limit} sets the maximum charging power limit considering the charging type $P^j$ and corresponding charging efficiency $\eta^j$. Constraint \eqref{const:charging_limit} regulates the charging time limit in each time window, which is greater than 5 mins and less than 30 mins. Finally, constraint \eqref{const:soc_limit} restricts battery SoC within the maximum and minimum values.

\subsubsection{Fast charger SoC restrictions}
To prevent trucks from using fast chargers when their SoC exceeds $80\%$ and to ensure that fast chargers do not charge the battery SoC beyond $80\%$, the following constraints are enforced when charging. Constraint (\ref{const:soc_threshold}) ensures that if \( b_{i,d,t} > 80 \% \), then \( \delta_{i,d,t}^{80} = 1 \), and \( y_{i,d,t}^1 = 0 \), preventing the use of the fast charger. Also, if \( b_{i,d,t} \leq 80 \% \), then \( \delta_{i,d,t}^{80} = 0 \), allowing the fast charger to be used. Constraint (\ref{const:fast_charger_cap}) ensures that when the fast charger is used, the battery level after charging (\( b_{i,d',t'} \)) is limited to at most 80\%.

\begin{align}
% & \text{(SoC cap before fast charging)} 
% & b_{i,d,t} - 80 \leq M \cdot  \delta_{i,d,t}^{80}
% && \forall i, d, t \in S_{i,d,t} \\
& \text{(SoC cap before fast charging)}
& b_{i,d,t} \leq \frac{80}{100} + M_3 \cdot (1- y_{i,d,t}^1) 
&& \forall i, \; d, \; t \in \mathcal{S} \label{const:soc_threshold} \\[0.3em]
& \text{(SoC cap during fast charging)} 
& b_{i,d',t'} \leq \frac{80}{100} + M_3 \cdot (1 - y_{i,d,t}^1) 
&& \forall i, \; d', \; t' \in \mathcal{S}\;
\label{const:fast_charger_cap}
\end{align}

Here, \( b_{i,d,t} \) is the SoC of truck \( i \); 
\( y_{i,d,t,1} \) is a binary variable equal to 1 if truck \( i \) uses a fast charger at time \( t \) on day \( d \); \( M_3 \) is a large constant; and \( \mathcal{S} \) is the set of time steps \( t \) on day \( d \) when truck \( i \) is parked. 
The pair \( (d', t') \) represents the next time step after \( (d, t) \), defined as \( (d, t+1) \) if \( t < T-1 \), or \( (d+1, 0) \) if \( t = T-1 \) and \( d < D-1 \), 
where \( T \) is the number of time steps per day and \( D \) is the number of days.

\subsubsection{Total number of chargers in each type and zone}
The number of trucks charging with charger type \( j \) cannot exceed the number of available chargers of type \( j \) in zone \( z \) at any time.

\begin{equation}
\sum_{i \in \mathcal{I}} y_{i,d,t}^j \mathbbm{1}\{\text{truck $i$ at zone $z$ at time $t$, day $d$}\} \leq x_z^j \quad \forall z \in \mathcal{Z}, \forall j \in \mathcal{J}, d \in \mathcal{D}, t \in \mathcal{T}
\end{equation}

Where $\mathbbm{1}$ is a binary indicator to represent whether truck $i$ is at zone $z$ at time $t$ of day $d$. 

\subsection{Decision-dependent robust formulation considering parking duration uncertainty}
\label{sec:decision_dependent}

The stochastic effective parking duration, denoted by the random variable $\hat{pp}_{i,d,t}$, is unknown at the time of scheduling. We assume that this duration is endogenous: it depends on the charging decision $y_{i,d,t}^j$. Specifically, we model the realized parking duration $(\hat{pp})_{i,d,t}'$ as an affine transformation of the intrinsic (base) random duration $\hat{pp}_{i,d,t}$.

Let $k^j \geq 1$ be a scaling constant associated with charger type $j$ (e.g., representing the reduction in charging time due to fast charging). The decision-dependent duration is defined as:
\begin{equation}
\label{eq:affine_dependency}
(\hat{pp})_{i,d,t}' := \hat{pp}_{i,d,t} \cdot \left( 1 - \sum_{j \in \mathcal{J}} \frac{y_{i,d,t}^j}{k^j} \right)
\end{equation}
Since we enforce single charger usage ($\sum y^j \le 1$), the term in the parentheses acts as a scalar scaling factor. 

\subsubsection{Feature-driven cluster-wise uncertainty sets}
\label{feature_based}
Assuming that we have a total number of $N$ historical records of parking duration, each is indexed by $\omega \in \Omega$, $\Omega=\{1, 2,..., N\}$, and records a truck's parking duration $pp^{\omega}$ and its feature vector $v^\omega\in \mathcal V$, where $\mathcal V$ is the feature space. Each feature vector encodes information for each parking duration, including contextual and numerical variables such as half-hour of day, day of year, and zone. The fleet information describes the trucks' operational purposes and is categorized into eight types, such as drill-and-blast or production activities. For instance, the feature vector of a parking duration $\omega$ can be represented as,

\begin{align*}
\boldsymbol{v}^{\omega} = &\{1: (00:00), 20: (\text{The 20th day of the year 2023}),\\&  4: (\text{Zone 4}), 3: (\text{Production team}), 255: (\text{Truck ID 255}) \}
\end{align*}

We found that parking duration is strongly correlated with fleet groups, half-hour intervals, and zones, but only slightly related to the day of the year. This motivates us to partition the parking duration into $c \in \mathbb{N}$ clusters to represent its stochasticity.

Given the calendar effect of parking events, we propose using hierarchical clustering by reordering the feature vectors for each truck. We compute the average parking duration for each half-hour zone combination, aggregated across all days of the year. These averaged parking durations under half-hour zone combinations serve as the feature vectors for individual trucks. Let the newly arranged feature vector be $(\boldsymbol{v}^{i})^{'}$. Using these time–zone features alongside fleet information, we perform hierarchical clustering to group trucks, using Euclidean distance as the similarity measure.

\begin{align*}
(\boldsymbol{v}^{i})^{'} = &\{1: (00:00; \text{Zone 1}), 2: (00:30; \text{Zone 1}),..., N_t\cdot N_z: (24:00; \text{Zone 8})\}
\end{align*}

We next model the distribution of uncertain parking duration for all trucks or truck groups and construct the corresponding feature-driven uncertainty set. Assuming that the parking duration of each cluster is identically and independently distributed, we could calculate the empirical mean $\mu_c$ and variance $\sigma_c^2 = \frac{\sum_{\omega\in\Omega_c}(\hat{pp}^{\omega}-\mu_c)^2}{N_c}$. Considering the inherent bounds of parking duration, we build a box-shaped uncertainty set as follows,  

\begin{equation}
\mathbb P :=
\Bigg\{
\hat{pp} \;\Big|\;
\begin{array}{l}
\mathbb E_{\mathbb P}[\hat{pp}] = \boldsymbol{\mu}_{pp}, \\[2pt]
\mathbb E_{\mathbb P}\!\left[
(\hat{pp}-\boldsymbol{\mu}_{pp})
(\hat{pp}-\boldsymbol{\mu}_{pp})^{\mathsf T}
\right]
= \boldsymbol{\Sigma}_{pp}, \\[2pt]
\lVert \hat{pp} \rVert_\infty \le \Delta_c
\end{array}
\Bigg\}
\label{Uncertainty_set}
\end{equation}

where $\boldsymbol{\mu}_{pp} = (\mu_c(\boldsymbol{v}))_{c\in\mathcal{C}}$, and the covariance matrix $\boldsymbol\Sigma_{pp} = \text{diag}(\sigma_c^2 (\boldsymbol{v})_{c\in \mathcal{C}})$ are feature-driven. The values of $\Delta_c$ are based on $\boldsymbol{\mu}_{pp}$ and $\boldsymbol\Sigma_{pp}$, which control the conservativeness of our formulation.

\subsubsection{Decision-dependent uncertainty sets and moment definitions}
\label{two_ddu}
We could construct a decision-dependent uncertainty set $\mathbb{P}$ (\ref{Uncertainty_set}) based on the first and second moments. Let $\boldsymbol{\mu}_c$ and $\boldsymbol{\Sigma}_c$ denote the empirical mean vector and covariance matrix of the base parking duration for a cluster $c$, derived from historical data without charging interventions. We set $\Delta_c$ to be $\boldsymbol{\mu}_c \pm \boldsymbol{\sigma}_c$.

The mean and variance of the \textit{updated} duration $(\hat{pp})'$, denoted as $\mu_c(y)$ and $\Sigma_c(y)$, depend on the decision variables. We consider two cases for this dependency:

\paragraph{Example 1: Affine dependency on instantaneous decisions.}
The moments are scaled directly by the specific decision for that parking slot. As $y_{i,d,t}^j$ is a ``here-and-now'' decision variable fixed prior to uncertainty realization, the adjusted moments are:
\begin{align}
\mu_{c}(y_{i,d,t}) &= \hat{\mu}_c \left( 1 - \sum_{j \in \mathcal{J}} \frac{y_{i,d,t}^j}{k^j} \right) \\
\Sigma_{c}(y_{i,d,t}) &= \hat{\Sigma}_c \left( 1 - \sum_{j \in \mathcal{J}} \frac{y_{i,d,t}^j}{k^j} \right)^2
\end{align}

\paragraph{Example 2: Affine dependency on cluster-level expected decisions.}
Here, the scaling factor depends on the aggregate behavior of the fleet within a cluster $c$. Let $N_c$ be the total number of parking slots in cluster $c$. We define the \textbf{empirical expected value} of charging decision $\bar{Y}_c$ as:
\begin{equation} 
\label{eq:cluster_empirical_mean}
\bar{Y}_c^j := \frac{1}{N_c} \sum_{(i,d,t) \in c} y_{i,d,t}^j
\end{equation}
Note that $\bar{Y}_c$ is a linear function of decision variables, not a stochastic expectation.

\textbf{Proposition 1} Given that the empirical expected value $\bar{Y}_c$ for a cluster of parking duration $c$, the moments for any instance in this cluster are scaled uniformly:
\begin{align}
\mu_{c}(\bar{Y}_c^j) &= \hat{\mu}_c \left( 1 - \sum_{j \in \mathcal{J}}\frac{\bar{Y}_c^j}{k^j} \right) \label{eq:cluster_mean} \\
\Sigma_{c}(\bar{Y}_c^j) &= \hat{\Sigma}_c \left( 1 - \sum_{j \in \mathcal{J}}\frac{\bar{Y}_c^j}{k^j} \right)^2 \label{eq:cluster_variance}
\end{align}

\textit{Proof of Proposition 1} Please see \ref{proof:moments} of the Supporting information.

\subsubsection{RO-DDU constraint and reformulations}
To ensure schedule feasibility, we enforce a RO-DDU constraint on the charging power limit:
\begin{equation}
\label{eq:DRO_constraint}
p^{ch}_{i,d,t} \le \sum_{j \in \mathcal{J}} \eta^j P^j \Delta t \cdot y_{i,d,t}^j \cdot (\hat{pp})'
\end{equation}

Based on the uncertainty variable $(\hat{pp})'$, we construct a box-shaped uncertainty set with bounds $\Delta_c = \boldsymbol{\mu}_c \pm \boldsymbol{\sigma}_c$ and derive tractable reformulations of the charging power constraint for two cases.

\textbf{Theorem 1}. Example 1: Affine dependency on instantaneous decisions. Constraint (\ref{eq:affine_final_main}) is the reformulation of the linear power limit constraint (\ref{eq:DRO_constraint}) under the robust uncertainty set with affine dependency on instantaneous decisions $\boldsymbol{y}^j$. This formulation preserves the linearity of binary charging decisions, keeping the problem tractable.
\begin{align}
\label{eq:affine_final_main}
p^{ch}_{i,d,t} \leq \sum_{j \in \mathcal{J}} \eta^j P^j \Delta t \left[ \boldsymbol{\hat{\mu}}_{c}^\intercal \left( 1 - \frac{1}{k^j} \right) \boldsymbol{y}^j - \left( 1 - \frac{1}{k^j} \right) \boldsymbol{\hat{\Sigma}}^{1/2}_{c} \boldsymbol{y}^j \right]
\end{align}

Example 2: Affine dependency on cluster-level expected decisions. After applying decision-dependent moments (\ref{eq:cluster_mean}) and (\ref{eq:cluster_variance}), constraint (\ref{eq:affine_final_main_cluster}) is the reformulation of the linear power limit constraint (\ref{eq:DRO_constraint}) under the robust uncertainty set with affine dependency on cluster-level expected decisions $\bar{Y}_c$.

\begin{align}
\label{eq:affine_final_main_cluster}
p^{ch}_{i,d,t} \leq \sum_{j \in \mathcal{J}} \eta^j P^j \Delta t \left[ \boldsymbol{\hat{\mu}}_{c}^\intercal \left( 1 - \frac{\bar{Y}_c^j}{k^j} \right) \boldsymbol{y}^j -  \left( 1 - \frac{\bar{Y}_c^j}{k^j} \right) \boldsymbol{\hat{\Sigma}}^{1/2}_{c} \boldsymbol{y}^j \right]
\end{align}

This formulation is nonlinear because it includes the product of the binary variable $\boldsymbol{y}^j$ and the continuous variable $\bar{Y}_c$.

\textit{Proof of Theorem 1.} Please see \ref{ro_ddu} of the Supporting information.

\subsection{Cluster-based Linearization (Example 2)}
\label{cluster_based}

While statistically rigorous, constraint \eqref{eq:affine_final_main_cluster} involves the product of the continuous variable $\bar{Y}_c$ (the sum of binary variables normalized by~$N_c$) and the binary decision variable $y^j$, introducing bilinear terms. To solve this within a MILP framework, we introduce an auxiliary variable $Z_{i,d,t}^j = \bar{Y}_c^j \cdot y_{i,d,t}^j$. Since $\bar{Y}_c \in [0, 1]$ and $y_{i,d,t}^j \in \{0, 1\}$, this product can be exactly linearized using the McCormick envelopes \cite{mccormick_computability_1976}:
\begin{align}
& Z_{i,d,t}^j \leq \bar{Y}_c^j \label{eq:mccormick1} \\
& Z_{i,d,t}^j \leq y_{i,d,t}^j \label{eq:mccormick2} \\
& Z_{i,d,t}^j \geq \bar{Y}_c^j - (1 - y_{i,d,t}^j) \label{eq:mccormick3} \\
& Z_{i,d,t}^j \geq 0 \label{eq:mccormick4}
\end{align}

Substituting $\boldsymbol{Z}^j$ into~\eqref{eq:affine_final_main_cluster} renders the constraint linear in the optimization variables, albeit at the cost of increasing the problem size. We obtain the constraint~\eqref{eq:cluster_linearized}.

\begin{align}
\label{eq:cluster_linearized}
p^{ch}_{i,d,t} \leq \sum_{j \in \mathcal{J}} \eta^j P^j \Delta t \left[ \boldsymbol{\hat{\mu}}_{c}^\intercal \left( 1 - \frac{1}{k^j} \right) \boldsymbol{Z}^j - \left( 1 - \frac{1}{k^j} \right) \boldsymbol{\hat{\Sigma}}^{1/2}_{c} \boldsymbol{Z}^j \right]
\end{align}

\section{Solution Algorithm}
\subsection{Robust planning using the whole year dataset}
\label{robust algorithm}

The planned EV infrastructure must remain robust across year-round fleet operations, that is, capable of meeting trip-related energy consumption, given the limited parking duration. We compare two planning results under different scheduling operations, based on the selective or full dataset: 

In terms of data usage, data can be categorized as \textit{day selection} (DS) or \textit{full year} (FY). For DS, we selected a subset of representative days based on statistical criteria, including total fleet distance, number of moving trucks, total fleet energy consumption, and total stopping time per day. We also include days around the representative days to preserve the trip and battery-cycle continuity. In contrast, the FY method uses the full-year dataset directly.

For scheduling operations, we consider the \textit{hindsight} (HS) and \textit{iterative planning and scheduling} (IPS) approaches. The HS method assumes perfect foresight and performs one-shot scheduling. The IPS method is similar to rolling-horizon scheduling. The full-period dataset is first decomposed into monthly subsets, allowing several overlapping days between periods. The algorithm then plans charger deployment for the entire period by iteratively integrating scheduling and planning at the sub-period level. It starts with an initial charger allocation derived from the dataset and processes each sub-period sequentially. If the existing infrastructure can satisfy the charging demand of this sub-period, only charging scheduling is performed. Otherwise, the algorithm jointly determines the charging schedule and the additional chargers required, updates the infrastructure, and carries the new configuration forward to subsequent sub-periods.

\begin{figure}[!htbp]
    \centering
    \includegraphics[width=4.5in]{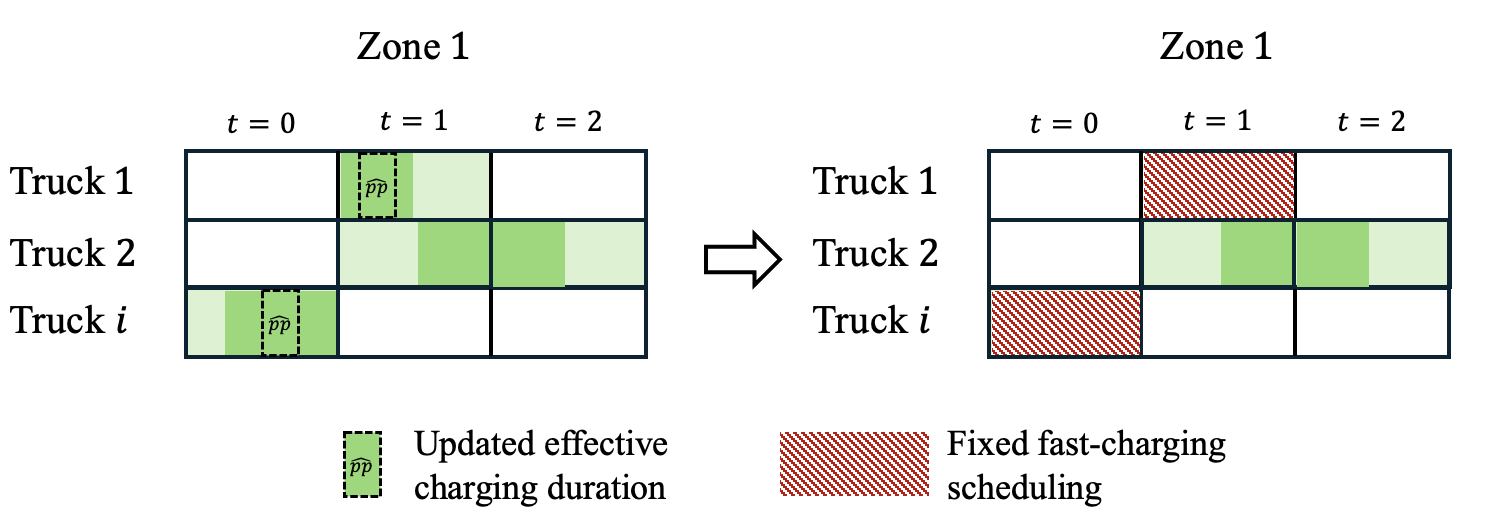}
    \caption{The fix-and-optimize heuristic where fast-charging scheduling is assigned and fixed based on the forecast high charging demands}
    \label{charging_forecast_diagram}
\end{figure}

The proposed planning and scheduling model is difficult to solve when the optimization horizon is long, as in hindsight scheduling. We propose a fix-and-optimize heuristic for the scheduling problem based on high-charging-demand forecasts. It has been applied to the timetable \cite{fonseca_fix-and-optimize_2024} and job-shape scheduling \cite{li_learning-guided_2025}, where part of the solution is fixed while the remaining variables are free to form subproblems. As shown in Fig. \ref{charging_forecast_diagram}, we first identified parking intervals with high charging demand and pre-assigned fast-charging schedules, thereby accelerating the overall optimization process.

\subsection{Fix-and-optimize heuristic based on high charging demand forecasts}
\label{fix_and_optimize}

\begin{algorithm}[h]
\label{alg:planning_scheduling}
\KwIn{Master problem $\mathcal{M}$ with deterministic parking duration $pp_{i,d,t}$; Parking duration clusters $c (i,t,z) \in \mathcal{C}$}
\KwOut{BEV charger installation decisions $x_z^j$}

  \Repeat{The maximum iteration limit is reached}{
    % Step 1: Initialization
    Solve the master problem $\mathcal{M}$; \Comment{\textit{Warm start at the first step if the solution available}}\; 
    Obtain scheduling outcomes $\mathbf{y} = (y^j_{i,d,t})_{(i,d,t)\in\mathcal{S}, j\in \mathcal{J}}$ and charging power $\mathbf{p}^{ch} = (p^{ch})_{(i,d,t)\in\mathcal{S}}$\;

    % Step 2: Generate uncertain demand

    % Step 3: Evaluate charging demand
    \For{$(i,d,t)\in c\;, j \in \mathcal{J}$}{
        Compute the lower bound of parking durations $\hat{pp}_{i,d,t} = \mu_c - \sigma_c $ for the cluster $c$\; \Comment{\textit{Applying scaling factors $k^j$ if RO-DDU}} \;
      Compute the updated upper bound of charging power $p^{ch'}_{i,j,d,t}$ 
      \[
      p^{ch'}_{i,j,d,t} \gets \sum_{j\in\mathcal{J}}\Delta t \cdot P^j \cdot y^j_{i,d,t} \cdot \hat{pp}_{i,d,t} \cdot \eta^j\;
      \]

      % Step 4: Robust constraint enforcement
      \If{\text{Violations are detected}}{
        Impose the fast-charging constraint $y_{i,1,d,t} = 1$ in the master problem\;
      }
    }
  }

  % Step 5: Post-processing
Post-process the solution by discarding part of the solution and re-solve the problem with $\hat{pp}_{i,d,t}$\;
\caption{Fix-and-optimize heuristic under parking duration uncertainty}
\label{alg:planning_scheduling}
\end{algorithm}

As illustrated in Algorithm \ref{alg:planning_scheduling}, we first initialize the master problem with the starting values $pp_{i,d,t}$. Next, we compute the lower bound of parking duration $\hat{pp}_{i,d,t}$ based on the cluster $c$ to replace the initial values and compute the updated bounds of the charging power. We then compare the updated upper bounds to the required charging power in the master problem $\mathcal M$ to identify any violations. A violation occurs if the required charging power exceeds the upper bound by more than a threshold $\kappa^v$ and the parking window does not belong to a special zone. The threshold $\kappa^v$ is initialized at 2 and decays exponentially across iterations. When a violation is detected, a fast-charging scheduling constraint $y_{i,1,d,t} = 1$ is added for that parking window. These constraints are subsequently incorporated into the master problem, after which the model is resolved from the first step.

This iterative process continues until the maximum number of iterations is reached. Afterward, we post-processed the optimization results by discarding part of the solution, for instance, a high number of fast chargers in one zone. The remaining solutions for fast and slow chargers are then used as lower bounds to resolve the model. Ultimately, the installation plan for all zones and charger types was determined.

\section{Application}

We applied our model to an open-pit mining site with 203 light-duty trucks in operation, planned for replacement with BEVs. Seven special-use trucks, such as ambulances, that will not be electrified, along with another seven trucks dispatched fewer than 10 days, are excluded from the study. The remaining 189 trucks will be replaced with electric trucks equipped with 100 kWh batteries. We consider the following two types of BEV chargers, as listed in Table \ref{tab:charger_comparison}. The capital cost of fast chargers does not include installation costs, as installation is location-dependent. We set the penalties for low SoC and charging time to be equal in the objective function. Although the charging efficiency could vary with the charging type, outdoor temperature, and charging power \cite{sevdari_experimental_2023}, this paper uses a fixed value for charging efficiency. Overall, fast charging is more efficient than slow charging because it uses a high-efficiency rectifier located outside the truck \cite{arena_comprehensive_2024}. We therefore set the charging efficiencies for slow and fast charging to 90\% and 95\%. 

\begin{table}[h!]
\centering
\begin{tabular}{l c c}
\hline
\textbf{Specifications} & \textbf{Slow charger} & \textbf{Fast charger} \\
\hline
Capital costs (\$/charger) & 1,500 \cite{ford_ford_2025} & 38,000 \cite{primecomtech_primecomtech_2025} \\
Maximum charging rates (kW) & 11.2 & 150 \\
%Charging Duration & 10 hours & $\leq$ 1 hour  \\
SoC ranges & 10--100\% & 10--80\% \\
Charging efficiency & 90\% \cite{sevdari_experimental_2023} & 95\% \cite{arena_comprehensive_2024} \\
\hline
\end{tabular}
\caption{Technical parameters of two types of EV chargers}
\label{tab:charger_comparison}
\end{table}

The charging planning process begins by identifying potential charging sites based on the stop frequency of all trucks (Section \ref{identify_charging}). Next, the parking window, effective parking duration, and energy consumption of each truck are extracted (Section \ref{summary_inputs}). Then, hierarchical clustering based on the rearranged time-zone features is applied to the parking-duration samples to construct feature-based uncertainty sets (Section \ref{robust}). High-consumption vehicles (HCVs) with infeasible trips are then filtered out (Section \ref{set_up}). Finally, the proposed model plans chargers using different scheduling operations (Section \ref{results}).

\subsection{Identifying potential EV charging points}
\label{identify_charging}

\begin{figure}[!h]
    \centering
    \includegraphics[width=4.5in]{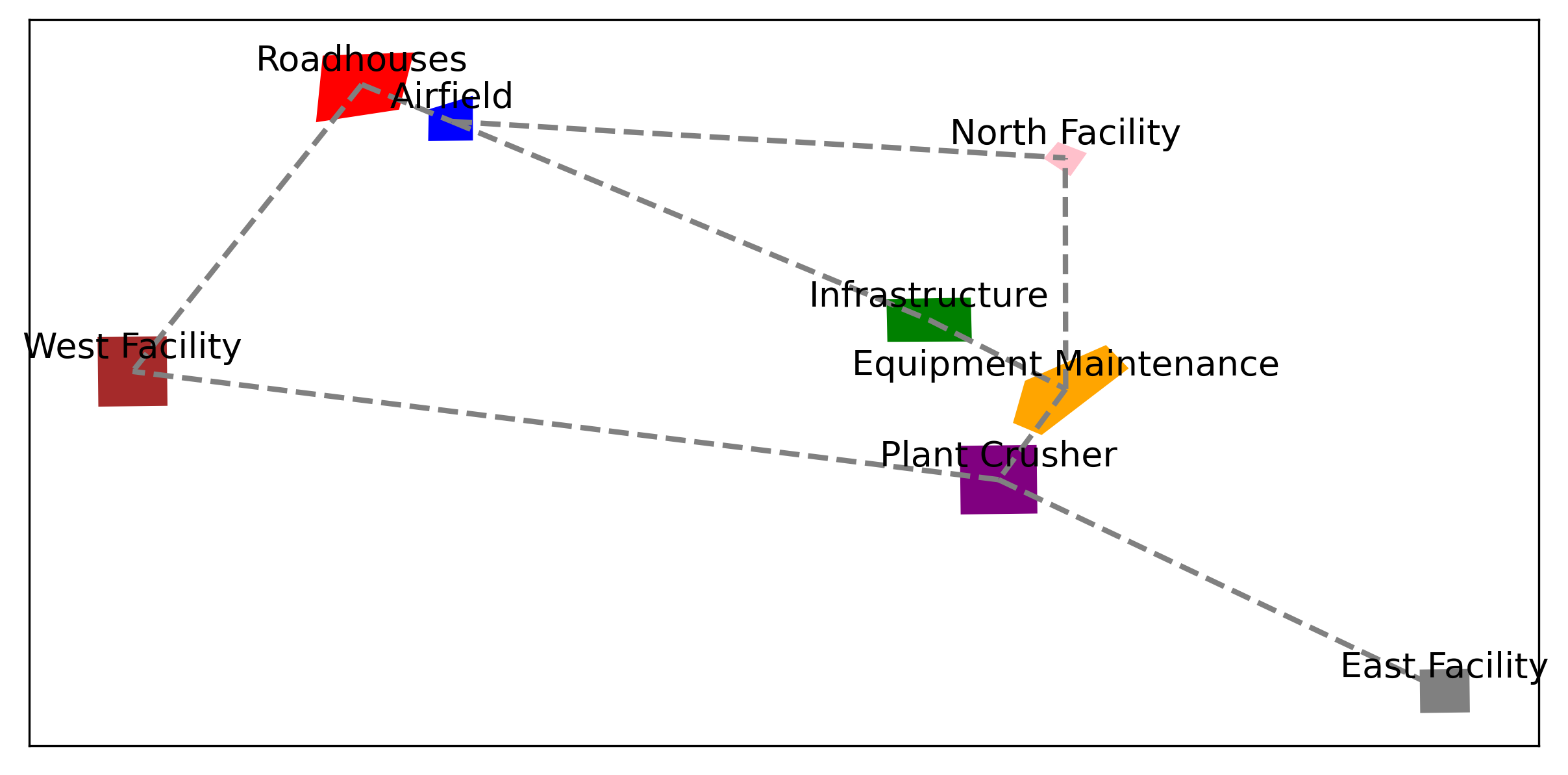}
    \caption{Identified eight charging zones and roads connected to eight charging zones (indicated as gray dashed lines)}
    \label{eight_charging_zones}
\end{figure}

\begin{figure}[!h]
    \centering
    \includegraphics[width=5in]{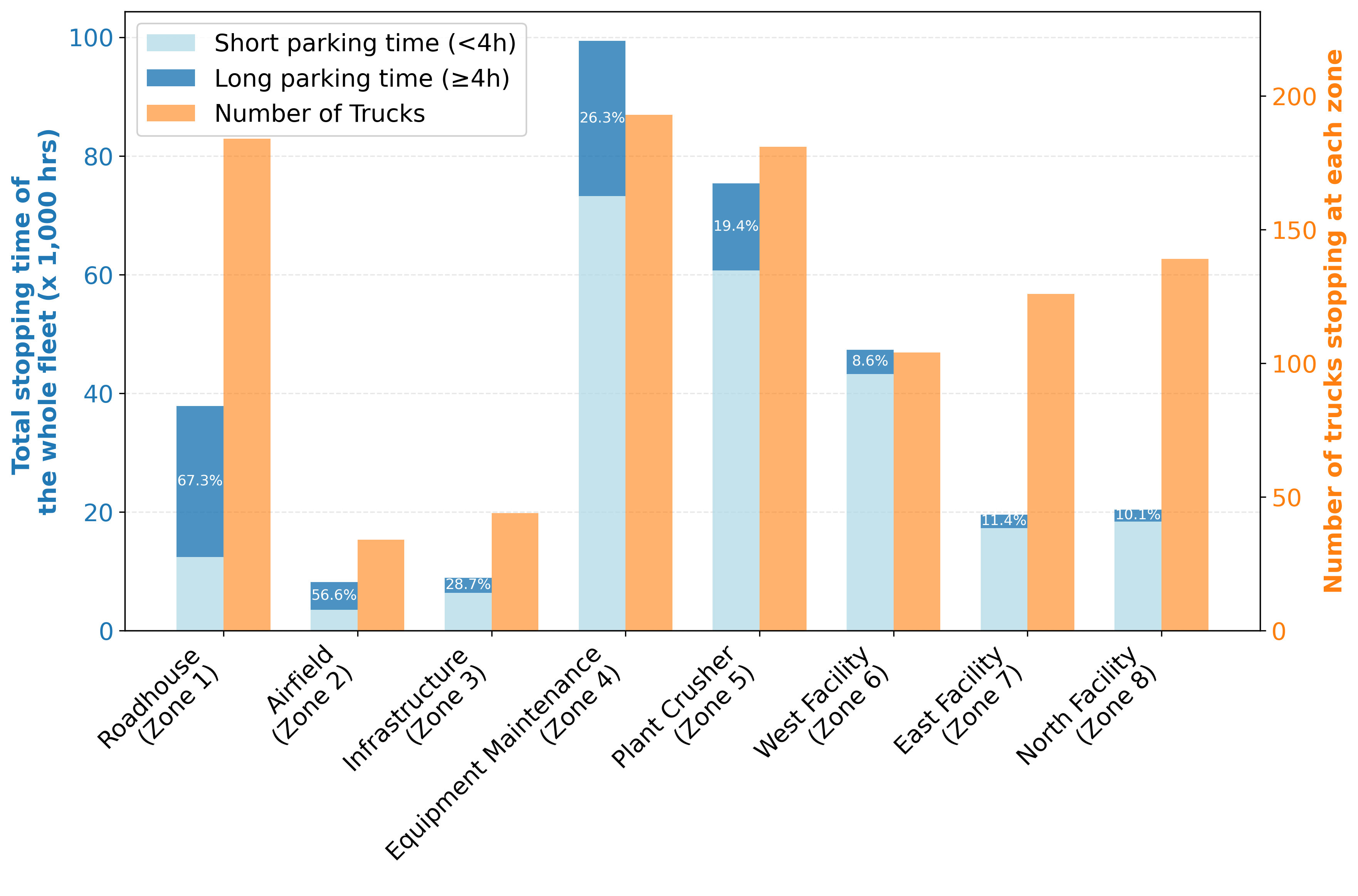}
    \caption{The total stopping time of the entire fleet ($\times$ 1,000 hours), including both short parking durations ($<$ 4 hours) and long parking durations ($\geq$ 4 hours), along with the annual number of trucks stopping at each of eight zones}
    \label{stopping_time}
\end{figure}

We used GPS data from 189 light-duty trucks at a mining site over one year. Their GPS data records include coordinates (i.e., longitude and latitude) and speed at each time step.  We then grid-divided the entire mining site into 100-meter $\times$ 100-meter square regions and counted the stop frequency in each square. We assumed that a truck stops if it moves less than 50 meters within a square region for at least five minutes. By comparing the number of stop frequencies per square, we identified eight EV charging points with the highest annual stop frequency. Fig. \ref{eight_charging_zones} shows eight EV charging points identified, and the dashed gray lines indicate the roads connecting them. Fig. \ref{stopping_time} shows the total stopping time of the entire fleet ($\times$ 1,000 hours), including both short parking durations ($<$ 4 hours) and long parking durations ($\geq$ 4 hours), along with the annual number of trucks stopping at each zone. The detailed statistics of eight charging zones are listed in Table \ref{tab:traffic_stops} in Appendix \ref{charging_points}. Zones 1 (roadhouse) and 4 (equipment maintenance) account for a large share of long-duration time greater than 4 hours, indicating that most overnight parking occurs in zones 1 and 4. 

\subsection{Summary of input datasets}

\label{summary_inputs}

We extracted three inputs for each truck in each half-hour interval based on their GPS records: (i) the travel distance of each truck in miles, (ii) the parking time windows as the input of the set $\mathcal{S}_{i,d,t}$, and (iii) the effective parking duration in each time window in minutes as the input of $pp_{i,d,t}$. The travel distance is then converted to energy consumption for each half-hour parking time window based on the second-order model of the temperature-dependent fuel economy for U.S. trucks \cite{goodall_feasibility_2024}, as the input of $\rho_{i,d,t}$. The detailed procedures to extract all three kinds of inputs are illustrated in Appendix \ref{inputs}.

 \begin{figure}[!h]
    \centering
    \includegraphics[width=\linewidth]{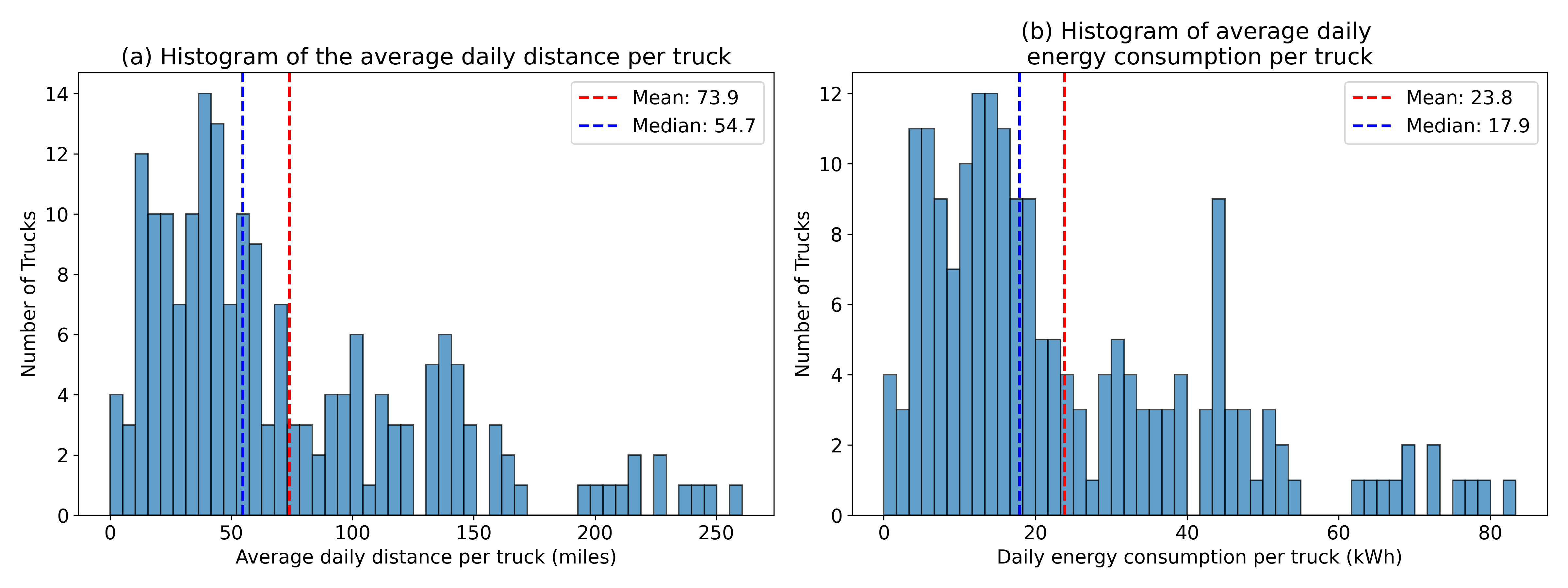}
    \caption{(a) The histogram of average daily distance per truck (miles) and (b) the histogram of average daily energy consumption per truck (kWh)}
    \label{histogram}
\end{figure}

Figs. \ref{histogram} (a) and (b) illustrate the distributions of average daily distance traveled per truck and the corresponding average daily energy consumption. On average, each truck traveled 73.9 miles per day. With a fuel economy of roughly 3–3.4 miles/kWh, this translates to a daily energy use of 22.4–24.6 kWh per truck, or about 30\% of the battery’s capacity. This implies that a full battery cycle would typically last 2–3 days. It is important to note that travel distance and energy consumption within charging zones are excluded from the model. We expect, however, that this additional consumption could be offset with minor operational adjustments, such as modifying parking durations.

\subsection{ML clustering for constructing feature-based robust sets}
\label{robust}

  \begin{figure}[!h]
    \centering
    \includegraphics[width=5in]{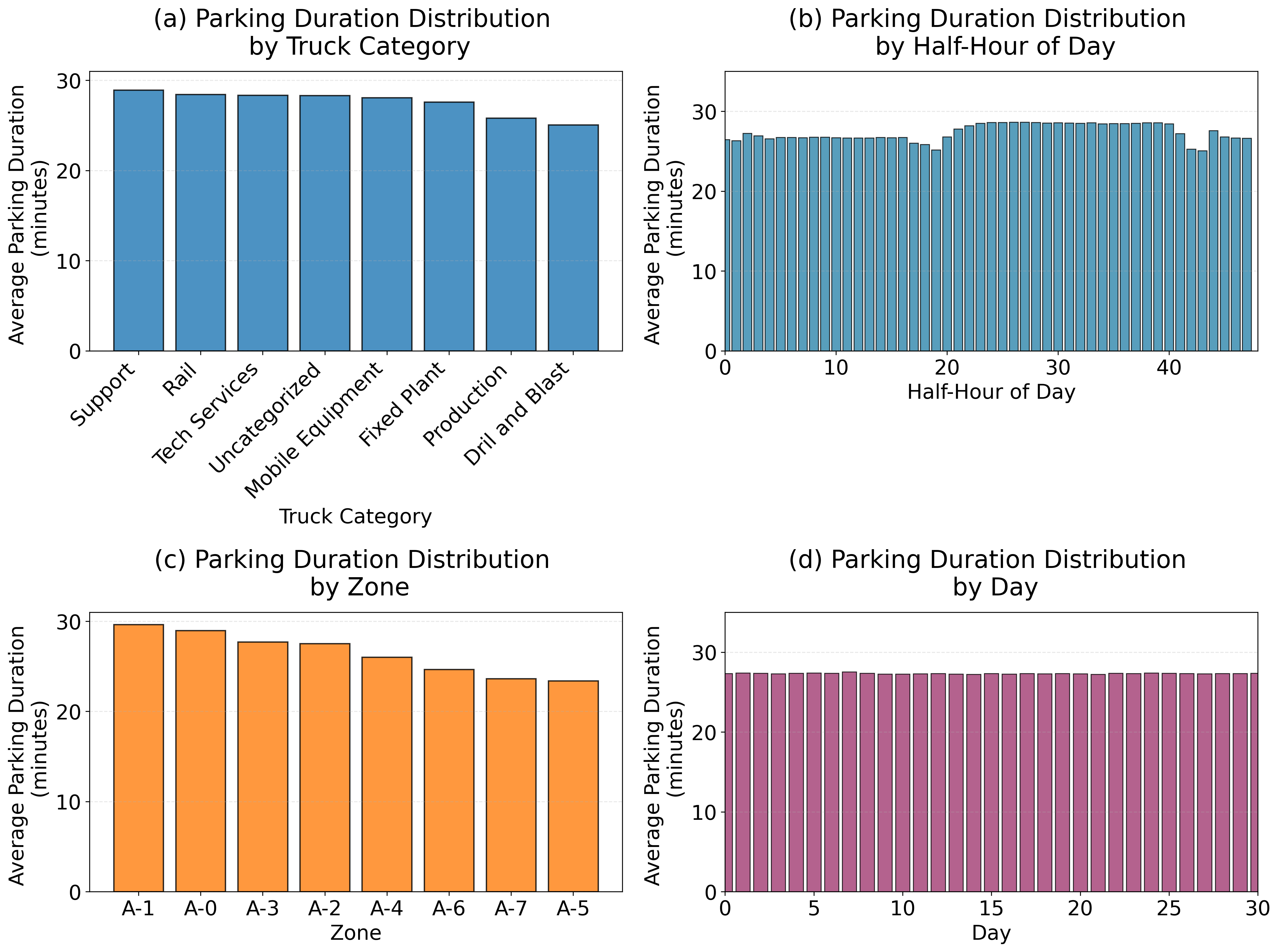}
    \caption{Parking duration distributions in a year in terms of four features: (a) truck category, (b) half-hour of a day, (c) zones, and (d) day of the month}
    \label{parking_minutes_histogram}
\end{figure}

  \begin{figure}[!h]
    \centering
    \includegraphics[width=\linewidth]{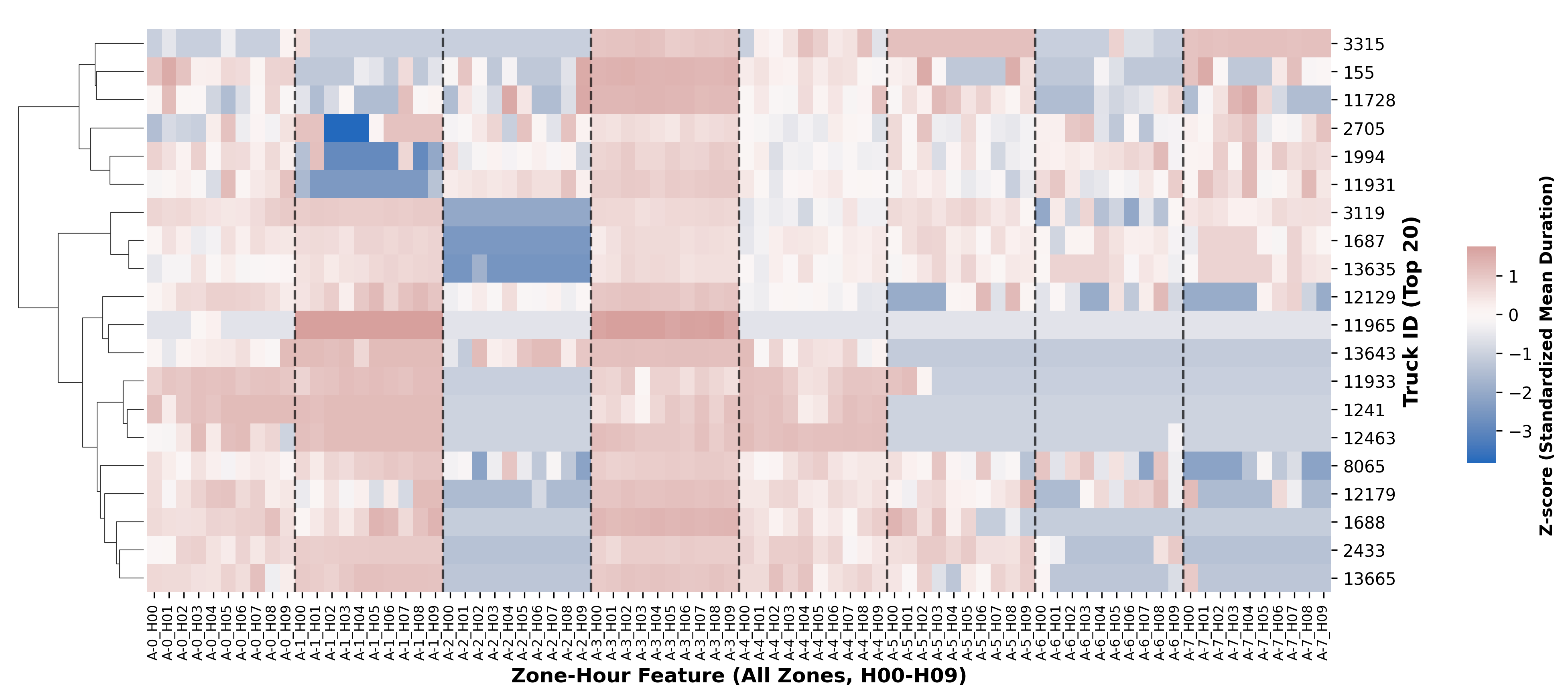}
    \caption{Average parking duration heatmap for the selected 20 trucks, constructed using rearranged zone–hour features. The heatmap displays only the first 10 hours for each of the eight zones, with zone boundaries indicated by black dashed lines. The hierarchical clustering dendrogram, based on feature distances, is shown on the left.}
    \label{feature_heatmap}
\end{figure}

To construct a feature-based, robust uncertainty set for heterogeneous parking duration, as illustrated in Section \ref{feature_based}, we first collected all parking durations in a year. Fig.~\ref{parking_minutes_histogram} (a) - (d) illustrate parking duration distributions in terms of four features, truck category, half-hour of the day, zone, and day of the month. We observed correlations between truck categories, half-hour of the day, and zones and parking duration. We then rearranged the time-zone features based on their combinations, and this gives a total number of $8 \text{zones} \times 48 \text{time slots} = 384$ features.

We then apply the proposed hierarchy clustering to group trucks based on the re-arranged time-zone features, as demonstrated in Section \ref{feature_based}. Before clustering, we normalized the parking duration of these trucks using z-scores. The z-score standardizes the data for each truck to a mean of zero with a standard deviation of one. In other words, the average parking duration of a truck is normalized to one, and the duration that is higher or lower than the average value is a positive or negative value, respectively.  Fig.~\ref{feature_heatmap} shows the normalized parking duration heatmap for the selected 20 trucks, constructed using rearranged zone–hour
features, with the hierarchical clustering dendrogram on the left. We first grouped trucks by category. Using the category information, hierarchical clustering yields three subgroups per category, for a total of 24 subgroups. We also compare these methods with the other three: hierarchical clustering without category information, k-means, and grouping each truck separately. These comparisons and clustering results are detailed in Appendix \ref{machine_learning_clustering}.

\subsection{Set up}
\label{set_up}

We prepared three inputs for the selective and full-year datasets for 2023. For the selective day approach, we defined 10 statistical criteria to select 39 days, including total fleet distance, number of active trucks, and total fleet energy consumption. Based on these criteria, we ranked all 365 days in a year and selected the top-ranking days. We tested different numbers of selected days, from 10 to 40, and observed that charger installation results stabilized at 30 days. Accordingly, we selected 39 days, including days adjacent to the 10 top-ranked days, because some vehicles charge every 2 or 3 days, and representative sub-periods must extend beyond battery cycles. The detailed selection criteria and procedures are provided in Appendix \ref{representative_day}. For the full year using rolling-horizon scheduling, we divided the full-year dataset into monthly subsets, with overlapping days between consecutive months to maintain trip continuity. This produces 12 monthly sub-datasets, each spanning 34 days.

 We conduct a preliminary simulation before optimization, in which each EV uses straightforward fast charging throughout the optimization horizon. In each simulation, a truck starts with a random SoC, and is then fully charged at each parking site. At the end of the optimization horizon, we track whether they could restore their original SoC. If the original SoC cannot be recovered, the truck is identified as a HCV, which is not feasible to be electrified if they stick to the current operational scheduling. This also includes vehicles that do not pass eight charging zones. The optimization is subsequently performed exclusively on the remaining trucks, and we recorded the number of HCVs excluded from the optimization model for each simulation. All experiments were solved using the Gurobi commercial solver. We set the accepted optimal gap for this MILP to be 1\%, less than the capital cost of one slow charger. All problems were solved on a computer equipped with an 8-core CPU running at 3.4 GHz and 160 GB of RAM.

\subsection{Result evaluation}
\label{results}

\subsubsection{Performance metrics} Apart from the number of fast and slow chargers installed, the following performance metrics are introduced to evaluate performance: 

\begin{itemize}

\item Fast charging hours (FC Hrs, per truck per day): The aggregated duration, measured in half-hour intervals, that a truck uses a fast charger per day.
\item Slow charging hours (SC Hrs, per truck per day): The aggregated duration, measured in half-hour intervals, that a truck uses a slow charger per day.
\item Abandonment hours (Ab Hrs, per truck per day):  The aggregated duration, measured in half-hour intervals, that a truck is abandoned and charged per day.
\item Average charging power (Avg. CP, kW): The mean power charged during all charging time slots.
\item Percentage of the number of periods when battery SoC is lower than 30\% relative to the total time slots (Pct. of Low SoC, \%): The total number of half-hour periods when the battery SoC is lower than 30\%.
\item Average battery SoC of each truck (Avg. SoC, \%): The average battery SoC of each truck throughout the trip.
\item Charging utilization rate (CUR, \%): The average number of charging duration, measured in half-hour intervals, each charger experiences compared to the optimization horizon. 

\end{itemize}

\subsubsection{The impacts of abandonment, range anxiety, and effective parking duration}
\label{impacts}

Using 39 representative days, we evaluate the impacts of abandonment, range anxiety, and effective parking duration using the HS method.

\begin{itemize}
    \item \textbf{Case 1 - Benchmark}: Trucks are prohibited from waiting in any zone under normal circumstances. However, when a vehicle's SoC drops to below 30\%, it is permitted to pause for 30 minutes to recharge. Two designated zones (i.e., zones 1 and 4) accommodate the overnight charging with unlimited waiting time. The model incorporates the specific parking duration for each parking window (i.e., $ \frac{5}{30} \leq pp \leq 1$).  
    \item \textbf{Case 2 - Full parking time}: This case maintains the same as the benchmark case, except that it has no specific parking duration constraints for each parking window (i.e., $ pp = 1$).
    \item \textbf{Case 3 - No overnight charging}: This case maintains the same conditions as the benchmark case, except that it has no special zones for the overnight charging.
    \item \textbf{Case 4 - No range anxiety}: This case maintains the same conditions as the benchmark case, except that it has no related constraints or penalty related to the range anxiety, in order words, there is no penalty for low SoC (i.e., $ P^{low} = 0$). Trucks are prohibited from waiting in any zone under normal circumstances, regardless of their SoC; in other words, the maximum waiting time is set to zero (i.e., $ T^{max}_{i,d,t} = 0$).

\end{itemize}

\begin{figure}[!h]
    \centering
    \includegraphics[width=5in]{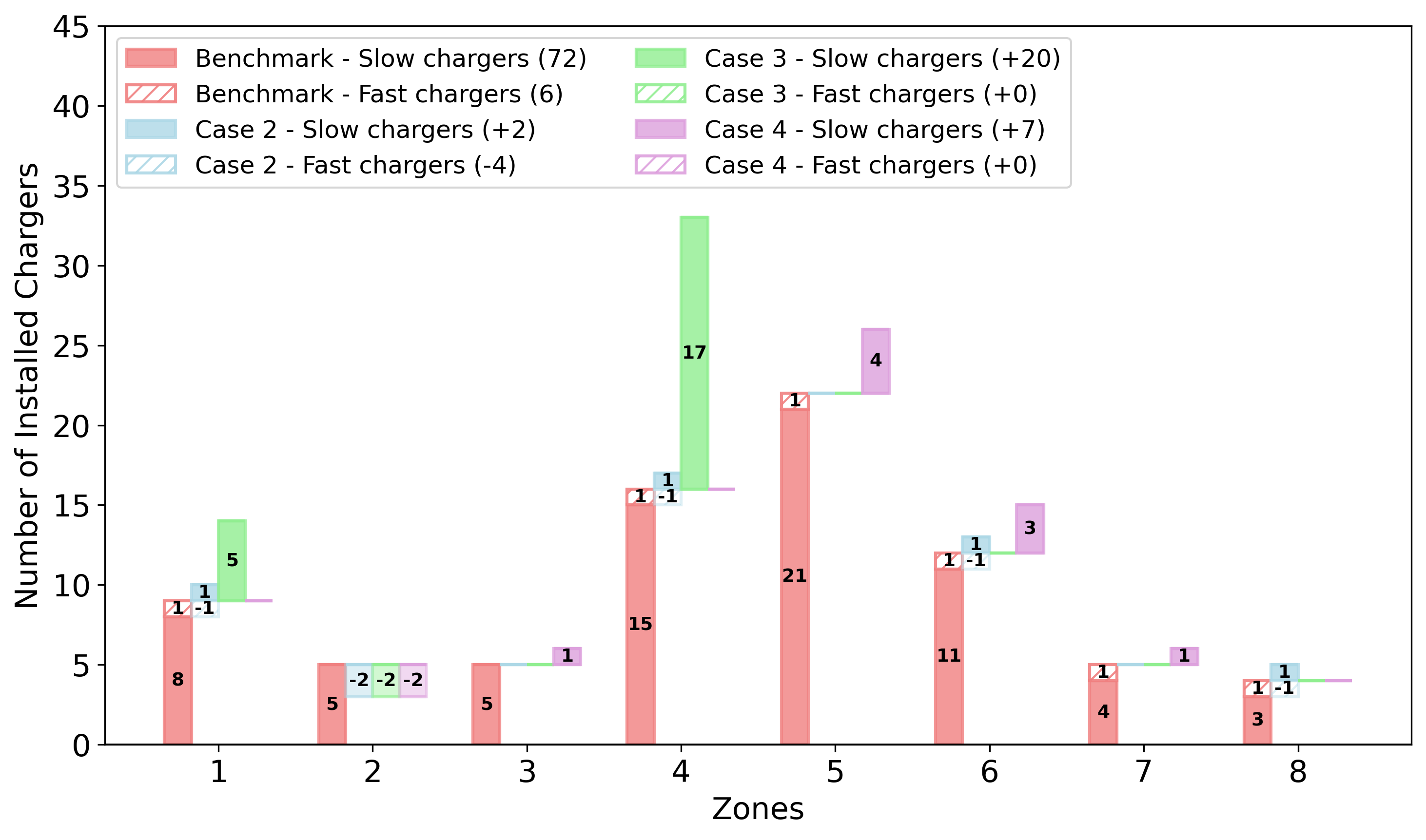}
\caption{Zone-level differences in the numbers of slow and fast chargers for Cases 2–4 relative to the benchmark. Dashed bars represent fast chargers, while solid bars represent slow chargers.}
    \label{differences_in_charger_number}
\end{figure}

Table \ref{tab:scenario_comparison} summarizes optimization results in four deterministic cases under different assumptions about abandonment, range anxiety, and effective parking duration. The benchmark case (case 1) has a mixed type of chargers, which are 72 of slow chargers and 6 of fast chargers. Fig. \ref{differences_in_charger_number} illustrates differences in the numbers of fast and slow chargers in cases 2–4, compared to the benchmark. In the benchmark case, the number of installed chargers is proportional to the total duration of vehicle stops at each of the eight charging zones, as presented in Fig. \ref{parking_minutes_histogram} (a), while zones 4 and 5 have the highest number of chargers among the eight zones. We then examine three cases in which overnight charging, range anxiety, and limited parking duration are not considered. Extending the parking duration to the full window reduces the number of fast chargers to three. Removing the designated overnight zones would require an additional 20 slow chargers to be installed in zones 1 and 4. In case 4, no range anxiety requires seven more slow chargers to be installed.

\begin{table}[h!]
\renewcommand{\arraystretch}{1.1} % Adjust row height
\centering
\begin{tabular}{c l c c c c c}
\hline
Cases
  & \makecell{Names} 
  & \makecell{\# of slow / \\ fast chargers} 
  & \makecell{Installation \\ costs (\$)} 
  & \makecell{Scheduling\\ penalty (\$) } 
  & \makecell{ Gap \\ (\%)}  
  & \makecell{ Time (Hrs)} \\
\hline
1 & Benchmark & 72/6 & 336,000 & 38,468 & 0.084 & 2.98\\ \hline
2 & Full parking time & 74/3 & 187,000 & 39,622 & 0.116 & 57.32 \\ \hline
3 & No overnight charging & 92/6 & 366,000 & 85,200 & 0.042 & 15.77 \\ \hline
4 & No range anxiety & 79/6 & 346,500 & 38,077 & 0.079 & 2.94 \\ 
\hline
\end{tabular}
\caption{Summary of charging planning outcomes and overall scheduling penalties in four deterministic cases: Benchmark, Full parking time, No overnight charging, and No range anxiety}
\label{tab:scenario_comparison}
\end{table}

\begin{table}[h!]
\renewcommand{\arraystretch}{1.1} % Adjust row height
\centering
\begin{tabular}{c c c c c c c c}
\hline
\makecell{Cases} 
  & \makecell{FC Hrs \\ (truck $\cdot$ day)} 
    & \makecell{FCUR \\ (\%)} 
    & \makecell{SC Hrs \\ (truck $\cdot$ day)} 
      & \makecell{SCUR \\ (\%)} 
  & \makecell{Ab Hrs \\ (truck $\cdot$ day)} 
    & \makecell{Pct. low SoC \\ (\%)}
& \makecell{Avg. CP\\ (kW)} \\
\hline
1 & 0.18 & 20.64 & 0.50 & 5.46 & 3.60 & 6.01 & 16.51 \\
\hline
2 & 0.08 & 19.78 & \textbf{0.79} & \textbf{8.34} & 3.60 & 4.91 & 13.11 \\
\hline
3 & \textbf{0.19} & \textbf{21.65} & 0.48 & 4.06 & \textbf{9.46} & 2.12 & 16.91 \\
\hline
4 & 0.17 & 19.06 & 0.43 & 4.21 & 3.68 & \textbf{62.57} & \textbf{19.03} \\
\hline
\end{tabular}
\caption{Summary of all the performance metrics in four deterministic cases; The number in \textbf{bold} highlights the highest values among four cases}
\label{scheduled_results_deterministic}
\end{table}

\begin{figure}[!h]
    \centering
    \includegraphics[width=\linewidth]{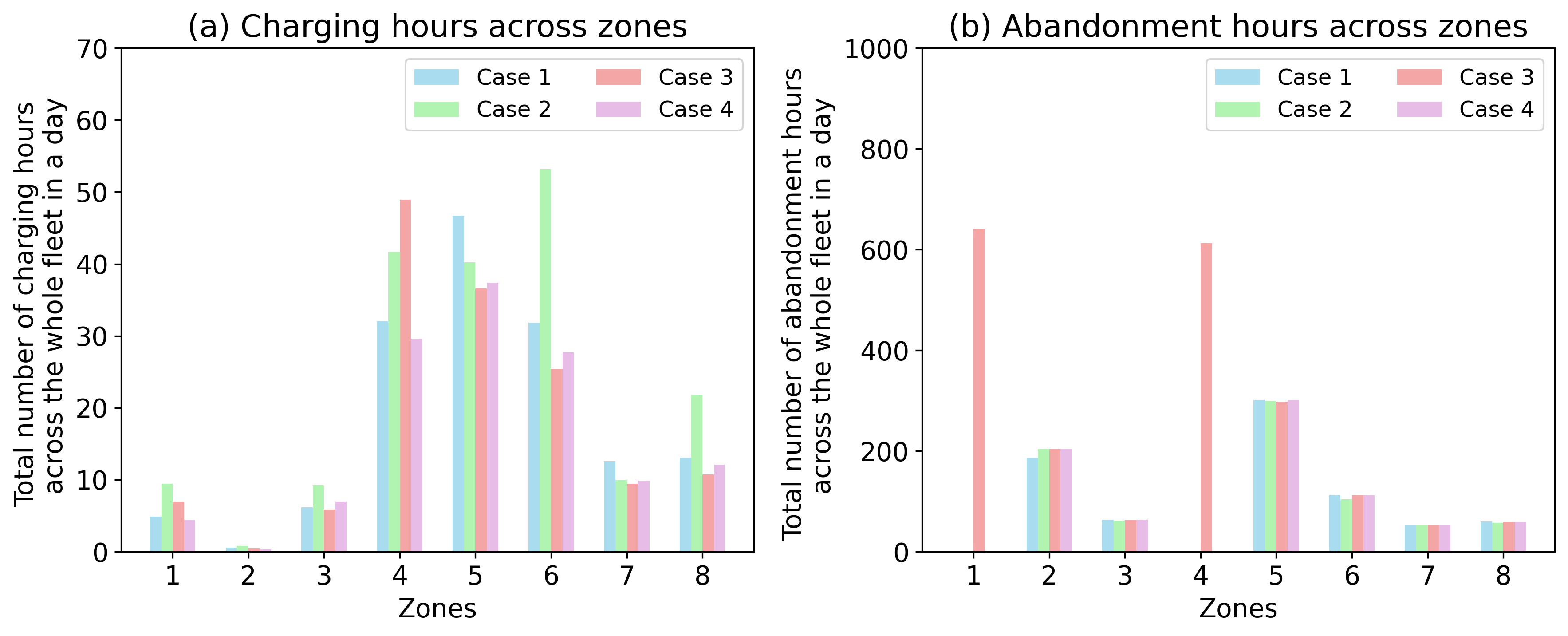}
\caption{The total number of (a) charging hours and (b) abandonment hours of the truck fleet in different zones in four deterministic cases}
    \label{different_in_zone}
\end{figure}

\begin{figure}[!h]
    \centering
    \includegraphics[width=\linewidth]{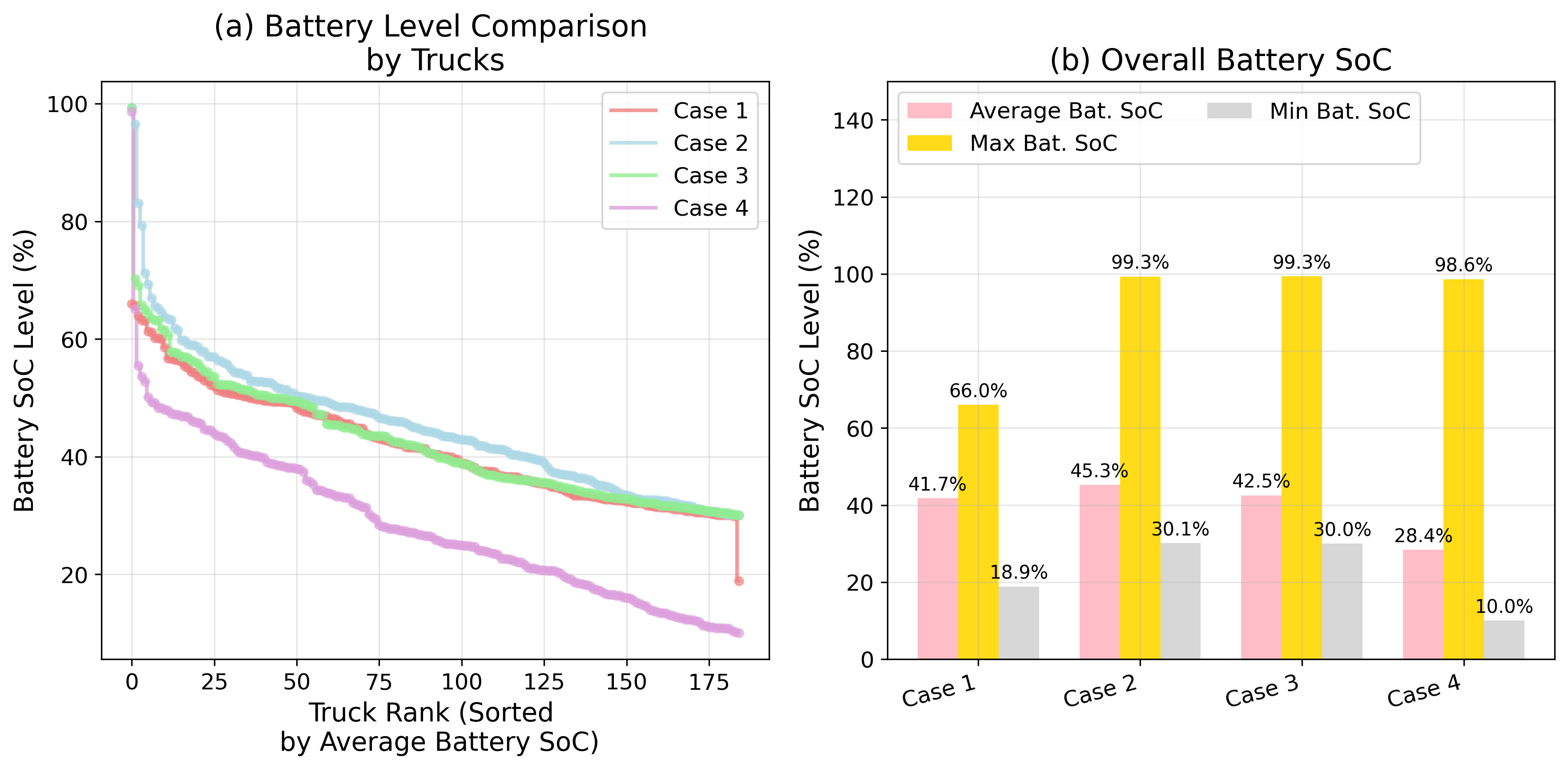}
\caption{(a) Average SoC of each truck and (b) the overall battery SoC for the entire fleet in four deterministic cases}
    \label{Average_SoC}
\end{figure}

Table \ref{scheduled_results_deterministic} summarizes all the performance metrics in four deterministic cases, and the detailed scheduling solution of four cases are presented in Appendix \ref{detailed_results_deterministic_cases}. As expected, case 2, which assumes the full parking duration, has the highest fast-charging hours and utilization rate among the four cases. Case 3, which excludes overnight charging, has the highest number of fast-charging and abandonment hours. Case 4, with no range anxiety, leads to the highest percentage of low SoC events and the average charging power.

Figs. \ref{different_in_zone} (a) and (b) show the zone-wise comparison of charging and abandonment time, respectively. In case 2, with the full parking duration, zones 6 and 8 have longer charging times.  In case 3, without the overnight charging, zones 1 and 4 have a significantly high number of abandonment hours. Fig. \ref{Average_SoC} (a) illustrates the average SoC of all trucks over 39 days, ranked from the highest to the lowest, and Fig. \ref{Average_SoC} (b) presents the overall battery SoC (i.e., the maximum, minimum, and average) of the entire fleet in four cases. Among four cases, case 4 (no range anxiety) yields the lowest average SoC ($28.4\%$), whereas case 2 with full parking duration achieves the highest average SoC ($45.3\%$). When the effective charging time constraint is applied, in other words, comparing the benchmark to case 2, part of the fleet in the benchmark case experiences a sharp drop in the average SoC to very low levels ($\leq 20\%$), due to the limited charging duration in the parking windows.

\subsubsection{The impacts of different RO-DDU formulations for parking duration uncertainty}
\label{decision-dependent}

We next evaluate the performance of two RO-DDU formulations for the charging power constraint under parking duration uncertainty as in Section \ref{two_ddu}. These formulations differ in whether the uncertainty set depends affinely on instantaneous charging decisions (hereafter referred to as `\textit{Instantaneous}') or on cluster-level expected charging decisions (hereafter referred to as `\textit{Expect}'). We adopt fleet groups obtained from hierarchical clustering that incorporates truck category information, as described in Section~\ref{robust}. We then apply different scaling factors for slow chargers ($k_{\text{slow}} = 4$) and fast chargers ($k_{\text{fast}} = 2$) in the instantaneous and cluster-level RO-DDU formulations. Here, we envisage a pessimistic scenario in which parking duration will be reduced after BEV chargers are installed. In addition, we include an RO independent of charging decisions (`RO-DI') as a benchmark. All three cases are evaluated using 39 selected days, as defined by the IPS method. After obtaining the charger planning results, we use the sample average approximation (SAA) to compare the scheduling performance of the \textit{Instantaneous} and \textit{Expect} cases.

\begin{table}[h!]
\renewcommand{\arraystretch}{1.1}
\centering
\begin{tabular}{l l c c c c c c}
\hline
Cases
  & \makecell{Formulations} 
  & \makecell{\# of \\ HCVs} 
  & \makecell{\# of slow / \\ fast chargers} 
  & \makecell{Installation \\ costs (\$)} 
  & \makecell{Scheduling\\ penalty (\$)} 
  & \makecell{Gap \\(\%)}  
  & \makecell{Time \\ (Hrs)}\\
\hline
1 
& RO - DI
& 19
& 67 / 6 
& 328{,}500 
& 20{,}911
& 0.01 
& 0.48 \\ \hline

2 
& Instant.
& 19
& 67 / 6 
& 328{,}500 
& 30{,}400
& 0.01
& 0.52 \\ \hline

3 
& Expect
& 19
& 70 / 6 
& 333{,}000
& 31{,}306
& 0.01 
& 10.21 \\ \hline
\end{tabular}
\caption{Summary of charging planning outcomes and scheduling penalties under three RO formulations (low-penalty case)}
\label{Instantaneous_versus_cluster}
\end{table}

\begin{figure}[!h]
    \centering
    \includegraphics[width=5in]{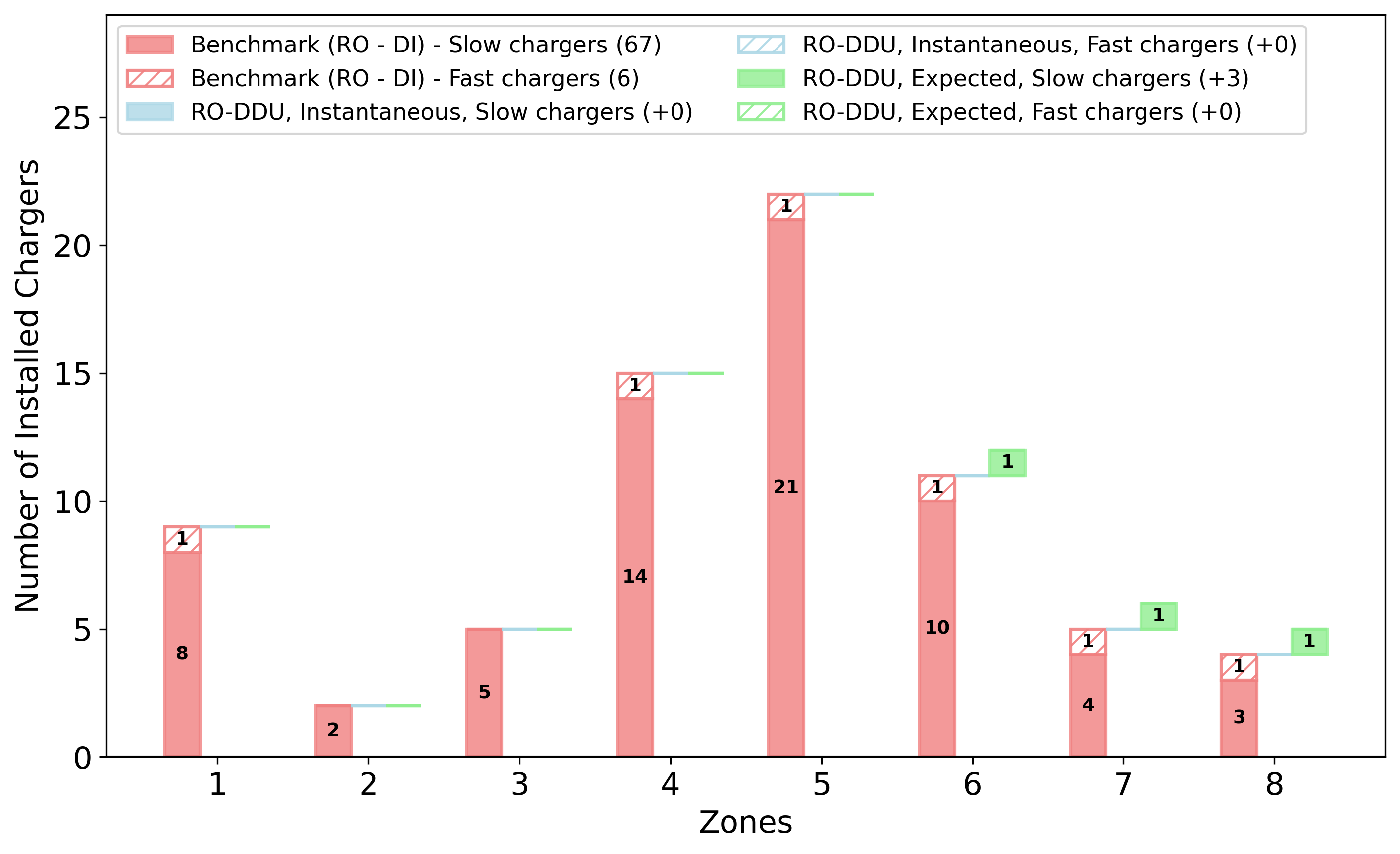}
\caption{Zone-wise differences in the numbers of slow and fast chargers for three RO cases: RO - DI, RO-DDU (Instantaneous), and RO-DDU (Expect). The benchmark case is the RO - DI case}
    \label{Installed_charger_different_cluster}
\end{figure}

\begin{table}[h!]
\renewcommand{\arraystretch}{1.1}
\centering
\small
\begin{tabular}{l l c c c c c c}
\hline
\makecell{Sample}
& \makecell{Cases} 
& \makecell{FC Hrs \\ (truck $\cdot$ day)} 
& \makecell{SC Hrs \\ (truck $\cdot$ day)} 
& \makecell{Ab Hrs \\ (truck $\cdot$ day)} 
& \makecell{Pct. of low \\ SoC (\%)} 
& \makecell{Avg. CP \\ (kW)} 
& \makecell{Tot. Penalty \\ (\$/truck $\cdot$ day)} \\
\hline
\multirow{3}{*}{\makecell{In\\sample}}
& RO-DI
& 0.16 & 0.45 & \textbf{4.14} & 1.52 & 9.47 & 4.43 \\
& Instant.
& \textbf{0.22} & \textbf{0.52} & 4.13 & 1.72 & 7.80 & \textbf{4.51} \\
& Expect.
& 0.16 & 0.46 & 4.10 & \textbf{2.35} & \textbf{9.48} & 4.40 \\
\hline
\multirow{2}{*}{\makecell{Out of\\sample}}
& Instant.
& 0.21 & 0.39 & 3.77 & 4.38 & 9.28 & \textbf{4.02} \\
& Expect.
& 0.21 & 0.39 & 3.77 & \textbf{4.90} & \textbf{9.33} & 4.01 \\
\hline
\end{tabular}
\caption{Summary of performance metrics for in-sample and out-of-sample evaluations under RO-DI, instantaneous DDU, and expectation-based DDU at the low-penalty case. Values are rounded to two decimal places. The values in \textbf{bold} indicate the highest value within each sample group and column.}
\label{tab:performance_comparison_combined}
\end{table}

Table \ref{Instantaneous_versus_cluster} summarizes the charger installation results of three RO cases obtained in the in-sample datasets. Using the IPS approach can decompose the problem and solve three instances quickly, compared to the HS method. It relaxes the battery energy balance over a long scheduling horizon. Fig. \ref{Installed_charger_different_cluster} shows their zone-wise differences in the numbers of installed slow and fast chargers. Only when using the cluster formulation, the required number of chargers increases, particularly in zones 6, 7, and 8, where the charging duration variability is higher than in other zones. In addition, the cluster formulation requires substantially longer solution times due to the introduction of nonlinear charging-power constraints.

Table \ref{tab:performance_comparison_combined} summarizes the scheduling performance metrics of in-sample and out-of-sample tests in three cases. During in-sample tests, the RO-DI formulation yields the highest charging power, followed by RO-DDU (Expect) and RO-DDU (Instantaneous). This aligns with the theoretical analysis of charging power bounds via cluster-level aggregation in Section \ref{subsec:efficiency_analysis}. The RO-DDU (expect) formulation accounts for decision dependence and the correlation among charging decisions within the same cluster. Therefore, the RO-DDU (expect) formulation has the highest scheduling penalty among the three cases. In out-of-sample tests, the scheduling penalty obtained from the RO-DDU (expect) formulation becomes less than the RO-DDU (Instantaneous) formulation. This means that, by accounting for correlations within the cluster, the expected formulation yields more efficient scheduling.

\begin{table}[h!]
\centering
\renewcommand{\arraystretch}{1.1} % Adjust row height
\begin{tabular}{c c l c c c c c c}
\hline
\makecell{Data \\ Range} 
& \makecell{$\Delta_c$} 
  & \makecell{Solving \\ methods} 
  & \makecell{\# of \\ HCVs} 
  & \makecell{\# of slow/fast \\ chargers}
  & \makecell{Installation \\ costs (\$)} 
    & \makecell{Scheduling \\ penalty  (\$)} 
  & \makecell{Gap \\(\%) } 
  & \makecell{Time \\(Hrs)} \\
  %& \makecell{\# of instances\\ reaching time \\ limits}\\
\hline
\multirow{2}{*}{DS} 
& \multirow{2}{*}{0}  
  & HS  & 19 & 68/6 & 330,000 & 35,678 &  \textbf{9.06} & 11.1 \\
  && HS - h & 19 & 68/6 & 330,000 & 35,679 & 0.07 & 2.40 \\
\hline
\multirow{2}{*}{DS} 
& \multirow{2}{*}{$\pm \sigma$}  
  & HS  & 19 & 68/6 & 330,000 & 35,658 & \textbf{9.04} & 11.1 \\
  && HS - h & 19 & 69/6 & 331,500 & 36,065 & 0.05 & 2.15 \\
\hline
\multirow{2}{*}{DS} 
&$\pm \sigma$   & HS  & 19 & 67/8 & 442,500 & 35,565 & \textbf{23.19} & 11.1 \\
&DDU& HS - h & 19 & 70/6 & 330,000 & 22,455 & 0.06 & 3.29 \\
  % && \textcolor{blue}{IPS} & 19 & 70/6 & 330,000 & 22,455 & 0.008 & 10.21 \\
\hline
\multirow{2}{*}{DS} 
& \multirow{2}{*}{$\pm 2 \sigma$}  
  & HS  & 19 & 68/6 & 330,000 & 35,736 & \textbf{9.02} & 11.1 \\
  && HS - h & 19 & 69/6 & 331,500  & 35,863 & 0.05 & 5.20 \\
  % && \textcolor{red}{IPS} & 19 & 76/8 & 418,000 & 33,402 & 0.05 & 23.8 \\
\hline
% \multirow{3}{*}{FY}
% &0 & IPS   & 48 & 77/5 & 305,500 & 33,466 & 0.09 & 11.3 \\
% & $\pm \sigma$ & IPS   & 48 & 75/7 & 378,500 & 33,331 & 0.05 & 12.0 \\
% & $\pm 2\sigma$ & IPS  & 48 & 76/8 & 418,000 & 33,402 & 0.05 & 23.8 \\
% \hline
\end{tabular}
\caption{Summary of optimization results with the varying size of uncertainty set based on the selected 39-day and full-year dataset using HS, HS with heuristic (HS - h)}
\label{tab:solving_types}
\end{table}

\textit{Tackling high computation costs with the `fix-and-optimize' heuristic:} Considering the high computation costs of the 39-day instance under HS scheduling, we leverage the fix-and-optimize heuristic to solve problems. We impose a maximum time limit of 40,000 seconds for each instance and report the optimality gap when the solver fails to converge within this limit. In addition, we initialize the master problem as the warm start.

As presented in Table \ref{tab:solving_types}, we found that solving these HS instances directly is difficult, with all four cases reaching the maximum time limit. Using the proposed heuristic method (HS - h), the problem can reach an acceptable optimality gap of less than 0.1\% while keeping the computation time within the prescribed limits. We used a warm start because the solution to the deterministic benchmark case is available, and completing 10 iterations takes approximately one hour. Across cases with varying uncertainty sets, the only difference between the HS and HS-h methods is the installation or placement of a single fast charger. This indicates that assigning specific fast chargers can hinder convergence. In contrast, the proposed heuristic prioritizes placing the fast charger, allowing the model to converge more efficiently. The detailed solution processes using the HS-h method are presented in Appendix \ref{fix_and_optimize}.

\subsubsection{Robust infrastructure planning based on the full year dataset}

To ensure robust infrastructure planning over the year, we use the full-year dataset to plan the charging infrastructure under the IPS methods, as described in Section~\ref{robust algorithm}. These three full-year instances, corresponding to the size of uncertainty set increasing from 0 to $2\sigma$, provide the most robust planning results compared with the selective datasets.

\begin{table}[h!]
\centering
\renewcommand{\arraystretch}{1.1} % Adjust row height
\begin{tabular}{c c l c c c c c c}
\hline
\makecell{Data \\ Range} 
& \makecell{$\Delta_c$} 
  & \makecell{Solving \\ methods} 
  & \makecell{\# of \\ HCVs} 
  & \makecell{\# of slow/fast \\ chargers}
  & \makecell{Installation \\ costs (\$)} 
    & \makecell{Scheduling \\ penalty  (\$)} 
  & \makecell{Gap \\(\%) } 
  & \makecell{Time \\(Hrs)} \\
\hline
\multirow{3}{*}{FY}
&0 & IPS   & 48 & 77/5 & 305,500 & 33,466 & 0.09 & 11.3 \\
& $\pm \sigma$ & IPS   & 48 & 75/7 & 378,500 & 33,331 & 0.05 & 12.0 \\
& $\pm 2\sigma$ & IPS  & 48 & 76/8 & 418,000 & 33,402 & 0.05 & 23.8 \\
\hline
\end{tabular}
\caption{Summary of optimization results with the full-year dataset using the IPS method. For the IPS method, the reported gap is the optimality gap of the final period in the iterative planning.}
\label{tab:full_years}
\end{table}

\begin{figure}[!h]
    \centering
    \includegraphics[width=5in]{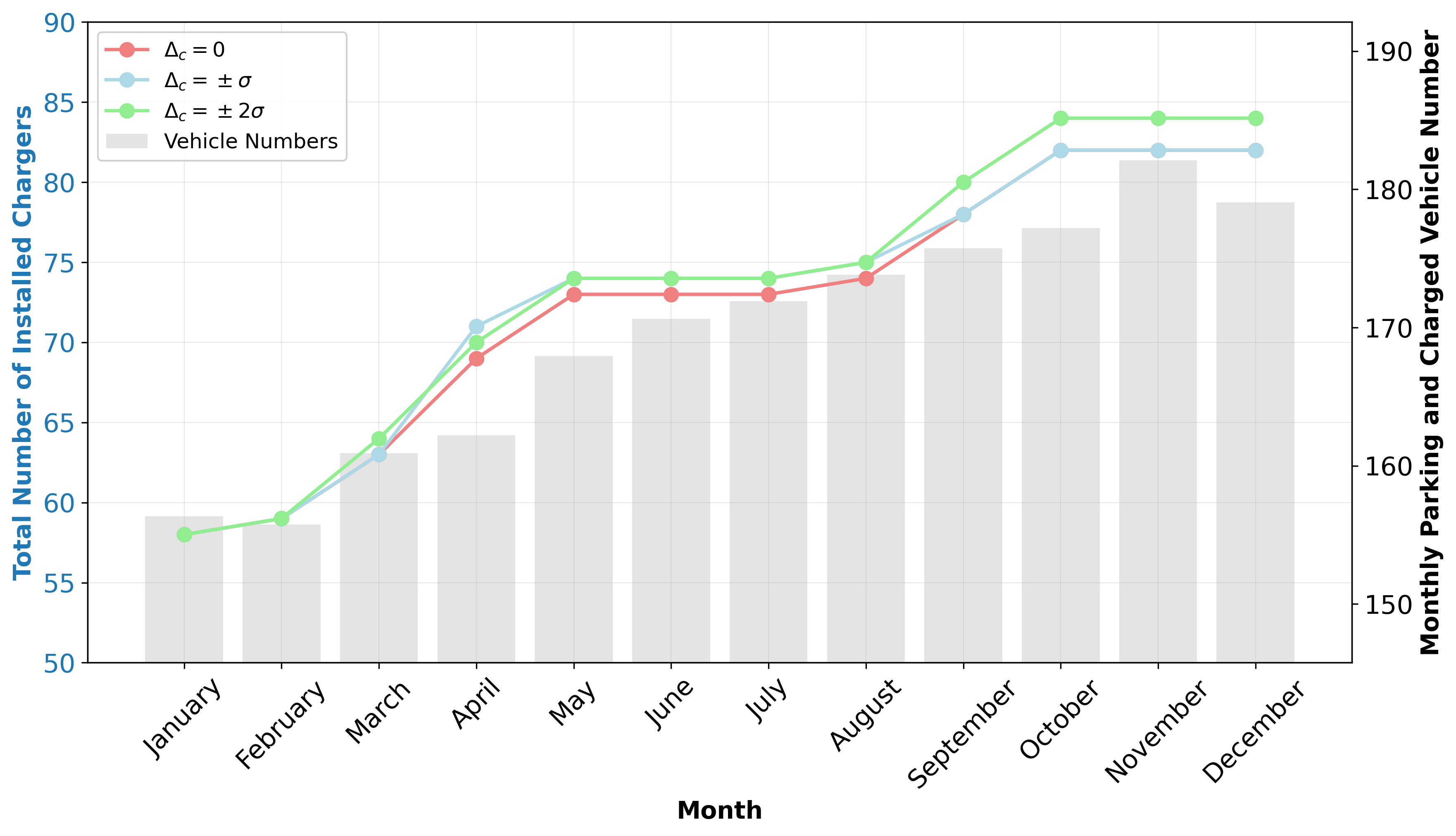}
\caption{The total number of chargers in each month based on the full-year dataset using the IPS method in three instances with varying sizes of uncertainty sets (in colored lines) versus the total number of trucks parking and charged (in gray bars) in all eight zones}
    \label{four_instances}
\end{figure}

\begin{figure}[!h]
    \centering
    \includegraphics[width=5in]{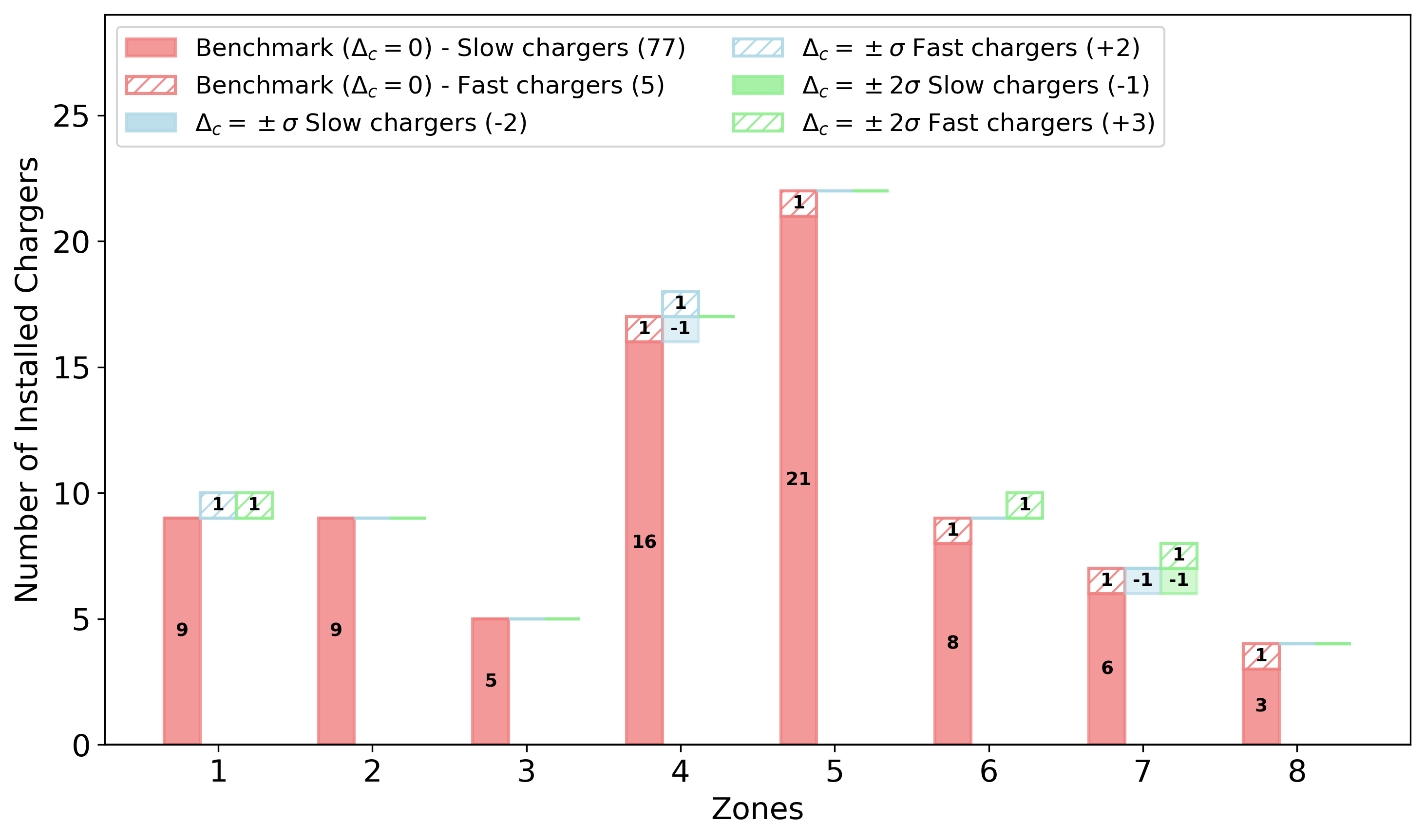}
\caption{Zone-level variations in the number of slow and fast chargers across three full-year cases are illustrated. The benchmark corresponds to the case with a zero-sized uncertainty set, in which parking duration is fixed at the mean value for each cluster.}
    \label{Installed_charger_different_alpha}
\end{figure}

Table~\ref{tab:solving_types} summarizes the optimization results. When extending the data range to the full year using the IPS method, between 1 and 3 iterations exceeded the time limit. However, all three full-year instances converge by the end of the 12-month horizon. In the final instance involving joint planning and scheduling, an optimality gap of less than 0.1\% is achieved. Fig.\ref{Installed_charger_different_alpha} shows zone-wise differences in the number of slow and fast chargers in three full-year cases with the increasing size of uncertainty sets. As uncertainty increases, two or three additional fast chargers are replaced with slow chargers.  

When comparing results from the 39-day dataset with those from the full-year dataset, the latter yields more chargers because the full-year data captures the end-of-year period with the highest vehicle counts. In contrast, the results derived from the 39-day dataset require even fewer chargers than the deterministic benchmark, because an additional 18 trucks are excluded from the HCV set due to their very short parking durations. Fig. \ref{four_instances} shows the total number of chargers (i.e., the sum of the number of fast and slow chargers) determined by the IPS procedure from monthly subproblems over 12 months in three instances. From January to December, the number of trucks parked and charged across the eight zones increases from 156 to 182, resulting in a steady rise in the number of required charger installations.

As the parking duration uncertainty grows, the total number of EV chargers increases, particularly for fast chargers. We then evaluated three full-year instances by calculating their performance metrics across 12 months. Table \ref{scheduled_results_stochastic} summarizes the performance metrics of all three cases. As the uncertainty increases, both fast and low charger utilization rates reduce. The average abandonment hour remains constant across the three stochastic instances. 

\begin{table}[h!]
\renewcommand{\arraystretch}{1.1}
\centering
\begin{tabular}{c c c c c c c c}
\hline
\makecell{$\Delta_c$} 
  & \makecell{FC Hrs \\ (truck $\cdot$ day)} 
  & \makecell{FCUR \\ (\%)} 
  & \makecell{SC Hrs \\ (truck $\cdot$ day)} 
  & \makecell{SCUR \\ (\%)} 
  & \makecell{Ab Hrs \\ (truck $\cdot$ day)} 
  & \makecell{Pct. low SoC \\ (\%)} 
  & \makecell{Avg. CP \\ (kW)} \\
\hline
$0$
& 0.21 
& \textbf{31.39} 
& \textbf{0.62} 
& \textbf{5.98} 
& 6.64 
& \textbf{6.97} 
& 12.44 \\
\hline
$\pm \sigma$
& 0.21 
& 21.88 
& 0.36 
& 3.54 
& 6.64 
& 2.46 
& \textbf{13.18} \\
\hline
$\pm2\sigma$
& \textbf{0.22} 
& 20.80 
& 0.35 
& 3.43 
& 6.64 
& 3.59 
& 12.80 \\
\hline
\end{tabular}
\caption{Summary of all performance metrics in three stochastic cases. The number in \textbf{bold} highlights the highest values among the three cases.}
\label{scheduled_results_stochastic}
\end{table}

\begin{figure}[!h]
    \centering
    \includegraphics[width=\linewidth]{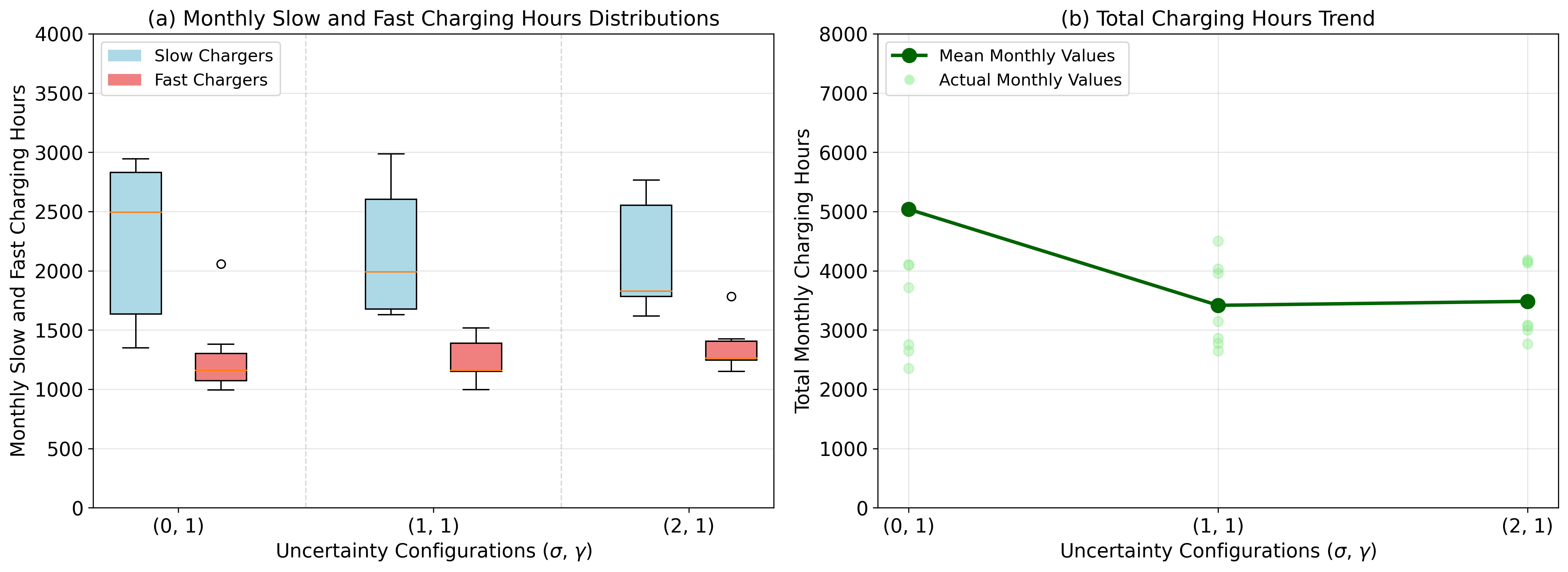}
\caption{(a) Distributions of monthly slow and fast charging hours of the entire truck fleet and (b) the monthly charging hour trends across a year in three uncertainty scenarios}
    \label{four_stochastic_instances_by_months_compare}
\end{figure}

Fig. \ref{four_stochastic_instances_by_months_compare} (a) shows distributions of monthly slow and fast charging hours of the entire truck fleet in the three stochastic instances, and Fig. \ref{four_stochastic_instances_by_months_compare} (b) shows the total monthly charging hour trends. By comparing three cases, we have the following observations: (1) With the increasing uncertainty of the charging duration, the total charging hours (the parking slots being decided to charge) reduce, but the fast charging hours increase. This indicates that the scheduling system seeks to avoid parking slots with highly uncertain parking durations and instead uses fast chargers more frequently. (2) If we further increase the uncertainty of charging duration, the total charging hours will increase, which is opposite to the previous trend, as more charging slots are required to meet the energy consumption.

\subsection{Conclusion and discussion}

We develop a robust model for planning and scheduling the charging infrastructure required to electrify light-duty trucks at industrial sites with mixed charger types. The model determines charging infrastructure deployment by selecting charging locations and charger types, and jointly optimizes charging schedules while accounting for waiting time, charging abandonment, and range anxiety. We conduct a case study at an open-pit mining site comprising eight charging zones. Three input datasets are constructed from year-long GPS records of approximately 200 operational trucks: (i) parking time slots discretized into half-hour intervals, (ii) realized parking durations within each half-hour window, and (iii) half-hourly energy consumption of all trucks. We assume fixed charging efficiencies for both slow and fast chargers and do not account for energy consumption during parking periods.  

\subsubsection{No overnight charging or no range anxiety will significantly increase the charger installations} Among eight zones, we assume that two designated zones allow overnight parking; otherwise, trucks will forgo charging if no chargers are available. The battery SoC status below 30\% is assumed to trigger the range anxiety, resulting in an additional half-hour waiting time. We found that neither overnight charging nor range anxiety significantly increases the number of charger installations, as both behaviors entail additional waiting time and reduce the need for installations. In addition, the charging hours account for only about 4\% of the total parking time. This means limited waiting time and resulting abandonment are the main reasons for the number of charger installations at industrial sites. 

\subsubsection{A greater uncertainty in charging duration leads to changes in total charging hours as well as shifts in charging type} We assume that parking duration in the parking window is stochastic. To identify patterns, hierarchical clustering is used to group trucks. Then, the uncertain parking duration of each truck can be represented as a robust, feature-based uncertainty set. As charging duration uncertainty increases, total charging hours decline, reducing utilization rates for both slow and fast chargers. Increased uncertainty also shifts charging patterns, with total fast-charging hours increasing while slow-charging hours decrease. However, when the lower bound of the robust parking-duration set is further reduced to its minimum value (i.e., 5 minutes), the total charging hours increase to meet each truck's energy requirements.

\subsubsection{Alternative decision-dependent formulations that rely on clustering produce different charger planning results} We consider that the uncertainty of charging duration is decision-dependent, which impacts and is impacted by charging decisions (i.e., whether or not charging and charging types).  We then investigate two decision-dependent formulations, assuming an affine dependence on instantaneous and cluster-level expected decisions. We found that cluster-level expected decisions yield tighter bounds on scheduling and a higher charging power level when we assume that charging decisions reduce parking duration. The instantaneous formulation always imposes a certain risk on individual parking time slots and neglects aggregated, correlated charging/parking behaviors across the fleet.

\subsubsection{Fix-and-optimize heuristic can assist the robust scheduling depending on the forecasts} The proposed planning and scheduling framework is formulated as a large-scale MILP model. For a 39-day instance with 188 trucks, the model contains over 1.5 million binary variables during pre-solving, making it computationally challenging. To address this, we introduce a fix-and-optimize heuristic for scheduling the charging operations under uncertainty. This is to forecast high charging demand due to reduced parking durations and to schedule fast-charging actions. This reduces the computation time by approximately 2-5 times for 39-day instances and yields the optimality gap of the problem below 0.1\%. 

Finally, several promising avenues remain for future research. For example, future work could incorporate additional charging options, such as multiple types of fast chargers with rated powers exceeding 150 kW, to further reduce the number of required HCVs. Moreover, charging behavior modeling could be improved by integrating empirical data from existing electrified fleets to account for factors such as variations in range anxiety across times of day.

\subsection{Acknowledgment}

We gratefully acknowledge the sponsor company for providing GPS data from the operating trucks at the mining site. We also thank Maria Isabel Castrillo Viguera for her valuable contributions in processing the GPS data and preparing the input datasets.

\bibliography{main}
\newpage
\begin{APPENDICES}
\section{Waiting time and abandonment: full illustrations}
\label{full_illustrations}

\paragraph{State evolution on a parked stretch.}
Recall that $S_{i,d,t}\in\{0,1\}$ indicate that $(i,d,t)$ is a parking slot and $z(i,d,t)$ be the zone index when $S_{i,d,t}=1$.
Define the \emph{continuity indicator}
\[
\kappa_{i,d,t} :=
\begin{cases}
1, & S_{i,d,t}=1,\ S_{i,\tilde d,\tilde t}=1,\ z(i,d,t)=z(i,\tilde d,\tilde t),\\
0, & \text{otherwise},
\end{cases}
\]
where $(\tilde d,\tilde t)$ is the previous admissible slot (day wrap as needed). Then the implemented rules are equivalent
\[
w_{i,d,t} = 
\begin{cases}
0, & S_{i,d,t}=0 \ \text{or}\ \kappa_{i,d,t}=0,\\[0.2em]
w_{i,\tilde d,\tilde t} + \Delta t\!\left(1-\sum_{j\in\mathcal{J}} y^{j}_{i,\tilde d,\tilde t}\right), & \text{otherwise},
\end{cases}
\qquad
a_{i,d',t'} \ge a_{i,d,t} \ \text{for the next slot }(d',t').
\]
Hence: (i) any charge in $(\tilde d,\tilde t)$ prevents growth of $w$ in that step; (ii) any zone change resets $w$; (iii) day-to-day carryover is allowed only when the zone is unchanged.

\paragraph{Worked example 1 (day boundary, same zone).}
Consider three consecutive admissible slots for truck $i$ in the same zone: $(d,t{=}1)$, $(d,t{=}2)$, and $(d{+}1,0)$, none charged.
Then $w_{i,d,1}=0$, $w_{i,d,2}=\Delta t$, and $w_{i,d{+}1,0}=2\Delta t$, matching Fig.~\ref{fig:waiting-day}.

\paragraph{Worked example 2 (zone switch).}
Let $(d,1)$ and $(d,2)$ be in Zone~1 and $(d,3)$ in Zone~2. If no charging occurs, $w_{i,d,1}=0$, $w_{i,d,2}=\Delta t$, but $w_{i,d,3}=0$ because the zone changed, as in Fig.~\ref{fig:waiting-zone}.

\paragraph{Range anxiety and special zones.}
In non-special zones, the cap $T^{\max}_{i,d,t}=\Delta t\cdot \delta^{30}_{i,d,t}$ implies:
\emph{(a)} if $\mathrm{SoC}\ge 30\%$, then $\delta^{30}_{i,d,t}=0$ and the cap is $0$; the truck either charges immediately or abandons;
\emph{(b)} if $\mathrm{SoC}<30\%$, then one slot of waiting is permitted before abandonment.
In special (overnight) zones, $T^{\max}_{i,d,t}=L$ so multi-slot waiting is admissible, and $w$ can grow across the night until charging starts.

\paragraph{Two operational variants.}
We support two abandonment policies that only affect when resets are triggered:
\begin{enumerate}\setlength\itemsep{0.2em}
\item \textbf{Current-policy variant.} Resets occur only on parking gaps or zone changes. If the next slot is admissible and in the same zone, apply~\eqref{eq:waiting_time} and $a_{i,d',t'}\ge a_{i,d,t}$.
\item \textbf{SoC-based variant.} In addition to the above, designated \emph{special} slots (e.g., transition markers tied to SoC rules or overnight delineations) force a reset: if $(i,d,t)$ is special, set $w_{i,d,t}=0$ and do not apply carryover from~\eqref{eq:waiting_time} to the next slot.
\end{enumerate}

\paragraph{Consistency with effective charging time.}
Within each parked slot, the available charging time is the \emph{effective duration} $\hat p_{i,d,t}\in[5/30,1]$. Charging power is bounded by
\[
p^{\text{ch}}_{i,d,t} \le \sum_{j\in\mathcal{J}} \eta_j P_j\,\Delta t\, y^{j}_{i,d,t}\,\hat p_{i,d,t},
\]
So even when $w$ grows across a long stretch (e.g., in a special zone), the energy deliverable in any individual slot remains limited by $\hat p_{i,d,t}$.

\section{The value range of $M$ in the Big-M constraints}

\label{sec:bigM}

We specify data-driven, \emph{minimal sufficient} values for the Big-M constants used in
\eqref{const:low_soc}, \eqref{const:waiting_abandon}, and
\eqref{const:soc_threshold}--\eqref{const:fast_charger_cap}.
Throughout, the SoC bounds are those already imposed in the model,
$\,SoC^{\min}\le b_{i,d,t}\le SoC^{\max}\,$.

\paragraph{Low-SoC indicator (\texorpdfstring{$M_1$}{M1}) in \eqref{const:low_soc}.}
Constraint \eqref{const:low_soc} encodes the threshold at $30\%$ with a small $\epsilon>0$.
To make \eqref{const:low_soc} logically equivalent to the intended rule
$\delta^{30}_{i,d,t}=1 \Leftrightarrow b_{i,d,t}\le 0.3$ and
$\delta^{30}_{i,d,t}=0 \Leftrightarrow b_{i,d,t}\ge 0.3+\epsilon$,
it suffices to take
\[
M_1^\star=\max\bigl\{\,SoC^{\max}-0.3,\ 0.3+\epsilon-SoC^{\min}\,\bigr\}.
\]

\paragraph{Waiting-time cap with abandonment (\texorpdfstring{$M_2$}{M2}) in \eqref{const:waiting_abandon}.}
Write $T^{\max}_{i,d,t}$ as in \eqref{const_nl:tmax}.
Within non-special zones, the worst-case feasible waiting time over any contiguous admissible segment
$I\subseteq S^{NZ}_{i,d}$ is exactly its duration $|I|\Delta t$. Define
\[
W_{\max}:=\max_{i,d}\ \max_{\text{contiguous }I\subseteq S^{NZ}_{i,d}}\ (|I|\Delta t).
\]
Then taking $M_2^\star = W_{\max}$ makes \eqref{const:waiting_abandon} nonbinding whenever $a_{i,d,t}=1$;
when $a_{i,d,t}=0$, it reduces to the intended cap $w_{i,d,t}\le T^{\max}_{i,d,t}$.
If $M_2<W_{\max}$, feasibility would be cut off, so $M_2^\star$ is minimal.

% \paragraph{Special-zone constant (\texorpdfstring{$L$}{L}) in \eqref{const_l:tmax}.}
% Define the longest continuous stay in special zones by
% \[
% W^{Z}_{\max}:=\max_{i,d}\ \max_{\text{contiguous }I\subseteq S^{Z}_{i,d}}\ (|I|\Delta t).
% \]
% Any $L\ge W^{Z}_{\max}$ ensures $T^{\max}_{i,d,t}=L$ does not bind solely due to the special-zone duration.
% Choosing $L^\star = W^{Z}_{\max}$ is minimal.

\paragraph{Fast-charging SoC ceiling (\texorpdfstring{$M_3$}{M3}) in \eqref{const:soc_threshold}--\eqref{const:fast_charger_cap}.}
For the constraints
$b_{i,d,t}\le 0.8 + M_3(1-y_{i,d,t,1})$ and
$b_{i,d',t'}\le 0.8 + M_3(1-y_{i,d,t,1})$,
we require that when the fast charger is \emph{not} used ($y_{i,d,t,1}=0$) the right-hand side be at least $SoC^{\max}$.
Thus the minimal sufficient choice is
\[
M_3^\star = SoC^{\max}-0.8.
\]

\paragraph{Computation of $W_{\max}$ and $W^{Z}_{\max}$.}
The longest feasible waiting time in each zone type is the duration of the longest contiguous parked stretch.
Hence we compute
\[
  W_{\max}=\max_{i,d}\;\max_{\text{contiguous }I\subseteq S^{NZ}_{i,d}} (|I|\Delta t), 
  \qquad
  W^{Z}_{\max}=\max_{i,d}\;\max_{\text{contiguous }I\subseteq S^{Z}_{i,d}} (|I|\Delta t).
\]

% \begin{algorithm}[!t]
% \caption{\textsc{Compute-}$(W_{\max},\,W^{Z}_{\max})$}
% \DontPrintSemicolon
% \SetAlgoLined
% \KwIn{$\{\Delta t,\,\mathcal{I},\,\mathcal{D},\,\mathcal{T}_d,\,S_{i,d},\,\textsf{special}_{i,d,t}\}$}
% \KwOut{$W_{\max},\,W^{Z}_{\max}$}
% $W_{\max}\gets 0$;\; $W^{Z}_{\max}\gets 0$\;
% \ForEach{$i\in\mathcal{I}$}{
%   \ForEach{$d\in\mathcal{D}$}{
%     $\ell_{\mathrm{NZ}},\ell_{\mathrm{Z}}\gets 0$\;
%     \ForEach{$t\in \mathcal{T}_d$ in order}{
%       \uIf{$t\in S_{i,d}$}{
%         \uIf{$\textsf{special}_{i,d,t}=0$}{
%           $\ell_{\mathrm{NZ}} \gets \ell_{\mathrm{NZ}}+1$;\; $\ell_{\mathrm{Z}}\gets 0$\;
%           $W_{\max}\gets \max\{W_{\max},\, \ell_{\mathrm{NZ}}\cdot \Delta t\}$\;
%         }
%         \Else{
%           $\ell_{\mathrm{Z}} \gets \ell_{\mathrm{Z}}+1$;\; $\ell_{\mathrm{NZ}}\gets 0$\;
%           $W^{Z}_{\max}\gets \max\{W^{Z}_{\max},\, \ell_{\mathrm{Z}}\cdot \Delta t\}$\;
%         }
%       }
%       \Else{
%         $\ell_{\mathrm{NZ}},\ell_{\mathrm{Z}}\gets 0$\;
%       }
%     }
%   }
% }
% \Return{$(W_{\max},\,W^{Z}_{\max})$}
% \end{algorithm}

\paragraph{Summary.}
\[
\boxed{
\begin{aligned}
M_1^\star &= \max\{\,SoC^{\max}-0.3,\ 0.3+\epsilon-SoC^{\min}\,\},\\
M_2^\star &= W_{\max},\qquad
L^\star = W^{Z}_{\max},\\
M_3^\star &= SoC^{\max}-0.8.
\end{aligned}}
\]
In implementation, set $\epsilon$ on the order of the solver feasibility tolerance
(e.g., $10^{-6}$–$10^{-4}$) and, if desired, add a small margin of the same order to each $M^\star$.
When indicator constraints are available, these Big-M forms may be replaced by indicator constraints
to improve relaxation strength; we keep the Big-M form for consistency.

\section{Detailed Reformulation and Proofs for Decision-Dependent Constraints}
\label{addition_proof}

\subsection{Proof of Decision-Dependent Moments}
\label{proof:moments}

\begin{proposition} \textit{Let $\hat{pp}$ be a random variable with mean $\hat{\mu}$ and variance $\hat{\sigma}^2$. Let $y$ be a decision variable determined ex-ante, and $C(y) = (1 - \sum \frac{y^j}{k^j})$ be a scalar scaling factor. The random variable $(\hat{pp})' = C(y)\hat{pp}$ has mean $C(y)\hat{\mu}$ and variance $C(y)^2\hat{\sigma}^2$.}
\end{proposition}

\textbf{Proof:}
In the Robust Optimization framework, the decision variable $y$ is determined ``here-and-now,'' before the uncertainty $\hat{pp}$ is realized. Therefore, with respect to the probability measure $\mathbb{P}$ governing $\hat{pp}$, $y$ (and consequently $C(y)$) is a deterministic constant.

1. \textbf{Covariance Independence:}
The covariance between the random parameter and the decision variable is:
\begin{equation}
\text{cov}(\hat{pp}, y) = \mathbb{E}_{\mathbb{P}}[(\hat{pp} - \hat{\mu})(y - y)] = \mathbb{E}_{\mathbb{P}}[\hat{pp} - \hat{\mu}] \cdot 0 = 0
\end{equation}
Thus, there is no correlation between the realized noise and the decision variable.

2. \textbf{Mean Derivation:}
Using the linearity of the expectation operator $\mathbb{E}_{\mathbb{P}}$:
\begin{equation}
\mu' = \mathbb{E}_{\mathbb{P}}[C(y)\hat{pp}] = C(y)\mathbb{E}_{\mathbb{P}}[\hat{pp}] = C(y)\hat{\mu}
\end{equation}

3. \textbf{Variance Derivation:}
Using the property $\text{Var}(aX) = a^2\text{Var}(X)$ for any constant $a$:
\begin{equation}
(\sigma')^2 = \text{Var}_{\mathbb{P}}(C(y)\hat{pp}) = C(y)^2 \text{Var}_{\mathbb{P}}(\hat{pp}) = C(y)^2 \hat{\sigma}^2
\end{equation}

4. \textbf{Correlation Unchanged:} The Pearson correlation between $\hat{pp}$ and $\hat{pp}^{'}$ remains unchanged if they apply the same scaling factor.

\begin{equation}
\gamma' = \frac{(C(y))^2\sum_{i=1}^{n}(\hat{pp}_i-\bar{\hat{pp}})(\hat{pp}^{'}_i-\bar{\hat{pp}}^{'})}{(C(y))^2\sqrt{\sum_{i=1}^{n}(\hat{pp}_i-\bar{\hat{pp}})^2}\sqrt{\sum_{i=1}^{n}(\hat{pp}^{'}_i-\bar{\hat{pp}}^{'})^2}} = \gamma
\end{equation}

\qed

\subsection{RO-DDU reformulation}
\label{ro_ddu}

We use the RO-DDU formulation to model the charging-duration constraint based on a box-shaped uncertainty set. Specifically, we assume that the charging power is constrained by the lower bound of this set, defined as the cluster-level mean parking duration minus one standard deviation from the empirical samples. For Example 1: Affine dependency on instantaneous decisions, we have 

\begin{align}
\label{eq:affine_final_main_appendix}
p^{ch}_{i,d,t} \leq \sum_{j \in \mathcal{J}} \eta^j P^j \Delta t \left[ \boldsymbol{\hat{\mu}}_{c}^\intercal \left( 1 - \frac{\boldsymbol{y}^j}{k^j} \right) \boldsymbol{y}^j - \left( 1 - \frac{\boldsymbol{y}^j}{k^j} \right) \boldsymbol{\hat{\Sigma}}^{1/2}_{c} \boldsymbol{y}^j \right]
\end{align}

Since the $\boldsymbol{y}^j$ is a binary variable, we have 

\begin{equation} 
\label{binary}
(\boldsymbol{y}^j)^2 = \boldsymbol{y}^j
\end{equation}

Then, we can obtain constraint (\ref{eq:affine_final_main}), which is linear.

\subsection{Theoretical Analysis of Efficiency Gains via Cluster-Level Aggregation}
\label{subsec:efficiency_analysis}

Our numerical results indicate that the cluster-level formulation yields a significantly higher average charging power and a shorter total charging time than the instantaneous formulation. This efficiency gain is not merely an artifact of parameter tuning but a structural property of how uncertainty is aggregated. By treating the variance reduction from fast charging as a shared "public" effect within a cluster, the model creates a less conservative feasible region.

\begin{proposition}[Relaxation of Robust Energy Limits]
Let $\mathcal{F}_{inst}$ and $\mathcal{F}_{cluster}$ denote the feasible sets for the charging power $p^{ch}$ in the instantaneous and cluster-level formulations, respectively. Consider a specific parking slot where the solver selects a fast charger (decision variable $y=1$) with a duration reduction factor $k > 1$. If the associated cluster $c$ has an aggregate fast-charging adoption rate $\overline{Y}_c \in (0,1)$, then the robust upper bound on energy throughput in the cluster-level formulation is strictly greater than in the instantaneous formulation.
\end{proposition}

\begin{proof}
The robust energy limit $E^{max}$ is determined by the effective parking duration after accounting for the safety margin required by the RO formulation.

In the \textbf{Instantaneous Formulation}, the duration scaling factor depends exclusively on the binary decision $y$ for the current slot. When $y=1$, the scaling factor is $C_{inst} = 1 - 1/k$. The maximum feasible energy is:
\begin{equation}
    E^{max}_{inst} = \eta P \Delta t \left( 1 - \frac{1}{k} \right) (\hat{\mu}_c - \hat{\sigma}_c)
\end{equation}

In the \textbf{Cluster-Level Formulation}, the scaling factor depends on the aggregate adoption rate $\overline{Y}_c$. For the same slot where $y=1$:
\begin{equation}
    E^{max}_{cluster} = \eta P \Delta t \left( 1 - \frac{\overline{Y}_c}{k} \right)(\hat{\mu}_c - \hat{\sigma}_c)
\end{equation}

Since the adoption rate is an average over the cluster, typically $\overline{Y}_c < 1$. Given that fast charging reduces duration ($k > 1$), it follows that:
\begin{equation}
    \frac{\overline{Y}_c}{k} < \frac{1}{k} \implies \left( 1 - \frac{\overline{Y}_c}{k} \right) > \left( 1 - \frac{1}{k} \right)
\end{equation}
Assuming the baseline robust duration term (in square brackets) is positive—a necessary condition for any feasible charging—we conclude:
\begin{equation}
    E^{max}_{cluster} > E^{max}_{inst}
\end{equation}
Thus, the cluster-level formulation admits a strictly larger feasible region for energy throughput.
\end{proof}

\section{Data processing}

The overall information flow from the GPS data of all the year to the charger planning and scheduling results is presented in Fig. \ref{rio_tinto_procedure}. The whole procedure consists of five steps.

\begin{figure}[!h]
    \centering
    \includegraphics[width=5in]{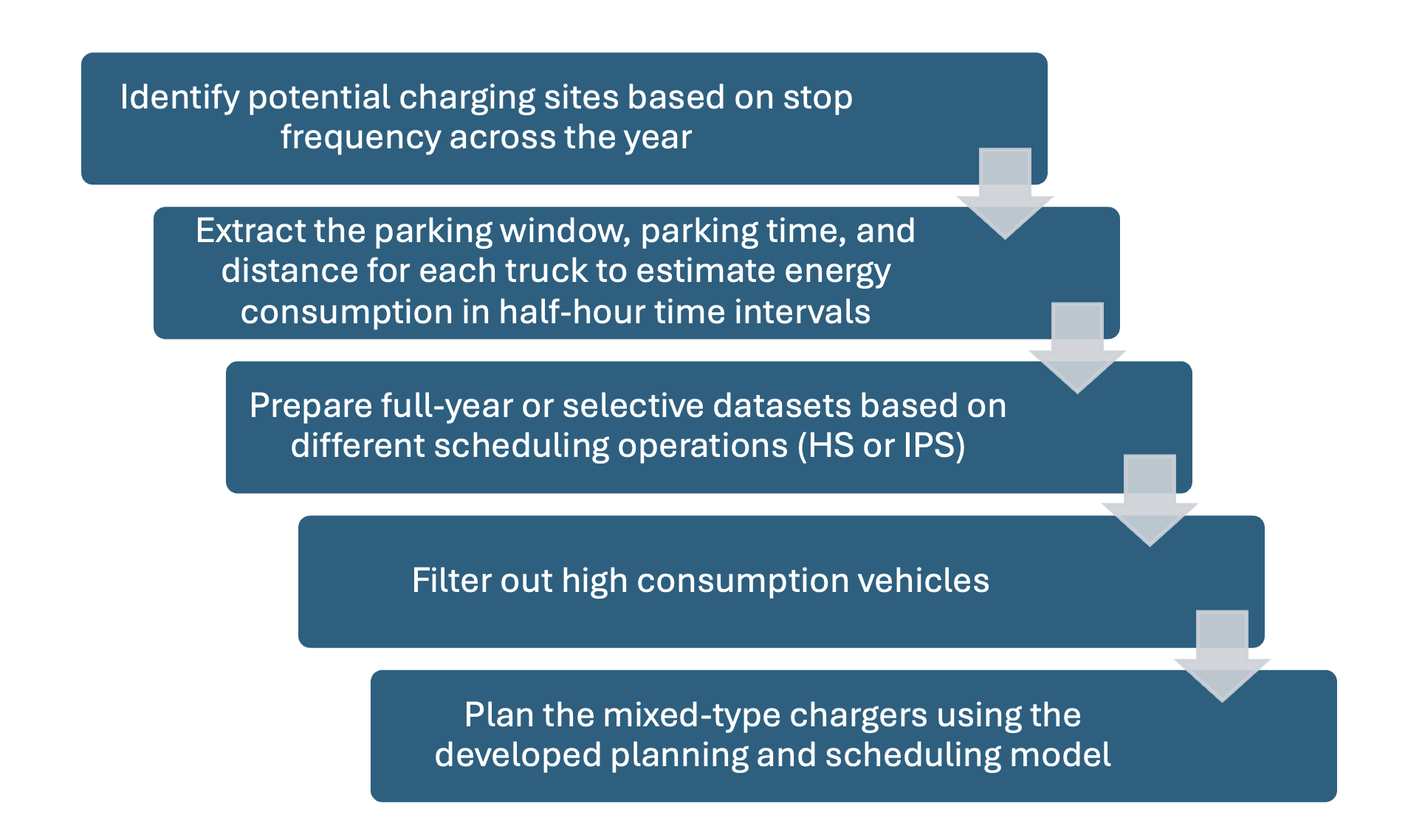}
    \caption{Information flow from the GPS data to the charger planning and scheduling results}
    \label{rio_tinto_procedure}
\end{figure}

\subsection{Detailed statistics of eight charging zones}
\label{charging_points}

We identified eight BEV charging zones for all trucks using GPS records, which include the latitude, longitude, and speed of trucks at each timestamp across a year. For each truck, the algorithm tracks its movement by converting GPS coordinates to UTM and measuring the distance between consecutive points. Please note that we do not consider elevation changes at the mining site due to insufficient geographic information. We grid divided the entire mining site into 100 meter $\times$ 100 meter square regions. If the truck moves less than 50 meters for longer than 5 minutes, this square is marked as a potential stop. Each stop point is recorded with the truck ID, coordinates, and stop duration, forming a list of candidate charging locations.

We ranked all candidate charging locations by total number of single stops over the year and selected the top eight charging zones with the highest counts. The eight charging zones cover 98.31\% of the parking time and 188 trucks of the entire fleet. A truck that does not pass any of the eight zones is considered a high consumption vehicle. Table \ref{tab:traffic_stops} lists the charging stop statistics of eight charging zones. Eight zones have different parking patterns: for instance, zones 1 and 4 are generally for overnight parking, with the average duration of a single stop exceeding 8 hours, while zones 5-8 are for driving through, with the average duration of a single stop less than 4 hours.

\begin{table}[h]
\centering
\begin{tabular}{|p{3.5cm}|r|r|r|r|r|r|r|r|r|}
\hline
\textbf{Metrics} & \textbf{SQ1} & \textbf{SQ2} & \textbf{SQ3} & \textbf{SQ4} & \textbf{SQ5} & \textbf{SQ6} & \textbf{SQ7} & \textbf{SQ8} & \textbf{\makecell{Avg. \\(total)}} \\ \hline
Number of single stops & 28,391 & 2,542 & 8,746 & 84,543 & 83,455 & 33,568 & 13,873 & 14,686  & 33,726\\
\hline
Relative to total (\%)& 10.41 & 0.93 & 3.21 & 30.99 & 30.60 & 12.31 & 5.09 & 5.38 & (98.92)\\
\hline
Total time stopped by whole fleet of trucks (in year) & 11,675 & 3,716 & 1,418 & 15,258 & 7,624 & 2,451 & 1,305 & 1,110 & 5,569.5\\
\hline
Relative to total (\%) & 25.76 & 8.20 & 3.13 & 33.66 & 16.82 & 5.41 & 2.88 & 2.45& (98.31) \\
\hline
Average duration of a single stop (hours) & 9.9 & 35.1 & 3.9 & 4.3 & 2.2 & 1.8 & 2.3 & 1.8 & 7.6\\
\hline
\# trucks stopping & 184 & 154 & 44 & 193 & 181 & 164 & 126 & 130 &  (188) \\
\hline
Average days stopped in a year (per truck) & 63.5 & 24.1 & 32.2 & 79.1 & 42.1 & 14.9 & 10.4 & 8.5 & 34.4\\
\hline
Average hours stopped in a day (hours per day) & 4.2 & 1.6 & 2.1 & 5.2 & 2.8 & 1.0 & 0.7 & 0.6 & 2.3\\
\hline
\end{tabular}
\caption{Traffic stop statistics of eight charging zones (SQ1: Road house, SQ2: Airfield, SQ3: Infrastructure, SQ4: Equipment maintenance, SQ5: Plant crusher, SQ6: West facility, SQ7: North facility, SQ8: East facility)}
\label{tab:traffic_stops}
\end{table}

\subsection{Input data generation}
\label{inputs}

 Three kinds of inputs are generated by discretizing the time horizon into half-hour time intervals: (1) the parking time window indication matrix to show if the truck parks at each of eight zones at each half-hour time interval (i.e., 1 means parking and 0 means not parking at one zone), (2) the actual parking time duration in minutes for each of eight zones ranging between 5 to 30 minutes, and (3) energy consumptions of each truck based on the temperature-dependent fuel-economy for light-duty trucks. Fig. \ref{parking_duration_distribution} shows (a) the parking time distribution of 189 trucks at Charging Zone 6 in the selected 39 days and (b) the histogram shows parking durations ranging from 5 to 30 minutes.

\begin{figure}[!h]
    \centering
    \includegraphics[width=\linewidth]{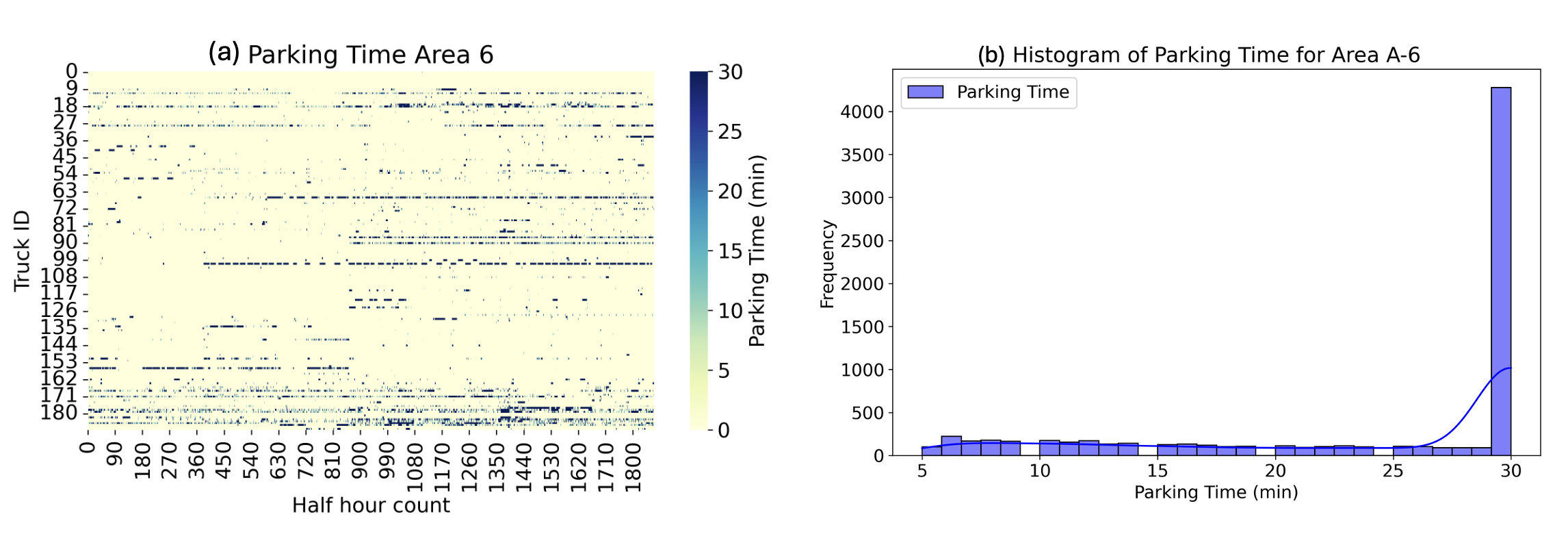}
\caption{(a) Parking time distribution of 189 trucks at Charging Zone 6 in the selected 39 days and (b) The histogram shows parking durations ranging from 5 to 30 minutes.}
    \label{parking_duration_distribution}
\end{figure}

\subsubsection{Temperature-dependent fuel-economy for light-duty trucks} 

For each truck's half-hour energy consumption, we calculate its travel distance (in miles) based on the coordination. Then, we convert these distances to energy consumption (in kWh) for each half-hour time window. We use a second-order, temperature-dependent fuel economy estimation model of U.S. electric trucks while moving \cite{goodall_feasibility_2024}. The fitting curve is presented as follows.

\begin{equation}
y = 6.602\times10^{-5} x^2 - 8.929\times10^{-3}x +7.745\times10^{-1} \quad x\in[20, 100]
\end{equation}

where $y$ is the fuel economy in kWh/mile, and $x$ is the outdoor temperature in F. The hourly temperature data of the mining site is taken from \cite{staffell_using_2016}. Fig. \ref{Temperature_FuelEconomy} shows the hourly temperature and the corresponding fuel economy across the year. Noted, we converted miles into kilometers in the plot.

\begin{figure}[!h]
    \centering
    \includegraphics[width=4in]{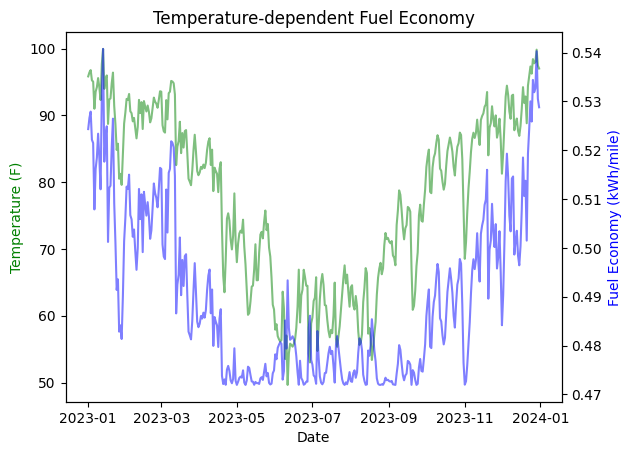}
\caption{The outdoor temperature (F) and corresponding temperature-dependent fuel economy (kWh/mile) across the year at the mining site}
    \label{Temperature_FuelEconomy}
\end{figure}

\subsubsection{Selecting the representative days from the full-year datasets}
\label{representative_day}

The 39 representative days based on ten criteria are listed in Table \ref{day_selection_criterion}. We ranked 365 days based on these ten criteria, and the top or bottom indicates the direction how these days get selected. We selected the 39 days listed in Table \ref{selected_day}. To maintain continuity in the truck-travel data, the days immediately adjacent to these top-ranked dates are also included. Additionally, the first and last two days of the whole year are incorporated to ensure completeness of the selected dataset.

\begin{table}[h!]
\centering
\caption{10 criterion for representative days}
\begin{tabular}{clcc}
\hline
\textbf{No.} & \textbf{Criteria Description} & \textbf{Ranking from} & \textbf{First day}  \\
\hline
1  & Total fleet distance                         & Top & 2023-12-14\\
2  & \# of moving trucks                          & Top & 2023-12-05\\
3  & Average distance per moving truck            & Top & 2023-01-07\\
4  & \# of individual trucks $>$ 296 km           & Top & 2023-06-16\\
5  & Total fleet energy consumption (EC)          & Top & 2023-12-14\\
6  & \# of consuming trucks                       & Top & 2023-11-21\\
7  & Average EC per consuming truck               & Top & 2023-01-07\\
8  & \# of individual trucks with EC $>$ 75 kWh   & Top & 2023-03-05\\
9  & Total stopping hours (whole fleet) in any area & Bottom & 2023-12-31\\
10 & \# of trucks stopped in any area along the day  & Bottom & 2023-01-08\\
\hline
\end{tabular}
\label{day_selection_criterion}
\end{table}

\begin{table}[h!]
\centering
\caption{The 39-day summary by ten criteria}
\begin{tabular}{|c|c|c|l|}
\hline
\textbf{From} & \textbf{To} & \textbf{Number of days} & \textbf{Criteria} \\
\hline
2023-01-01 & 2023-01-02 & 2  & Initial padding \\
2023-01-05 & 2023-01-10 & 6  & criteria 3, 7, 10 \\
2023-03-03 & 2023-03-07 & 5  & criteria 8 \\
2023-06-14 & 2023-06-18 & 5  & criteria 4 \\
2023-11-19 & 2023-11-23 & 5  & criteria 6 \\
2023-12-03 & 2023-12-07 & 5  & criteria 2 \\
2023-12-12 & 2023-12-19 & 8  & criteria 1, 5 \\
2023-12-29 & 2023-12-31 & 3  & criteria 9 \\
\hline
 & \textbf{total} & \textbf{39} &  \\
\hline
\end{tabular}
\label{selected_day}
\end{table}

\newpage

\section{ML clustering for the charging duration}
\label{machine_learning_clustering}

We begin by assembling features for clustering parking durations, incorporating both contextual attributes (fleet groups) and numerical variables (half-hour of day, day-of-month, day-of-week, and zone). The descriptions of full clustering features are presented in Table \ref{tab:cluster_features}, and Fig. \ref{full_feature_correlation} shows the correlation between these features and parking duration. The green bars represent the selected features and the gray ones represent the discarded one.

\begin{table}[htbp]
\centering
\caption{The descriptions of full clustering features}
\label{tab:cluster_features}
\begin{tabular}{r|l}
\toprule
Features & Numerical / Contextual values\\
\midrule
Zones & 1, 2, 3..., 8 \\
Fleet groups & \makecell{Support, Rail, Tech Services,\\ Uncategorized, 
                  Mobile Equipment, \\Fixed Plant, Production, \\
                  Drill and Blast}\\
Truck ID & 200 truck IDs\\
Half Hour of a day & 1, 2,...., 48 \\
Day of Month & 1, 2,..., 31 \\
Day of Week & 1, 2,..., 7 \\
\bottomrule
\end{tabular}
\end{table}

\begin{figure}[!h]
    \centering
    \includegraphics[width=5in]{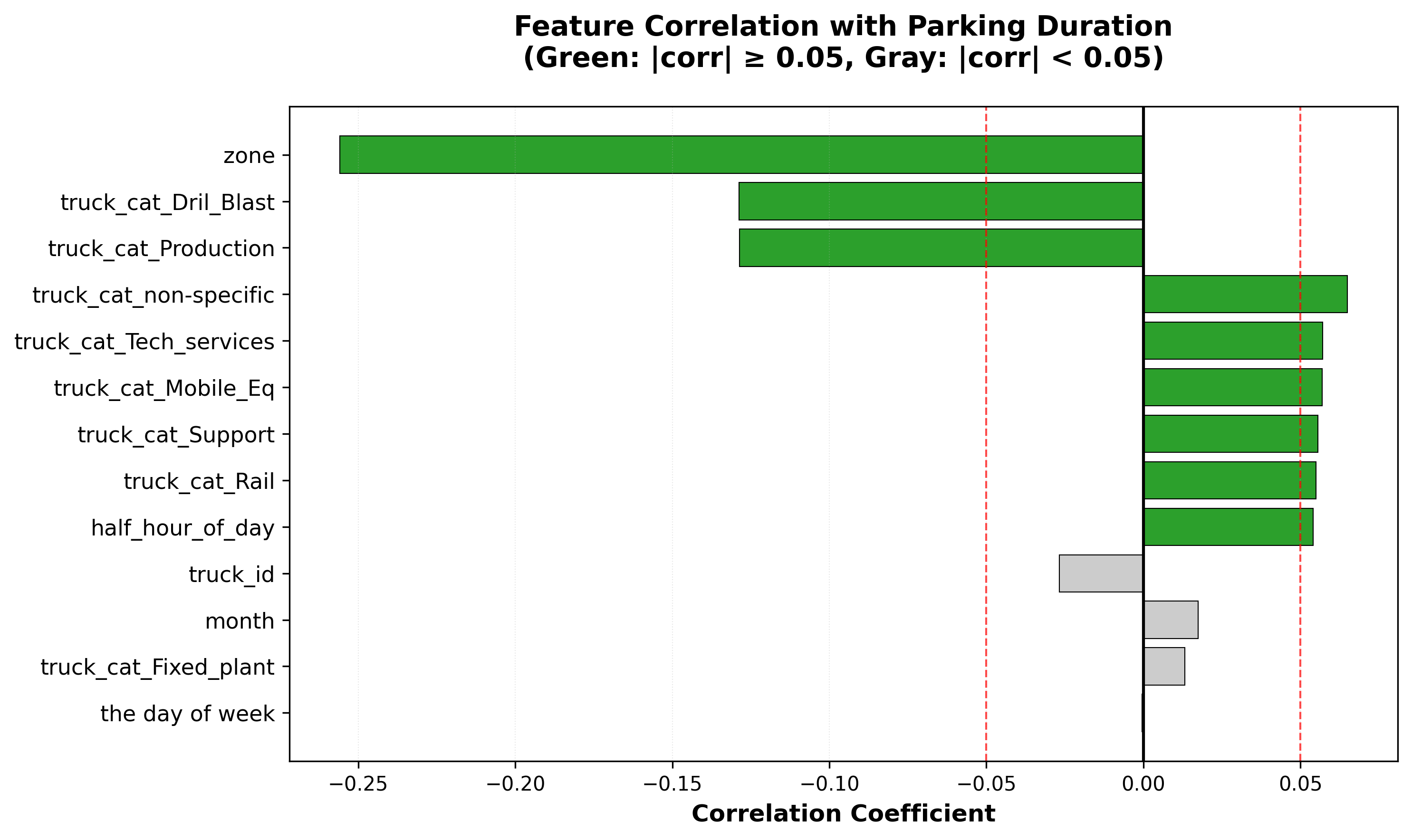}
\caption{The correlation of the full and selected features (The green bars represent the selected features and the gray ones represent the discarded ones)}
    \label{full_feature_correlation}
\end{figure}

Here, we compared three clustering methods and show that clustering by truck categories performs best. 

\begin{itemize}
    \item \textbf{K-means}: Cluster the charging duration into ten groups based on the k-means method, and the number of clusters is determined by the silhouette score.
    \item \textbf{Hierarchical clustering with truck category}: With the truck category information, the hierarchy clustering yields three subgroups per category, and yields a total number of 24 subgroups. 
    \item \textbf{Hierarchical clustering without truck category}: Without using the category information, the 200 trucks of the entire fleet are divided into 10 sub-groups by the hierarchical clustering, and each group has an average of 10 trucks.
    \item \textbf{Each truck as a cluster}: Without using the category information and hierarchy clustering, each truck is treated as an individual group as the baseline.
\end{itemize}

In total, there are 2,669,982 parking duration records throughout the year. We report the number of clusters obtained under each scenario as well as the theoretical numbers of clusters under four scenarios. Noted, the actual number of cluster are less than the theoretical numbers because some hour-zone combinations do not have records (i.e., trucks do not visit the zone at the hour). We assessed the clustering quality using a cluster-size-weighted Coefficient of Variation (CV). For each cluster, the CV is defined as the ratio of the standard deviation to the mean. An overall clustering quality metric is then computed as the cluster-size-weighted average of the CV across all clusters. This metric is introduced to quantify within-cluster heterogeneity. Since samples within each cluster are treated as i.i.d., lower within-cluster heterogeneity indicates better clustering performance. Accordingly, a smaller weighted-average CV indicates higher clustering quality. In terms of charging infrastructure planning, the size of the feature-based robust uncertainty set is defined as $\Delta_p = \pm \sigma$. Low-quality clustering will lead to overly conservative results with large uncertainty sets. Table~\ref{clustering_quality_compare} summarizes the number of clusters and the weighted CV (\%) using four methods.

  \begin{table}[h!]
  \renewcommand{\arraystretch}{1.1}
  \centering
  \begin{tabular}{c c c c c c }
  \hline
  Cases
    & \makecell{Names}
    & \makecell{\# of clusters \\ Theoretical no.}
    & \makecell{Average no.\\ of trucks}
    & \makecell{Average no.\\ of samples}
    & \makecell{Weighted \\ CV (\%)} \\
  \hline
  1
    & K-means
    & 10
    & -
    & 266,998
    & 22.00
\\ \hline

  2
    & \makecell{Hierarchical clustering \\w/ truck category}
    & 7,280 [9,216]
    & 7.8
    & 366
    & 5.01
\\ \hline

  3
    & \makecell{Hierarchical clustering \\w/o truck category}
    & 3,804 [3,804]
    & 18.8
    & 701
    & 19.68
 \\ \hline
  4
    & Each truck as a cluster
    & 42,251 [72,576]
    & -
    & 63
    & 4.24
\\ \hline
  \end{tabular}
  \caption{The number of clusters and the weighted CV (\%) of four methods}
  \label{clustering_quality_compare}
  \end{table}

We found that methods 2 and 4, hierarchical clustering with truck categories and each truck as a cluster, perform best in terms of the weighted CV. However, treating each truck as a cluster will yield only a few samples per cluster, which could be affected by outliers. We therefore use the second method. Figs.\ref{method_2} and \ref{method_3} show the hierarchical clustering process for one of the truck categories (Method 2) and for all trucks (Method 3), respectively.

\begin{figure}[!h]
    \centering
    \includegraphics[width=\linewidth]{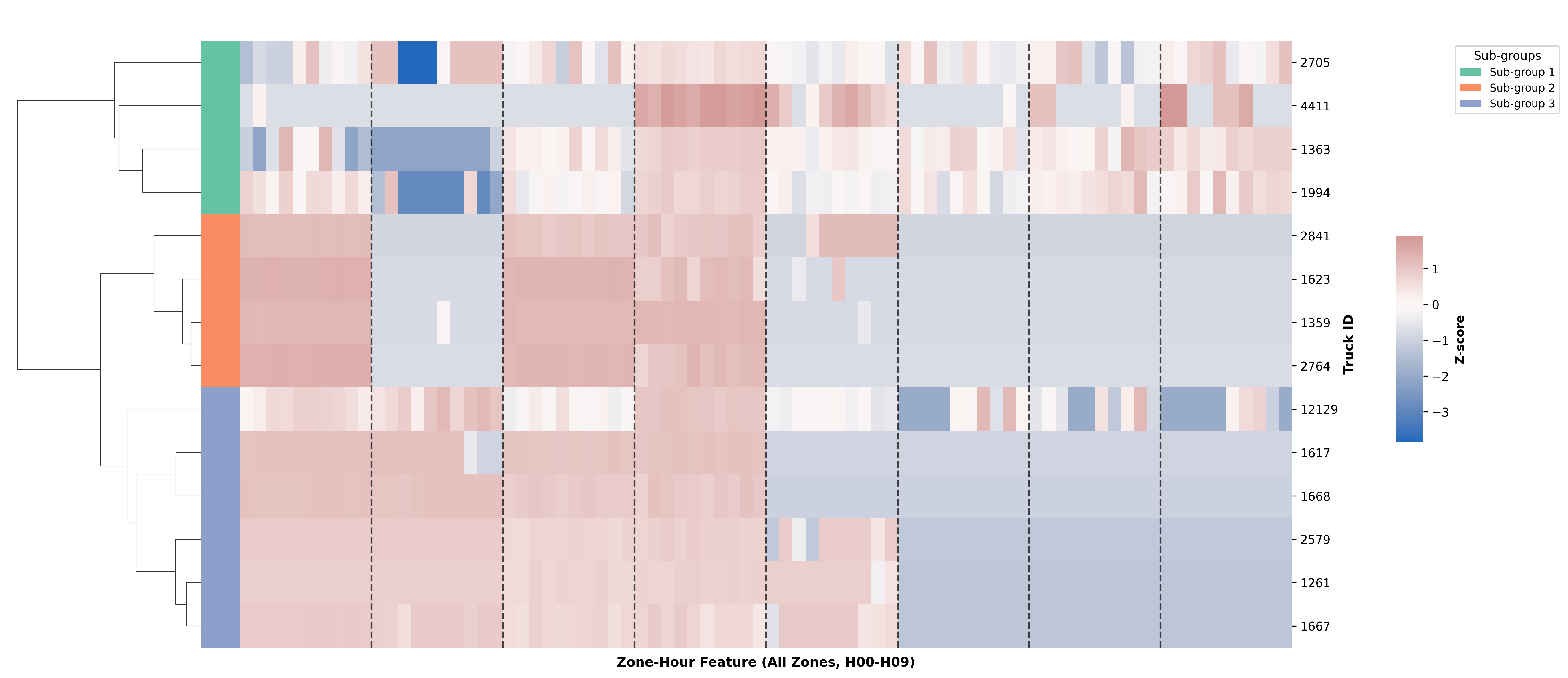}
\caption{Average parking duration heatmap for all trucks within the rail services, constructed using rearranged zone–hour
features. The heatmap displays only the first 10 hours for each of the eight zones, with zone boundaries
indicated by black dashed lines. The hierarchical clustering dendrogram, based on feature distances, is shown on the left.}
    \label{method_2}
\end{figure}

\begin{figure}[!h]
    \centering
    \includegraphics[width=\linewidth]{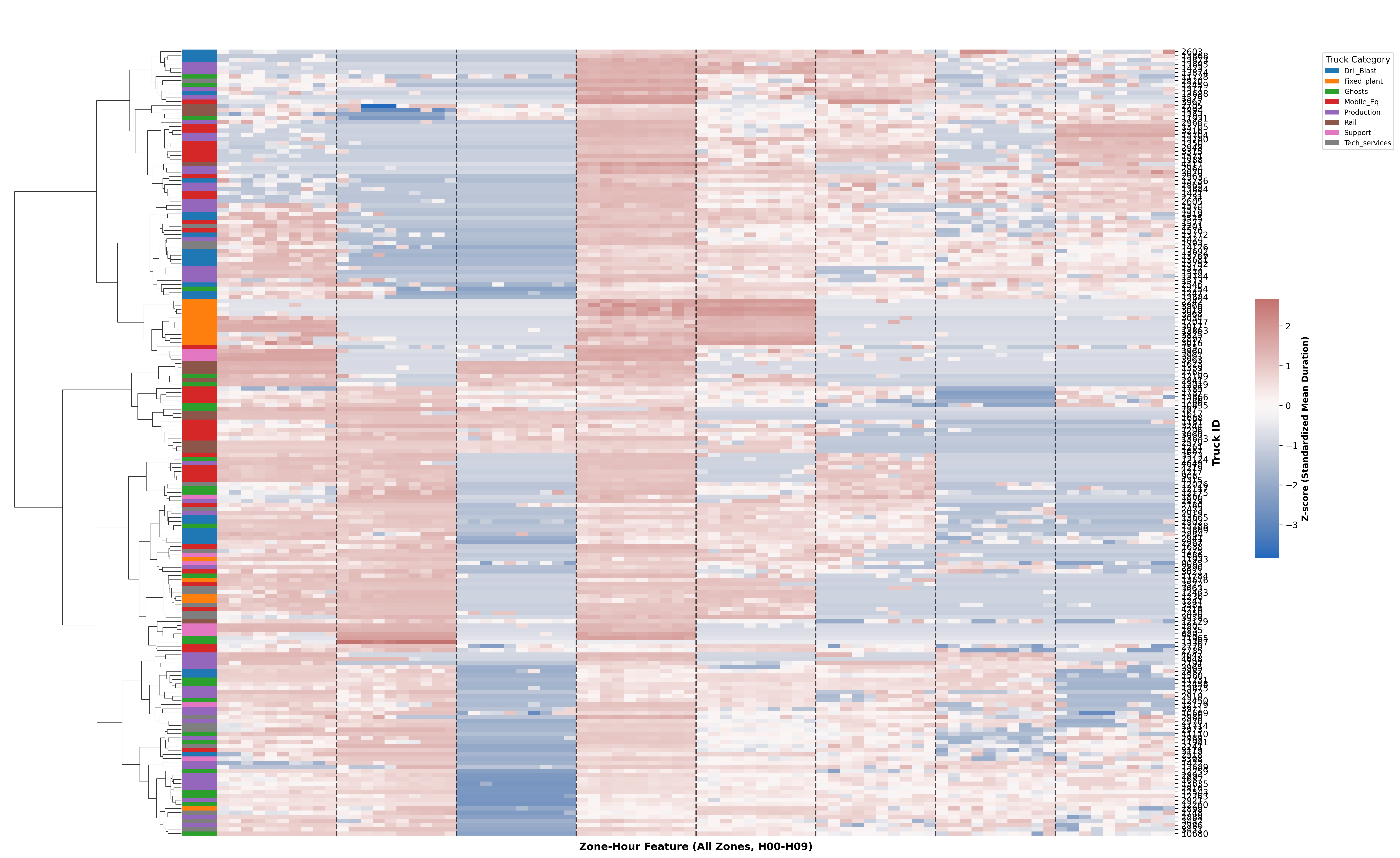}
\caption{Average parking duration heatmap for all trucks, constructed using rearranged zone–hour
features. The color bars on the very left indicate truck categories.}
    \label{method_3}
\end{figure}

\newpage

\section{Additional results of four deterministic cases}
\label{detailed_results_deterministic_cases}

This section presents the detailed optimization result of four deterministic cases: Case 1 (Benchmark), Case 2 (Full parking duration), Case 3 (No special zones), and Case 4 (No range anxiety). Figs. \ref{combined_parking_time_benchmark}, \ref{combined_parking_time_full_parking_duration}, \ref{combined_parking_times_no_special_zones}, and
\ref{combined_parking_times_no_range_anxiety} present the aggregated charging results of eight zones for four cases, and Figs. \ref{four_stochastic_instances_by_months_benchmark}, \ref{four_stochastic_instances_by_months_full_parking_duration},
\ref{four_stochastic_instances_by_months_no_special_zones}, 
and \ref{four_stochastic_instances_by_months_no_range_anxiety} present the individual charging results of 188 trucks at eight charging zones.

\begin{figure}[!h]
    \centering
    \includegraphics[width=5in]{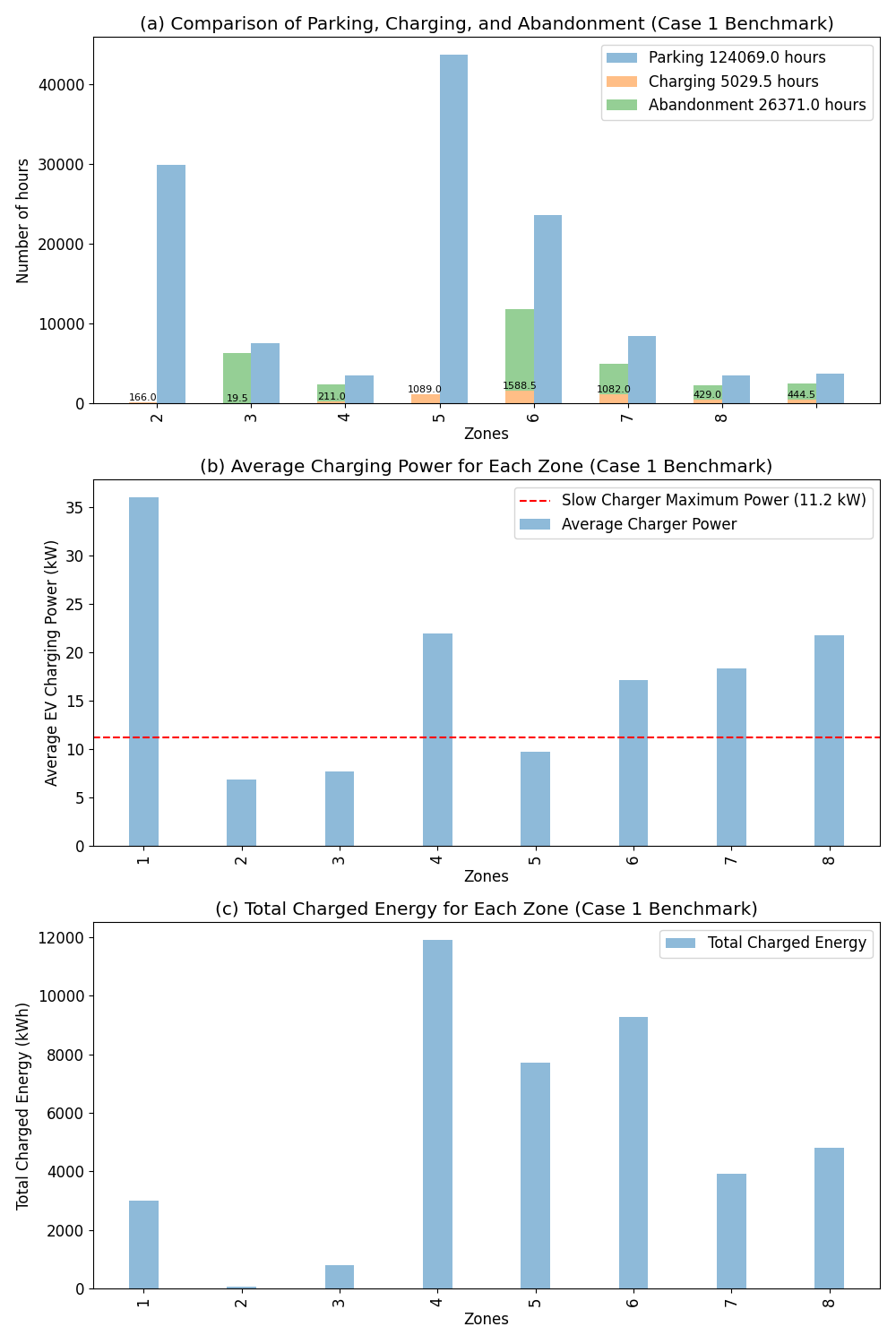}
\caption{The aggregated charging results of eight zones for case 1 benchmark: (a) The total hours of parking, charging, and abandonment during 39 representative days; (b) The average charging power for each zone during 39 representative days and (c) the total charged energy for each zone during 39 representative days}
    \label{combined_parking_time_benchmark}
\end{figure}

\begin{figure}[!h]
    \centering
    \includegraphics[width=\linewidth]{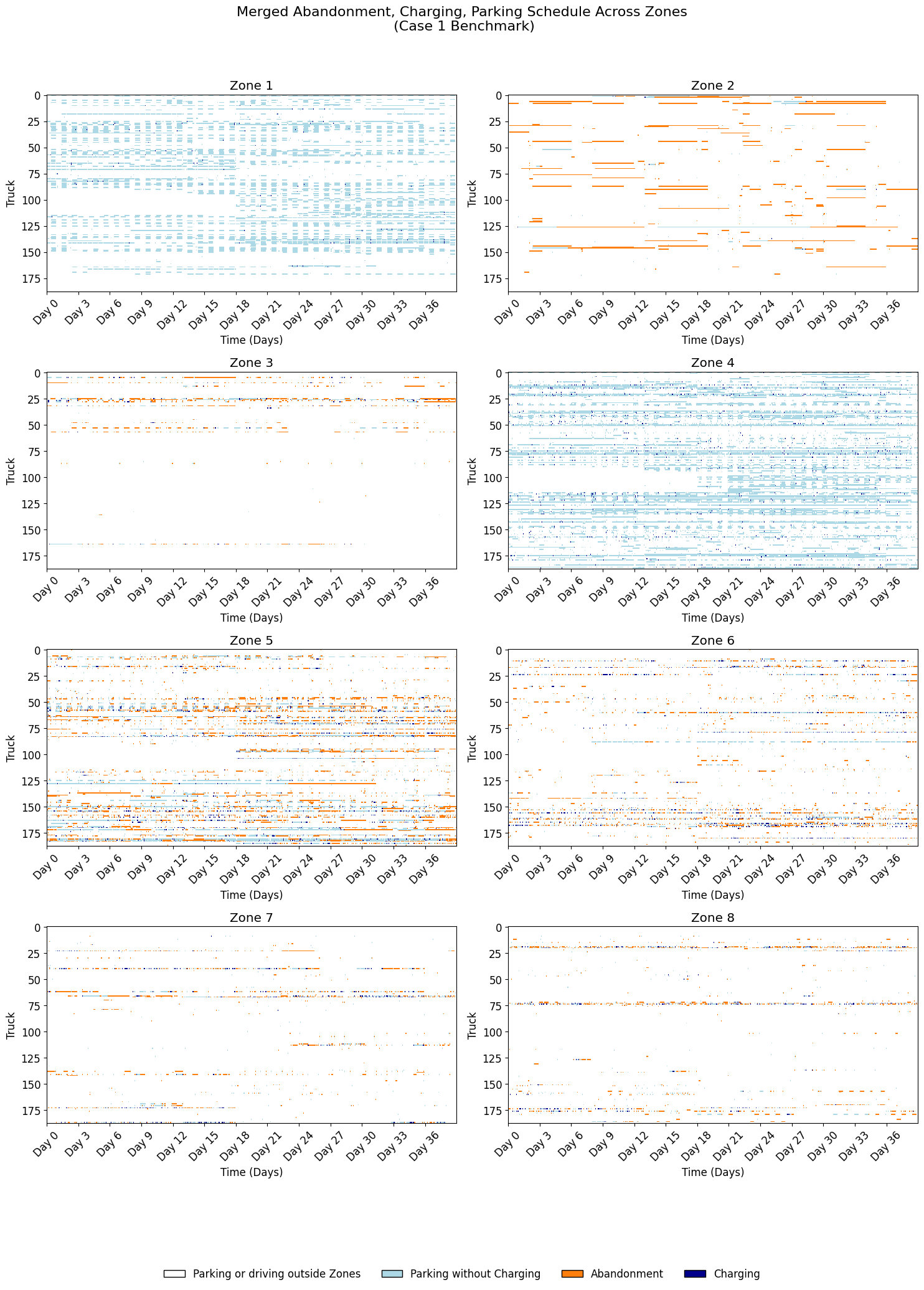}
\caption{The individual charging schedules of 188 trucks across eight charging zones, arranged from left to right and top to bottom, in the Case 1 benchmark}
    \label{four_stochastic_instances_by_months_benchmark}
\end{figure}

\begin{figure}[!h]
    \centering
    \includegraphics[width=5in]{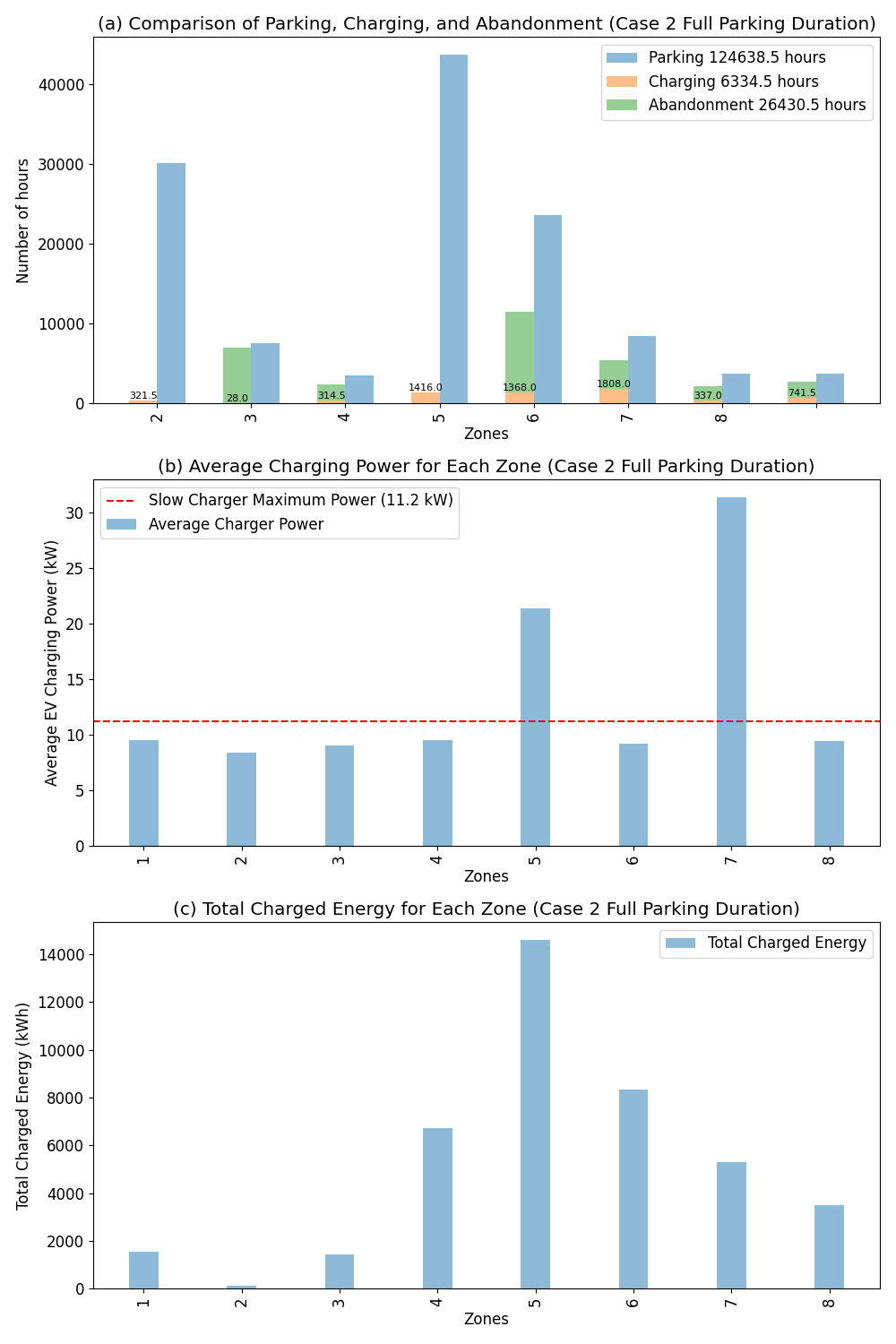}
\caption{The aggregated charging results of eight zones for case 2 full parking duration: (a) The total hours of parking, charging, and abandonment during 39 representative days; (b) The average charging power for each zone during 39 representative days and (c) the total charged energy for each zone during 39 representative days}
    \label{combined_parking_time_full_parking_duration}
\end{figure}

\begin{figure}[!h]
    \centering
    \includegraphics[width=\linewidth]{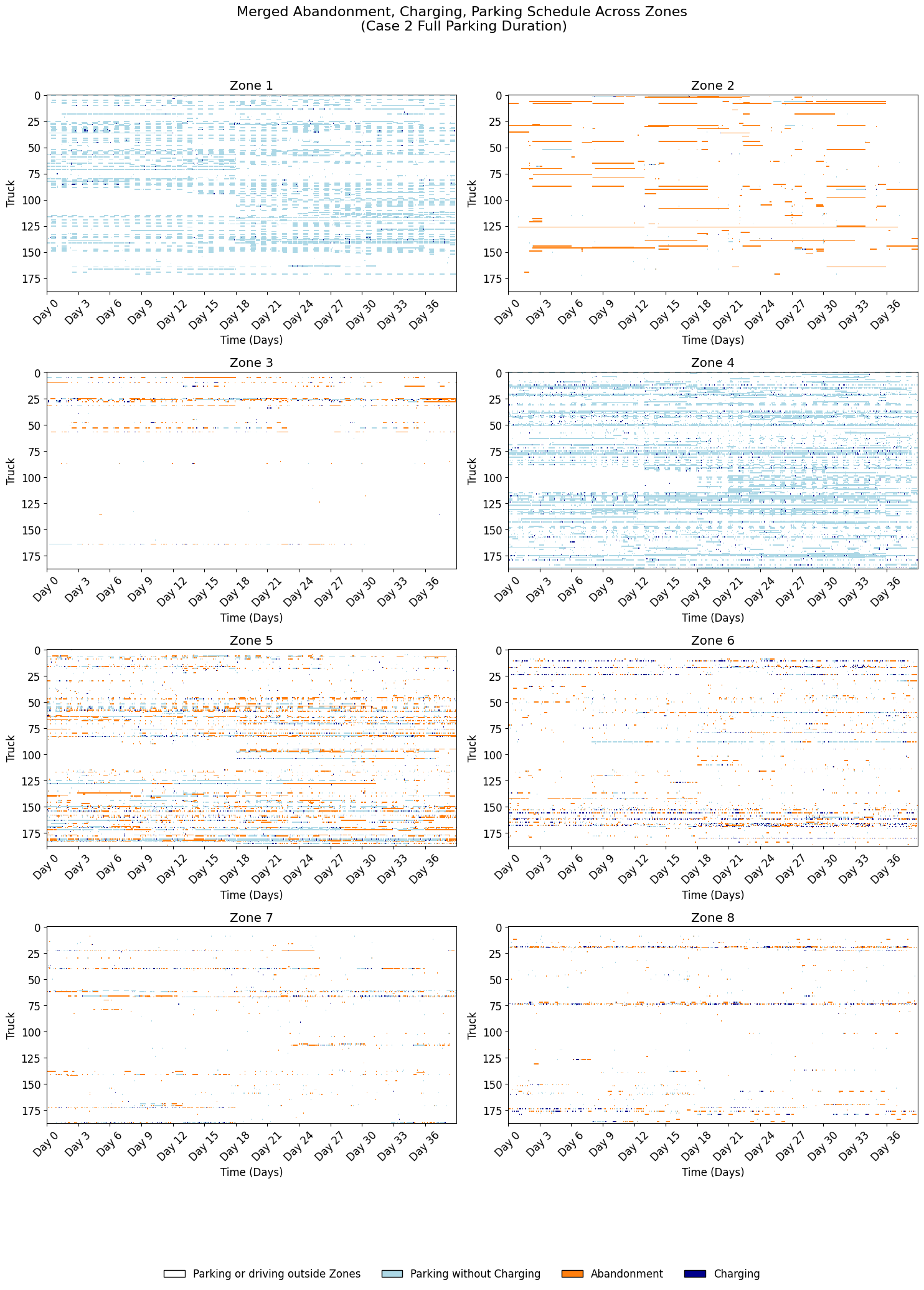}
\caption{The individual charging scheduling of 188 trucks at eight charging zones, arranged from left to right and top to bottom, in Case 2 full parking duration}
    \label{four_stochastic_instances_by_months_full_parking_duration}
\end{figure}

\begin{figure}[!h]
    \centering
    \includegraphics[width=5in]{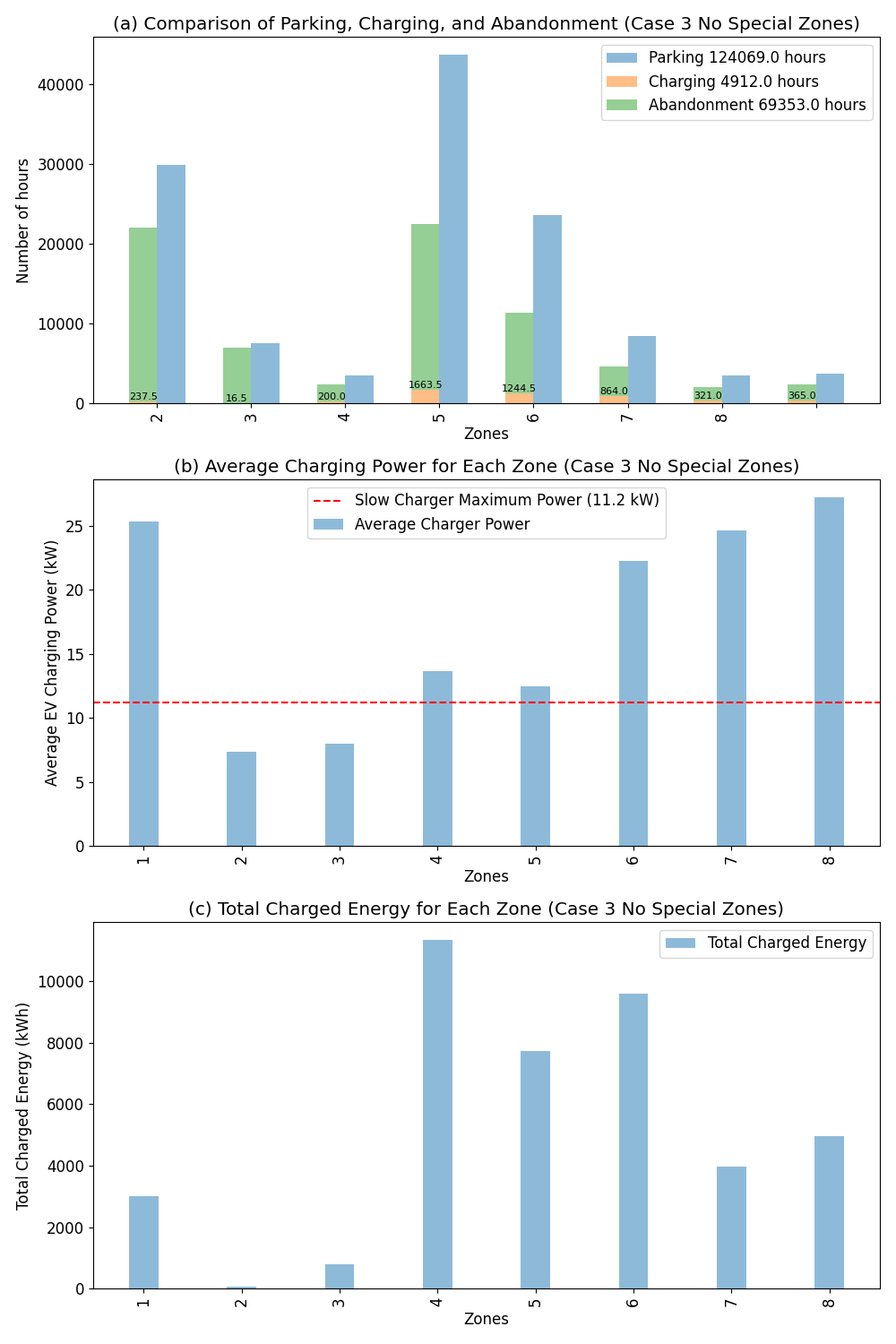}
\caption{The aggregated charging results of eight zones for case 3, no special zones: (a) The total hours of parking, charging, and abandonment during 39 representative days; (b) The average charging power for each zone during 39 representative days and (c) the total charged energy for each zone during 39 representative days}
    \label{combined_parking_times_no_special_zones}
\end{figure}

\begin{figure}[!h]
    \centering
    \includegraphics[width=\linewidth]{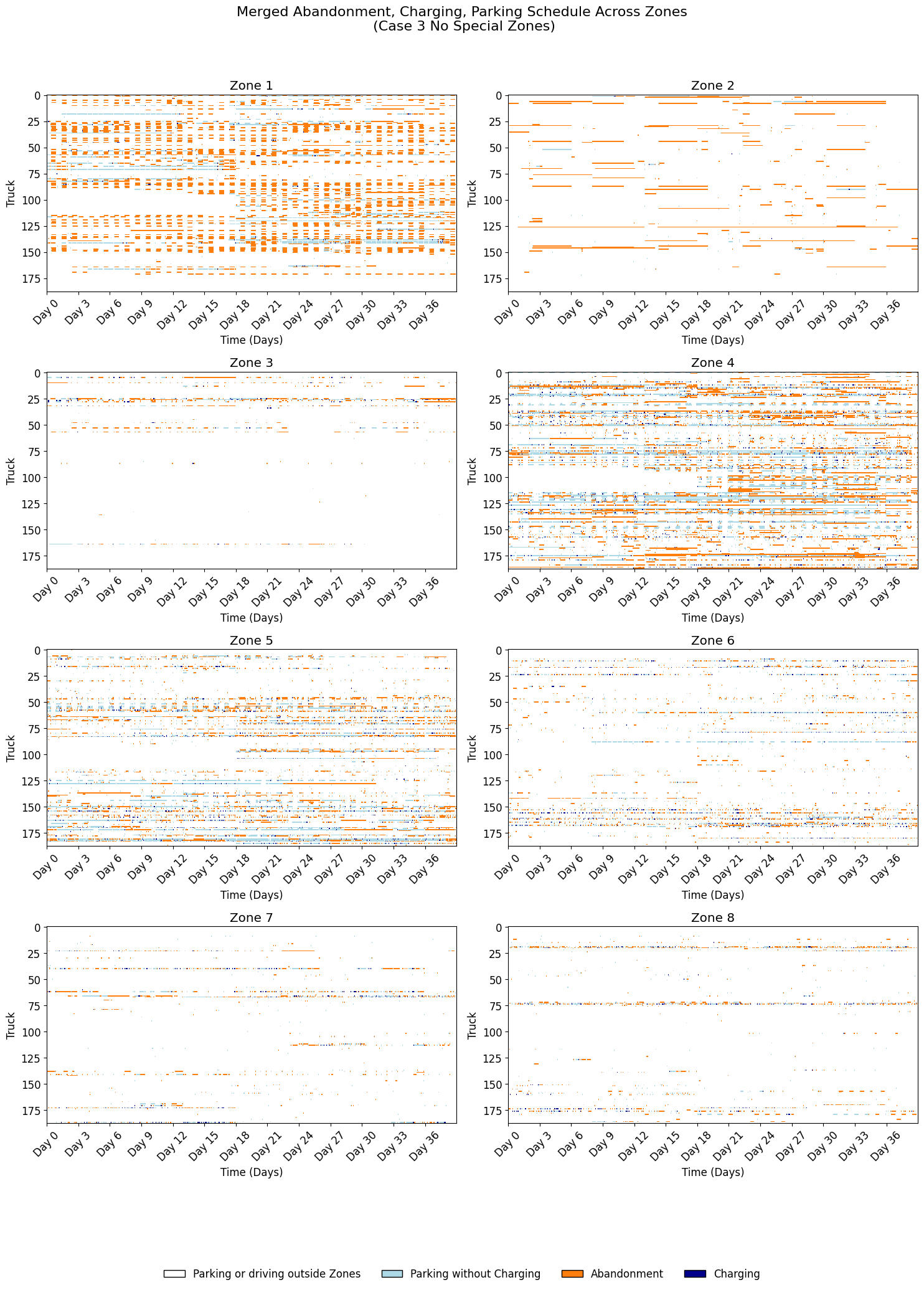}
\caption{The individual charging scheduling of 188 trucks at eight charging zones, arranged from left to right and top to bottom, in Case 3 no special zones}
    \label{four_stochastic_instances_by_months_no_special_zones}
\end{figure}

\begin{figure}[!h]
    \centering
    \includegraphics[width=5in]{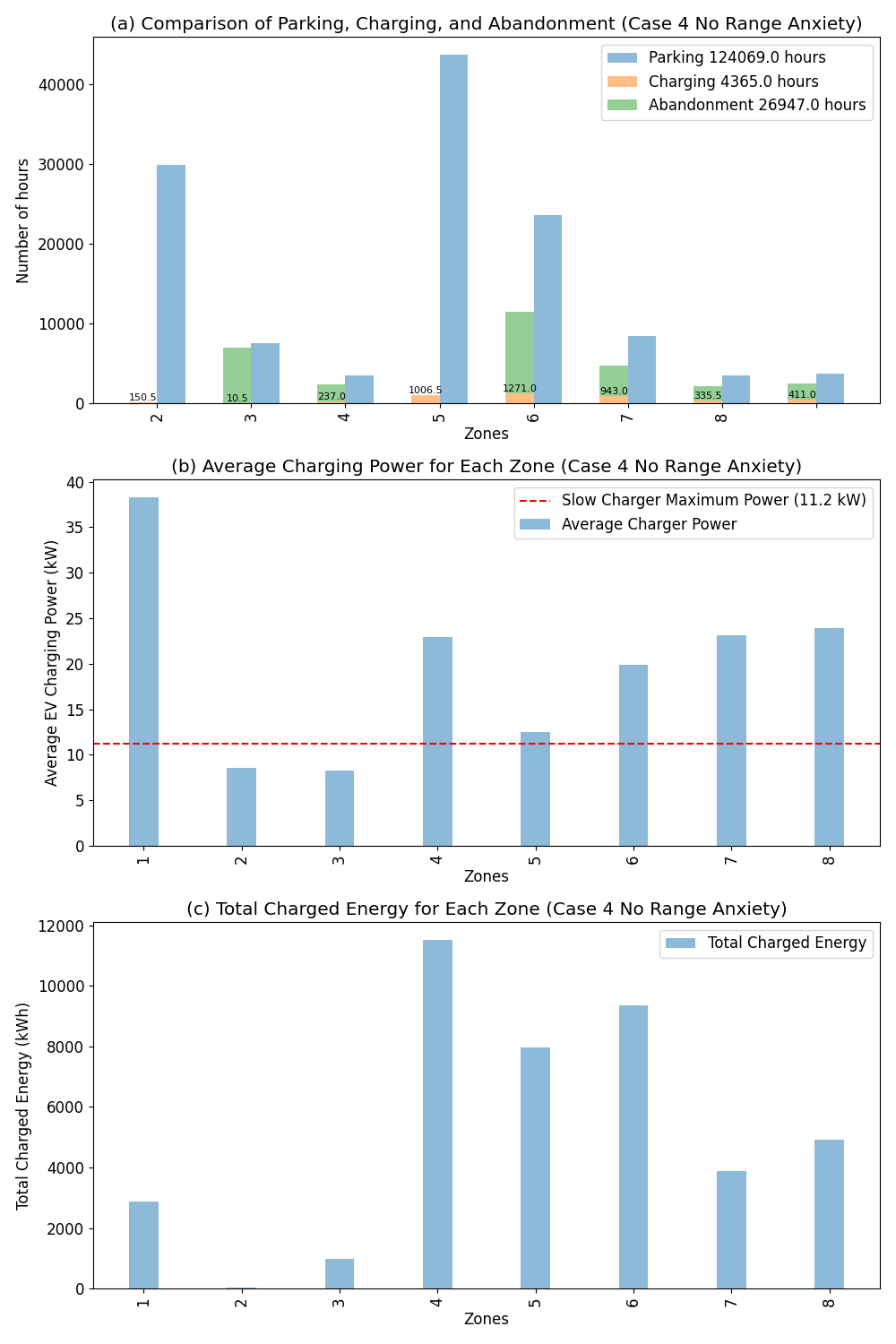}
\caption{The aggregated charging results of eight zones for case 4, no range anxiety: (a) The total hours of parking, charging, and abandonment of each zone during 39 representative days; (b) The average charging power for each zone during 39 representative days and (c) the total charged energy for each zone during 39 representative days}
    \label{combined_parking_times_no_range_anxiety}
\end{figure}

\begin{figure}[!h]
    \centering
    \includegraphics[width=\linewidth]{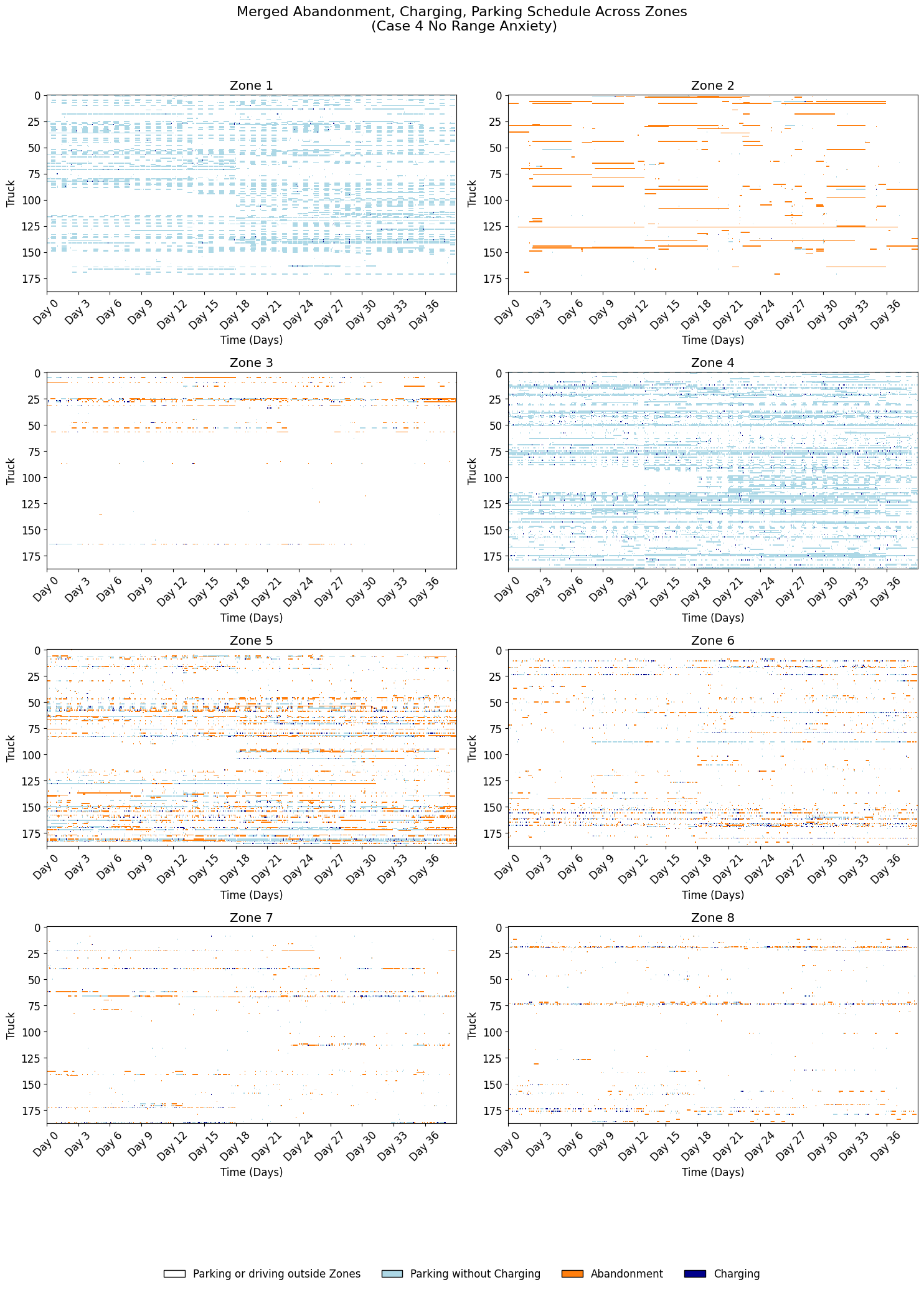}
\caption{The individual charging scheduling of 188 trucks at eight charging zones, arranged from left to right and top to bottom, in Case 4 no range anxiety}
    \label{four_stochastic_instances_by_months_no_range_anxiety}
\end{figure}

\FloatBarrier

\section{The detailed procedure of the fix-and-optimize algorithm}
\label{fix_and_optimize}

This section presents the detailed iteration procedures using the fix-and-optimize algorithm. Figs. \ref{fix_and_optimize_process_1} and \ref{fix_and_optimize_process_2} present changes to the enforced fast-charging constraints and the resulting solving time for the 39-day instance. Figs. \ref{charger_evolution_process_1} and \ref{charger_evolution_process_2} present the evolutions of the number of slow and fast chargers in each iteration. 

\begin{figure} [!htbp]
    \centering
    \includegraphics[width=4in]{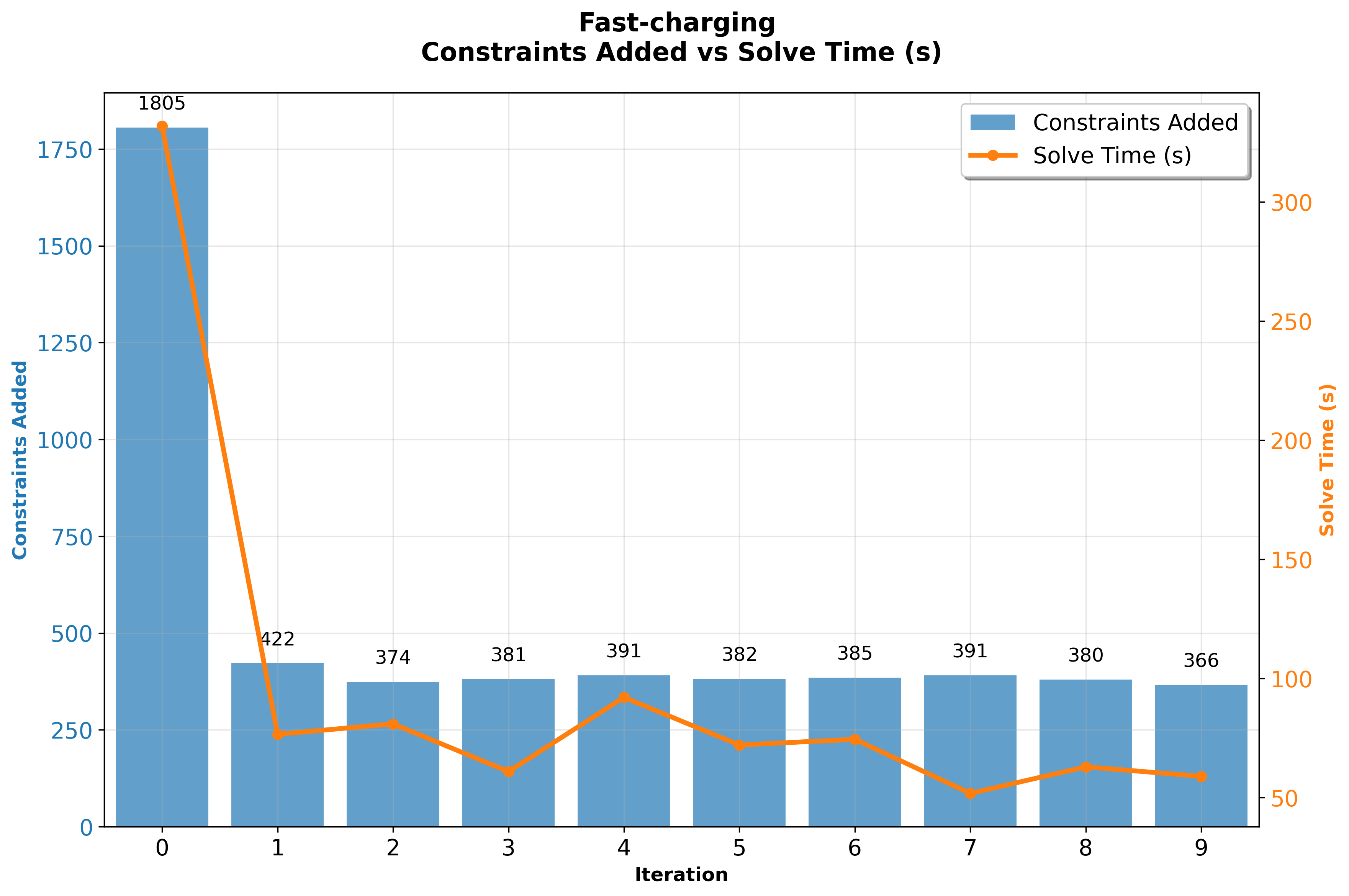}
\caption{The number of fast-charging constraints incorporated into the scheduling during each iteration and the corresponding solving time for the 39-day instance after applying the constraints with a warm start under the setting $\Delta_c = \pm2\sigma$ without DDU}
    \label{fix_and_optimize_process_1}
\end{figure}

\begin{figure} [!htbp]
    \centering
    \includegraphics[width=\linewidth]{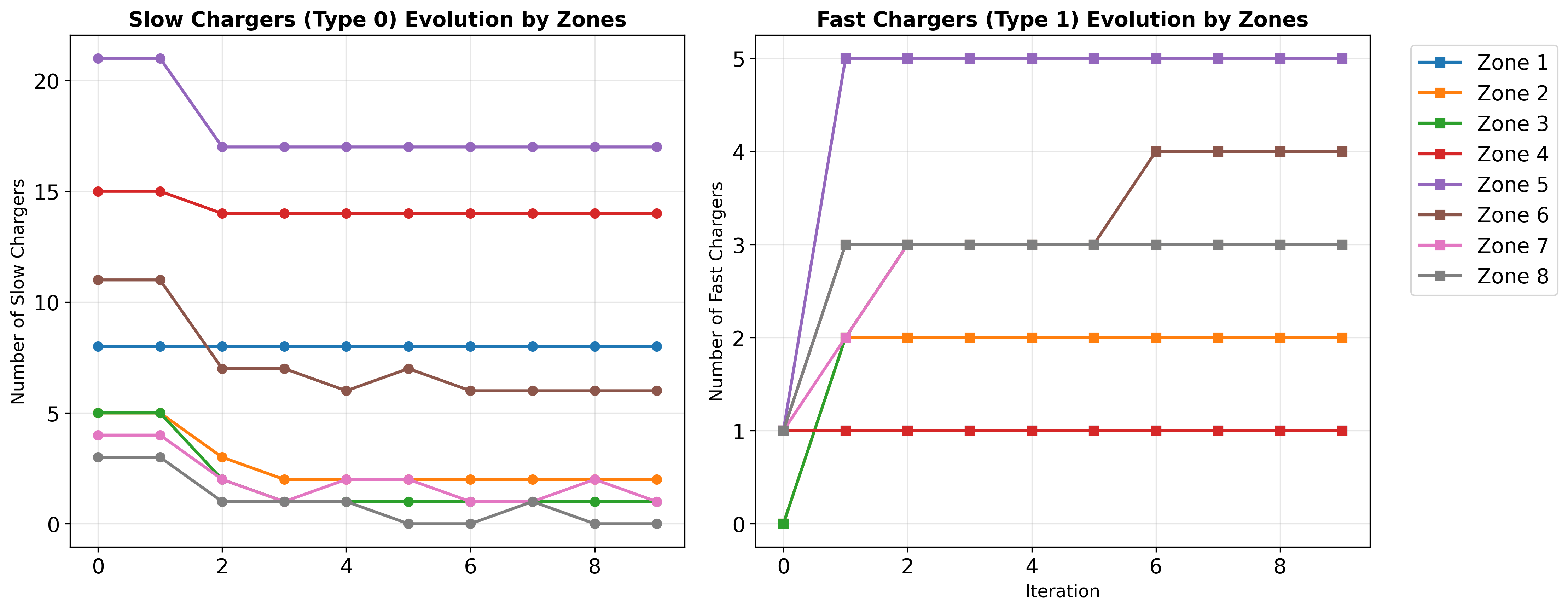}
\caption{Evolution of the number of slow and fast chargers in each iteration under the setting $\Delta_c = \pm2\sigma$ without DDU}
    \label{charger_evolution_process_1}
\end{figure}

\FloatBarrier

\begin{figure} [!htbp]
    \centering
    \includegraphics[width=4in]{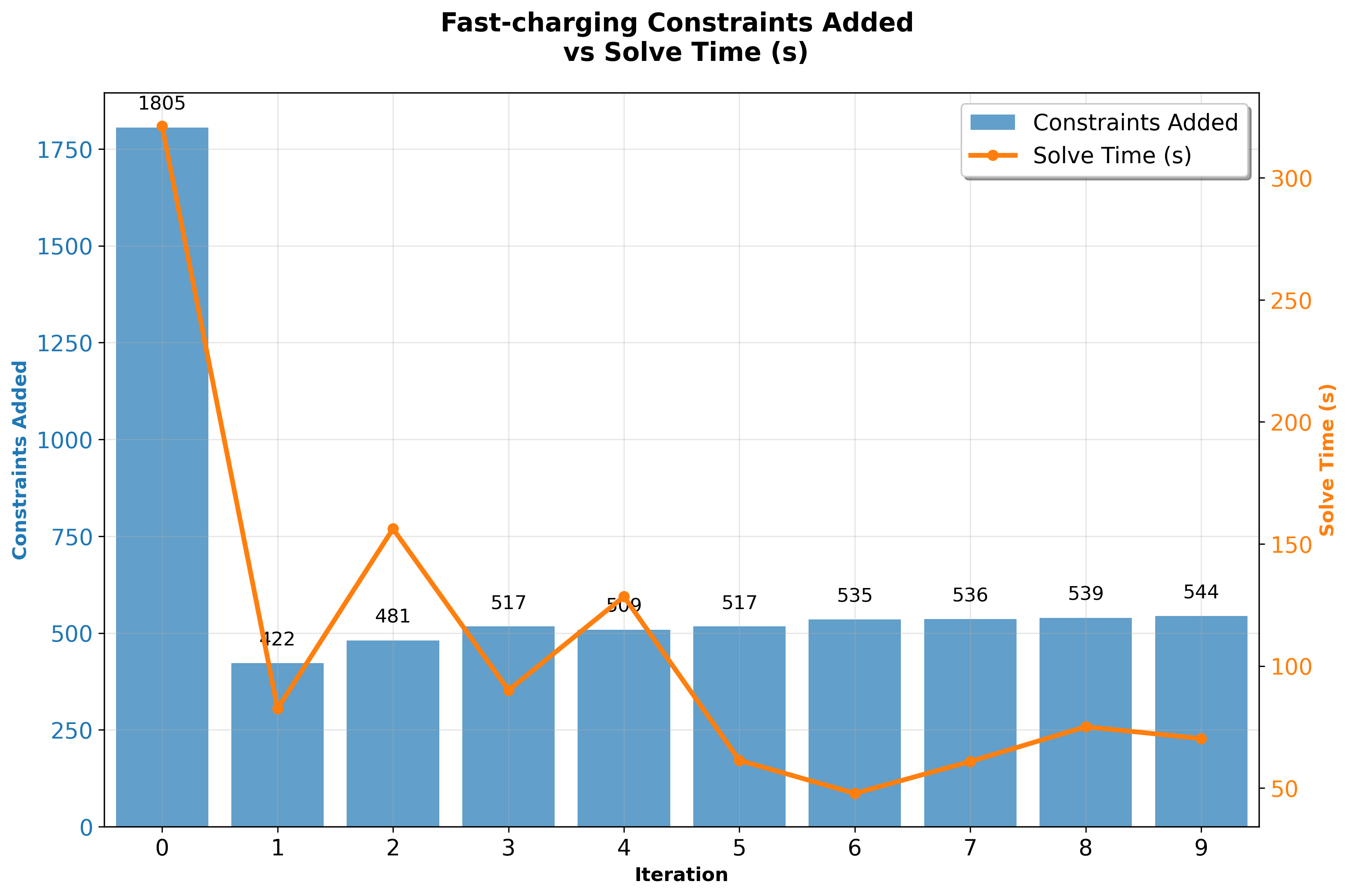}
\caption{The number of fast-charging constraints incorporated into the scheduling during each iteration and the corresponding solving time for the 39-day instance after applying the constraints with a warm start under the setting of $\Delta_c = \pm2\sigma$ with DDU (Expect formulation)}
    \label{fix_and_optimize_process_2}
\end{figure}

\begin{figure} [!htbp]
    \centering
    \includegraphics[width=\linewidth]{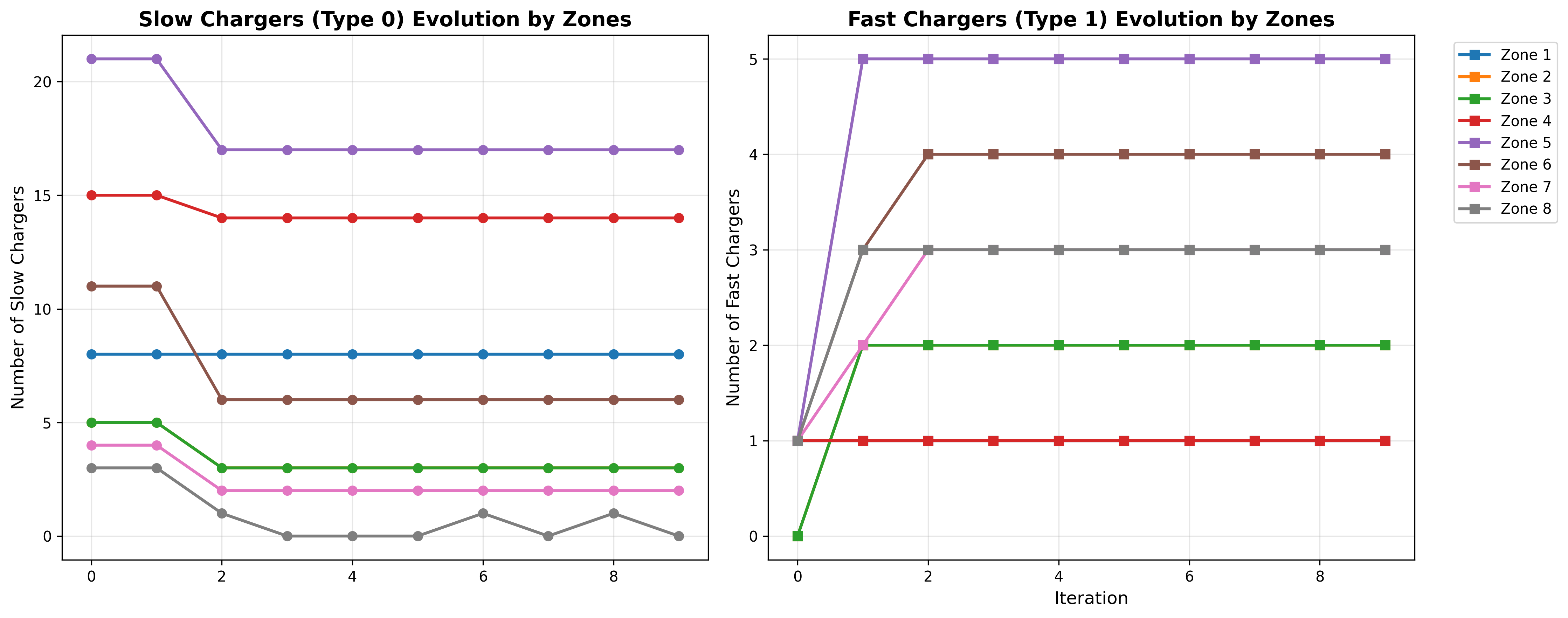}
\caption{Evolution of the number of slow and fast chargers in each iteration under the setting of $\Delta_c = \pm2\sigma$ with DDU (Expect formulation)}
    \label{charger_evolution_process_2}
\end{figure}

After obtaining the charger planning results in the final iteration, we post-processed the results by discarding the fast charger results greater than 2, and the rest of fast-charging installation results are taken as hard constraints. 
\end{APPENDICES}
\end{document}